%% file: doc_river_Langmuir.tex
\documentclass{jfm}

\RequirePackage{packages}
\RequirePackage{newcommands}

\newcommand{\arev}[1]{#1}

\ifCUPmtlplainloaded \else
\checkfont{eurm10}
\iffontfound
\IfFileExists{upmath.sty}
{\typeout{^^JFound AMS Euler Roman fonts on the system,
using the 'upmath' package.^^J}%
\usepackage{upmath}}
{\typeout{^^JFound AMS Euler Roman fonts on the system, but you
dont seem to have the}%
\typeout{'upmath' package installed. JFM.cls can take advantage
of these fonts,^^Jif you use 'upmath' package.^^J}%
\providecommand\upi{\upi}%
}
\else
\providecommand\upi{\upi}%
\fi
\fi

\ifCUPmtlplainloaded \else
\checkfont{msam10}
\iffontfound
\IfFileExists{amssymb.sty}
{\typeout{^^JFound AMS Symbol fonts on the system, using the
'amssymb' package.^^J}%
\usepackage{amssymb}%
  \let\leq=\leqslant
  
}{}
\fi
\fi

\ifCUPmtlplainloaded \else
\IfFileExists{amsbsy.sty}
{\typeout{^^JFound the 'amsbsy' package on the system, using it.^^J}%
\usepackage{amsbsy}}
{\providecommand\boldsymbol[1]{\mbox{\boldmath $##1$}}}
\fi

\title[]{
\arev{
Langmuir-type vortices in boundary layers driven by a 
criss-cross
 wavy wall topography
}
}
\author[A.~H. Akselsen and S.~\AA. Ellingsen]{Andreas H.\ Akselsen$^1$\thanks{Email address for correspondence: andreas.h.akselsen@ntnu.no}and Simen \AA. Ellingsen$^1$
}
\affiliation{$^1$Department of Energy and Process Engineering, Norwegian University of Science and Technology, N-7491 Trondheim, Norway}
\date{\today}           

\begin{document}

%
%

\maketitle

\begin{abstract}

\arev{
We investigate a mechanism to manipulate wall-bounded flows whereby wave-like undulations of the wall topography drives the creation of bespoke longitudinal vortices. A resonant interaction between the ambient vorticity of the undisturbed shear flow and the undulation of streamlines enforced by the wall topography serves to slightly rotate the spanwise vorticity of the mean flow into the streamwise direction, creating a swirling motion, in the form of regular streamwise rolls. The process is kinematic and  essentially identical to the `direct drive' CL1 mechanism for Langmuir circulation (LC)  proposed by \citet{Craik_1970_Langmuir_myidea}. Boundary layers are modelled by selecting suitable primary flow profiles. A simple, easily integrable expression for the cross-plane stream function is found in two asymptotic regimes: the resonant onset of the essentially inviscid instability at early times, and the fully developed steady state viscous flow. }%
Linear-order solutions for flow over undulating boundaries are obtained, fully analytical in the special case of a power-law profile. These solutions allow us to quickly map out the circulation response to boundary design parameters. The study is supplemented with direct numerical simulations which verify the manifestation of boundary induced Langmuir vortices in laminar flows with no-slip boundaries. Simulations show good qualitative agreement with theory. Quantitatively, the comparisons rest on a displacement length closure parameter adopted in the perturbation theory. While wall-driven LC appear to become unstable in turbulent flows, we propose that the mechanism can promote swirling motion in boundary layers, a flow feature which has been reported to reduce drag in some situations. 
\end{abstract}

\section{Introduction}
\label{sec:introduction}

The cellular vortex mechanism which causes the formation of `windrows' in open waters, commonly known as Langmuir circulation (LC) in honour of the pioneering observations by \citet{Langmuir_1938_original}, has come to be recognised as a central mechanism 
for the mixing of waters beneath the surface of the ocean and lakes,  and in particular the evolution of ocean thermoclines \citep{Li_1995_Langmuir_circulation_and_mixing_layer_Science,leibovich83,thorpe_2004_Langmuir_circulation_review}.
Recent studies strongly indicate that Langmuir rolls are more important to mixing in the upper ocean than previously thought, creating downward jets of surface water reaching far deeper than the surface wave motion itself \citep{belcher_2012_Langmuir_turbulence_and_the_ocean}. 

The underlying mechanics driving the phenomenon,
commonly know as Langmuir circulation in honour of the pioneering investigation by Langmuir(1938),
remained a mystery for decades until%
\citet{Craik_1970_Langmuir_myidea} proposed a nonlinear kinematic interaction surface waves and sub-surface mean  mechanism involving shear. 
Although greatly expanded upon over the years to follow,
particularly through the combined efforts of Craik and Leibovich \citep[notably][]{craik_leibovich_1976,leibovich_1977_averaveNS_IVP,craik_1977_CL2,craik_1982_GLM_O0_current,leibovich83}, the paper demonstrated a fundamental resonance mechanism occurring as nonlinear Lagrangian transport twists vortex lines in accordance to Helmholtz's vortex theorem. 
This redirects the orientation of vorticity in the flow to generate circular motion orthogonal to the current.
The model proposed in \citet{Craik_1970_Langmuir_myidea} assumed a multi-directional wave field in combination with a unidirectional current---an interaction which generates a `direct-drive' vortex forcing. 
Today, this interaction scenario is commonly referred to as the `CL1' mechanism.

In the research that followed Craik's 
proposition, an alternative instability mechanism, based on similar principles, was also revealed \citep{craik_1977_CL2}.
That mechanism, since termed the `CL2' mechanism, comes about as a result of a positive feedback where vortex motion increases a spanwise unevenness in the current profile, and 
is held to be the dominating Langmuir mechanism in the ocean \citep{thorpe_2004_Langmuir_circulation_review}.

More recently, the advancement of \citeauthor{andrews1978_GLM}'s (\citeyear{andrews1978_GLM}) theory of a generalised Lagrangian mean (GLM) has facilitated the study of the LC phenomenon in strongly sheared flows \citep{craik_1982_GLM_O0_current,phillips_1994_Langmuir_stability,phillips_1998_viscous_CLg}.
Langmuir vortices have been observed in hilly terrain by \citet{gong_CL2_hilly_terrain_experimetal}, their nature in connection with strong shear CL2  instability realised by \citet{phillips_1996_Langmuir_qoverz_profile}.
For clarity and brevity, we use the terms `Langmuir circulation' and `CL' mechanism loosely, also for strongly sheared flows; there exists a
near-perfect analogy  between conventional LC beneath a water surface, and a circulation phenomenon occurring above a rigid, wavy topography, where no-slip at the wall replaces wind above the ocean surface to create the necessary mean shear near the boundary.
\arev{  
It should however be noted that the scaling which underlay CL theory, as it is found in the LC of oceans and lakes, does not carry over to wall boundary layers.
}

\arev{
CL2-type instability in $O(1)$ shear flows, sometimes called CL2-O(1) instability, 
is often referred to as a wave catalysed instability, as opposed to wave driven,
as it is fed from the vorticity in the primary flow,
allowing the spanwise-periodic waves to outgrow the catalysing streamwise-periodic wave
\citep{phillips_2005_Langmuir_spacing}. 
The streamwise velocity perturbation thus generated becomes
strong enough to distort the primary flow field and modulate wave growth.
There is then a two-way coupling between wave field and shear flow which increases the complexity of the problem. 
We will in section~\ref{sec:GLM} consider 
how GLM theory and the phenomenon of wave distortion relates to strongly sheared CL1-type circulation.
We will discuss this in more detail section~\ref{sec:GLM},
relating prominent features of strong shear CL2 instability to the CL1 counterpart using GLM theory.
}

\arev{Although multiple studies have been conducted regarding 
the CL2-type instability near two-dimensional wavy boundaries \citep{phillips_1996_family_of_zq_CL2_instability,phillips_2005_Langmuir_spacing,phillips_2014_Langmuir_shallow_water},
no attention has to our knowledge
}%
been given to the potential of forced CL1-type circulation through constructional design.
This forms the backdrop for the present study; 
we will follow in the footsteps of \citet{Craik_1970_Langmuir_myidea}, demonstrating the shape and intensity of Langmuir vortices 
over bounding walls whose shape
provides a `direct-drive' mechanism kinematically identical  to CL1.

The mechanism we document suggests a new avenue to explore for 
manipulation of wall bounded flows.
Recent research indicates that the imposition of vortex structures and travelling waves can, in some situations, serve to reduce the drag experienced by the flow.
This is indeed an attractive prospect, particularly considering that friction loss in flow transport networks accounts for approximately 10\% of the global electric energy consumption \citep{kuhnen_2018_destabilizing_of_turbulence}.
\arev{
Research shows that intentionally imposed near wall flow structures, such as longitudinal `streaks' or `hairpins', 
can stabilise the overall flow regime and delay or prevent transition into turbulence 
\citep{Cossu_2002_stabilization_using_streaks,%
Cossu_2004_TS_waves,%
fransson_2005_TS_wave_stabilizaiton_experimental,%
Fransson_2006_delay_turbulence_transition_PhysRevLett,%
du_2000_suppressing_turbulence_by_means_of_traveling_wave_Science,
Pujals_2010_drag_reduction_experimental_wall,%
Pujals_2010_drag_reduction_experimental_body%
}.
A flow which has already become turbulent will generate coherent low-speed streak structures naturally, spawning and collapsing in the turbulent wall region.
As much as 80\% of the turbulent drag can be traced back to these streaks which
}
undergo a rapid cycle of events where they 
occasionally rupture from the wall (an event known as a `burst') causing the subsequent transport of fast-moving fluid back into the wall region 
\citep{corino_1969_burst_events,offen_1974_burst_events}.
Large scale vortices have been found capable of 
reducing the frequency and intensity of such burst events.
\citep{%
Schoppa_1998_drag_control%
,quadrio_2009_streamwisetravelling_waves_of_spanwise_wall_velocity_for_turbulent_drag_reduction%
,moarref_2010_traveling_wave_onset_of_turbulence_part1%
}.
\arev{
\citet{willis_2010_optimally_amplified_streak_drag_rediction_PhysRev}
demonstrates a mechanism for drag reduction in pipe flows where big vortices spanning the pipe radially transport turbulent streaks away from the wall region and into the bulk of the flow. 
}
Tested methods for generating vortices or travelling waves include
jets \citep{iuso_2002_drag_reduction},
blowing and suction \citep{Lieu_2010_traveling_wave_onset_of_turbulence_part2}
and
rotating pipe slabs \citep{auteri_2010_experimental_assesment_of_drag_reduction}.

Using specially designed surface roughness to modify
wall bounded flow,
often motivated by biomimicry, has been widely employed and has been shown to be capable of reducing
reducing turbulent drag \citep{dean_2010_shark_skin_surface_review,sirovich_1997_drag_reduction_passive_mech_Nature}.
Secondary vortex motion, aligned with the flow, can be generated via spanwise intermittent roughness patches \citep{anderson_2015_secondary_flow_due_to_roughness_patches,willingham_2015_secondary_flow_due_to_roughness_patches}
or streamwise-aligned obstacles \citep{yang_2018_topography_drive_secondary_flow,vanderwel_2015_spacing_of_lego_strips_and_secondary_flwo,kevin_2017_secondary_flow_herringbone_riblets,sirovich_1997_drag_reduction_passive_mech_Nature}.
\citet{anderson_2015_secondary_flow_due_to_roughness_patches} demonstrated that these structures are related to Prandtl's secondary flow of the second kind, driven and sustained by spatial gradients in the Reynolds-stress components.
\citet{chan_2015_pipe_egg_carton_roughness} and 
\citet{chan_2018_pipe_egg_carton_roughness_secondary_flow}
studied turbulent flows over an `egg carton' roughness configuration using direct numerical simulation, which can be seen as a special case of the wall geometry examined in the present paper. They too report circulation aligned with the flow attributed to Prandtl's secondary flow of the second kind.
This is a dynamic mechanism, as opposed to Langmuir circulation which is kinematic in nature.

Vortex generation can also be beneficial in the transport of  multiphase fluids.
For example, the pressure drop in a flow of two immiscible liquids of different viscosity (oil--water, say) will be sensitive to emulsion topography. 
A viscous phase entrained in a less viscous continuous phase will generate less wall friction than were it the other way around. 
\citep{schumann_2016_liquid_liquid_with_mixing,Piela_2008_phase_inversion_process,nadler_1997_phase_inversion_and_pressure_drop}.
\\

The paper is structured as follows: 
governing equations are presented in section~\ref{sec:governing_eq}.
A means for numerically computing a linearised solution for an arbitrary unperturbed shear current profile is given in section~\ref{sec:linear_sol}.
The particular second-order interaction of harmonics responsible for the generation of Langmuir vortices are considered in section~\ref{sec:Langmuir} and a solution for this motion derived in the limits of inviscid transient flow and viscous stationary flow. 
The computation procedure is summarized in section~\ref{sec:Langmuir:summary}.
Recent years has seen the use of GLM theory 
in the study of CL2-type stability.
In section~\ref{sec:GLM} we look into how this theory applies to the CL1-type mechanism, rediscovering the result equation derived in section~\ref{sec:Langmuir}. 
Results are given in  section~\ref{sec:results}.
Here we first look at free-surface flows in section~\ref{sec:results:free_surface} and then wall-bounded flows in section~\ref{sec:results:Langmuir_river}.
Simulation results are presented in section~\ref{sec:results:LBM} and a summary is provided in \ref{sec:summary}.

\section{Model Equations }
\label{sec:governing_eq}

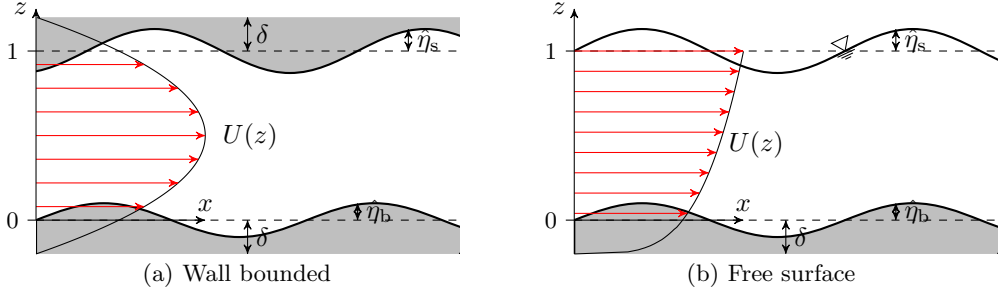
\begin{figure}%
\centering
\subfigure[Wall bounded]{\input{./figures/drawing_doublewall.tex}
\label{fig:schematic:doublewall}}
\hfill
\subfigure[Free surface]{
\input{./figures/drawing.tex}
\label{fig:schematic:free_surface}}
\caption{Sketch of the problem setup.
\arev{(Magnitudes of $\etas$ and $\etab$ are exaggerated.)}
}%
\label{fig:schematic}%
\end{figure}

The model considered in this work applies to incompressible flows where, whenever a free surface is present, surface tension is ignored. 
The problem is readily converted to nondimensional form using the depth or wall-to-wall distance $h$ and surface current peak velocity $\Us$, as follows:
\begin{align}
(x,y,z)&\mapsto (x,y,z)\,h,
&
\bk&\mapsto \bk\,/h,
&
t & \mapsto t\, h/U_0,
\nonumber\\
\p \bu\tot &\mapsto  \p \bu\tot\,\Us,
&
\p p\tot &\mapsto \p p\tot\, \rho \Us^2,
&&
\end{align}
$\bk=(k_x,k_y)$ being the wavenumber in the surface plane, $\p\bu\tot$ the fluid velocity and $\p p\tot$ the pressure.
Hatted quantities pertain to real space with the Fourier (wave-vector) space counterparts written without hats.
Flows over sinusoidally shaped boundaries of low steepness are in focus in the present study, as sketched in figure~\ref{fig:schematic}, yet any bathymetry shape can be analysed by superposition according to Fourier's theorem.
A slip velocity is assumed at the two boundary reference planes $z=0$ and $1$,
here achieved by stretching commonly used velocity profiles a displacement length $\delta$ such that the stagnation points of the bulk current $U(z)$ fall outside of the interior domain.
\arev{
Small undulations observed in streamlines near an actual corrugated wall
can then approximately be
modelled as the displaced wall sketched in figure~\ref{fig:schematic}.
Crude as this approximation may appear in terms of boundary layers,
it suffices for our purposes because the nature of
the vortex mechanism to be studied is kinematic, not dynamic, and does not rely  
on wall friction (other than that needed for generating the principal shear).
}
The boundaries, located at $z=\petab(\bmr)$ and $z=1+\petas(\bmr)$,
perturb and redirect current energy to undulate  the velocity field within.
Here, $\bmr=(x,y)$ is the horizontal coordinate.

The problem to be solved consists of the Navier-Stokes equations, along with a kinematic condition at the boundaries;
\begin{subequations}
\begin{alignat}{2}
\!\!\left.
\begin{aligned}
\sqbrac{\pdiff {\p\bu\tot}t}	+ 
( \p\bu\tot \cdot \p\nabla)\p \bu\tot + \p\nabla \p p\tot  &= 
\sqbrac{\Rey\inv \p\nabla^2\p\bu} -  \Fr^{-2} \bm e_z\\
\p\nabla \cdot \p \bu\tot &=0
\end{aligned}
\right\};&
\quad& \petab(\bmr) &\leq z\leq 1+\petas(\bmr), \label{eq:problem:Euler}\\
\p \bu\tot \cdot \p\nabla  \petab = \p w\; ;
& & z &= \petab(\bmr), \label{eq:problem:BC_b}\\
\p \bu\tot \cdot \p\nabla  \petas = \p w\; ;
& & z &= 1+\petas(\bmr), \label{eq:problem:BC_s}
\end{alignat}
\label{eq:problem}%
\end{subequations}%
with $\p\nabla=(\pp_x,\pp_y,\pp_z)$.
The Froude number is $\Fr=\Us/\sqrt{g h}$  and
$\Rey= \Us h/\nu$ is the Reynolds number
($g$ being the gravitational acceleration and $\nu$ the kinetic eddy viscosity).
The viscous and transient terms placed in square brackets are considered only in relation to the resonant second-order mode interaction---first-order solutions are assumed inviscid and at a steady state.
When the upper surface is free, $\p\eta$ is a function of the local flow and the inviscid dynamic boundary condition that pressure be constant at the surface is imposed. 

\section{Linearised solution}
\label{sec:linear_sol}

Assume that the motions attributed to the perturbed boundaries are small compared to the current velocity $U\of z$.
Separating these, we have
\begin{subequations}
\begin{align}
\p\bu\tot\of{\bmr,z} &= 
\bm U(z)
+  \p\bu\of{\bmr,z,t};
\quad \bm U = (U,0,0),
\label{eq:bu}
\\
\p p\tot\of{\bmr,z} &= \Fr^{-2}(1-z) +  \p p\of{\bmr,z,t}.
\label{eq:p}
\end{align}%
\label{eq:bu_p}%
\end{subequations}%
A constant reference pressure term has been set to zero. 
A steady state of the first-order motion can be reached without the influence of viscosity.
Assuming such a state has been reached and the Reynolds number  be large, 
we neglect the transient and viscous terms in the first-order motion.
(Note that viscous effects in the steady state are indirectly modelled via the shape of $U(z)$).
Let the boundary topography $\p\eta$ be described as a superposition of modes $\eta(\bk)\rme^{\rmi \bk\cdot\bmr}$ (real when summed), $\eta\in\{\etab,\etas\}$.
The flow field $\{\p\bu,\p p\}(\bm r,z)$ of the linear solution becomes a superposition of these wave modes $\{\bu,p\}(\bk,z)$.
After linearisation and insertion of \eqref{eq:bu_p},
the linearised Euler equations read
\begin{subequations}
\begin{align}
\rmi k_x U \bu 
+ \bm U'(z) w
+\nabla p &= 
0,
\label{eq:Euler:bu}
\\
\nabla\cdot \bu &= 0, \label{eq:Euler:cont}
\end{align}
\label{eq:Euler}%
\end{subequations}%
where
$\nabla=(\rmi k_x,\rmi k_y,\pp_z)$.
Eliminating $u$, $v$ and $p$ then yields the Rayleigh equation
\begin{equation}
w''-\br{k^2+U''/U}w = 0,
\label{eq:Rayleigh}
\end{equation}
$k=(k_x^2+k_y^2)^{\frac12}$.

The kinematic boundary conditions \eqref{eq:problem:BC_b} and \eqref{eq:p}, 
linearised about the reference planes, read
\begin{subequations}
\begin{align}
w(0) &= \rmi k_{x}U(0) \etab,
\\
w(1) &= \rmi k_{x}U(1) \etas.
\end{align}
\label{eq:BC_kin_b}%
\end{subequations}%
(We describe in \citet{akselsen_ellingsen_2019_river} a procedure for extending the lower boundary condition to bathymetries of finite amplitude, but we will here consider only the linear conditions.)
If free-surface flow is considered then the additional linear dynamic condition 
\begin{equation}
\etas = \Fr^{2} p\of 1
\label{eq:BC_dyn_s}
\end{equation}
governs the upper boundary, determining $\etas$.

\subsection{
Numerical solutions for arbitrary current profiles
}
\label{sec:Uzq_ODE}
An arsenal of methods are at our disposal for evaluating the boundary value problem \eqref{eq:Rayleigh}--\eqref{eq:BC_dyn_s}. 
For example, \citet{phillips_1996_Langmuir_qoverz_profile} adopted a Galerkin method at this stage. 
The problem \eqref{eq:Rayleigh}--\eqref{eq:BC_dyn_s} is however rather simple and so we rather opt for a simple 
shooting technique.
 
Writing the Rayleigh equation \eqref{eq:Rayleigh} in terms of the pressure and integrate once yields
\begin{equation}
p''-2\frac{U'}{U}p'-k^2 p = 0.
\label{eq:Rayleigh_p}
\end{equation}
Next, substituting the vertical velocity component by the pressure component using \eqref{eq:Euler:bu} with \eqref{eq:BC_kin_b}, 
the boundary conditions on $p$ read 
\begin{align}
p'(0) &= [k_x U(0)]^2 \etab,
&
&
\begin{cases}
p'(1) = [k_x U(1)]^2 \etas; & \text{[fixed boundaries],}
\\
p'(1)-[\Fr\, k_x U(1)]^2 p(1) =0; & \text{[free surface].}
\end{cases}
\label{eq:BC_p}
\end{align}
\arev{Equations \eqref{eq:Rayleigh_p} and \eqref{eq:BC_p} reveal that $p$ is independent of the sign of $k_x$ and $k_y$ if $\eta$ is.}
\arev{The velocity components in terms of $p$ are }
\begin{align}
\arev u & \arev{=-\frac{p}{U} - \frac{U' p'}{k_x^2 U^2},}
&
\arev v & \arev{= - \frac{k_y p}{k_x U},}
&
\arev w & \arev{= \frac{\rmi p'}{k_x U}.}
\label{eq:uvw_of_p}
\end{align}
Integration of \eqref{eq:Rayleigh_p} is performed numerically using a standard ODE solver
\arev{and the guess for $p(0)$ adjusted according to the error in \eqref{eq:BC_p}}.
Computation time has never been found to exceed a second on a regular laptop computer, and
the procedure yields $p(z)$ and its derivative, from which the velocity components are directly retrieved using \eqref{eq:Euler:bu}.

\subsection{Analytical solution for a power law current profile}
\label{sec:Uzq_lin_sol}

Following \citet{phillips_1996_family_of_zq_CL2_instability} and \citet{phillips_1996_Langmuir_qoverz_profile},
we now consider the particular family of power law current profiles  where $U$ is proportional to $z$ raised to a power $q$.
\arev{
The power law allows for analytical solutions while representing a wide variety of primary flow profiles.
}
We assume the free surface geometry of figure~\ref{fig:schematic:free_surface}.
Several forms of the commonly studied current profiles are obtained with different values of $q$; 
the uniform current most commonly considered is then recovered by setting $q$ to zero.
A linear current profile is recovered when $q=1$.
The linear profile has been investigated frequently in the literature since it is---in 2D---the only rotational flow for which potential theory is applicable.
In the intermediate range $0<q<1$, concave-up 
profiles 
of the kind sketched in figure~\ref{fig:schematic:free_surface},
resembling that observed in turbulent bottom boundary layer flows reside.
A range of intermediate exponent values have over the years been suggested for turbulent flows
\citep{Cheng_2007_power_laws,Chen_1991_power_laws}.
Flows with a surface shear layer may be modelled with $q > 1$, resulting in a class of concave-down profiles.

\citet{akselsen_ellingsen_2019_river} used a profile $U=z^q$ with the $z$ axis defined so that the lower boundary reference plane was at $z=\delta$.
This plane is in the current formalism located at $z=0$ and we instead stretch and shift the vertical axis;
\begin{equation}
U(z) = [\zeta(z)]^q, 
\quad 
\arev{
\zeta(z) = \frac{z+\delta}{1+\delta}; 
}
\qquad  [0\leq z\leq 1]. 
\label{eq:U}
\end{equation} 
\arev{
Inserting \eqref{eq:U}
and adopting the substitution $w=\sqrt\zeta W(Z)$; $Z=k(z+\delta)$,
the Rayleigh equation \eqref{eq:Rayleigh} is
remoulded into the modified Bessel equation
\begin{equation}
W''(Z)+\frac{W'(Z)}{Z}-\br{1+\frac{(1/2-q)^2}{Z^2}}W(Z)=0
\end{equation}
}
whose two linearly independent homogeneous solutions are known.
Written in terms of a generic flow variable $\phi$, we have
\begin{equation}
\phi =  \sum_\pm \cc^\pm  \phi^\pm 
\label{eq:velocity_field_O1}
\end{equation}
with
\begin{subequations}
\begin{align}
w^\pm(z) &= \rmi\sqrt \zeta I_{\pm(q-\frac12)}\of{\tilde k \zeta}, \label{eq:velocity_field_O1_pm:w}
\\
p^\pm(z) &=  \frac{k_x}{k}  \zeta^{q+\frac12} I_{\pm(q+\frac12)}\of{\tilde k \zeta}, \label{eq:velocity_field_O1_pm:p}
\\
\bu\_h^\pm(z) &= \frac{\rmi q \zeta^{q-1}  w^\pm\bm e_x-\tilde\bk p^\pm}{\tilde k_x \zeta^q},  \label{eq:velocity_field_O1_pm:uv}
\end{align}
\label{eq:velocity_field_O1_pm}%
\end{subequations}%
$I_\alpha$ being the modified Bessel function of the first kind of order $\alpha$ 
and $\bu\_h=(u,v)$ the horizontal velocity vector.
A stretched wave vector $\tilde\bk = \bk(1+\delta)$ has also been introduced.
The remaining constants $a^+$ and $a^-$ are determined by the two boundary conditions; \eqref{eq:BC_kin_b} and \eqref{eq:BC_dyn_s} yield the relationships
\begin{equation}
\cc^\pm = 
\begin{cases}
\rmi k_{x} 
\frac{\etas w^\mp(0)-U(0) \etab  w^\mp(1)
}{
w^\pm(1)w^\mp(0) -w^\pm(0)w^\mp(1)},
&\text{[fixed boundaries]},
\\
\rmi k_{x} U(0) \etab
\sqbrac{
w^\pm(0)
-\frac{w^\pm\of 1-\rmi k_x  \Fr^2 p^\pm\of 1}{w^\mp\of 1 - \rmi k_x  \Fr^2  p^\mp\of 1}w^\mp(0)
}^{-1},
&\text{[free surface]}.
\end{cases}
\label{eq:coeffs_O1}
\end{equation}
Further details on the analytical free surface solution is given in \citet{akselsen_ellingsen_2019_river}.

\section{
The resonant second order wave interaction
}
\label{sec:Langmuir}

\begin{figure}%
\centering
\includegraphics[width=.4\columnwidth]{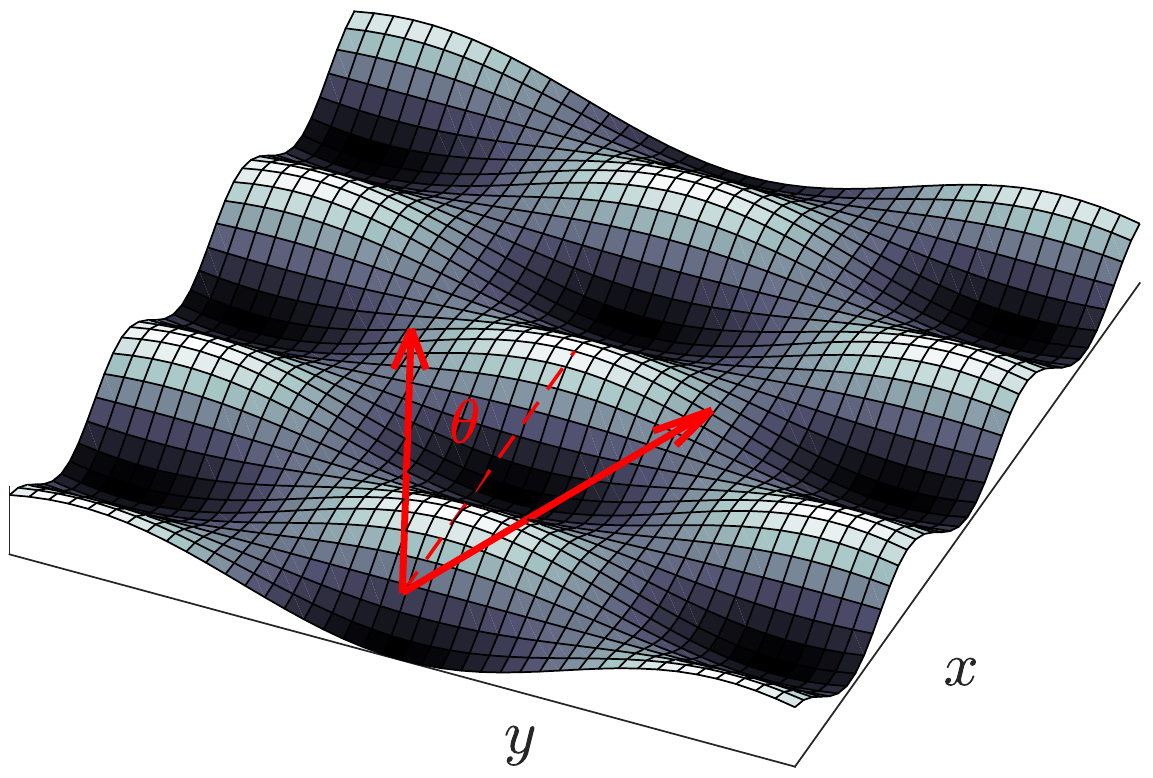}%
\includegraphics[width=.5\columnwidth]{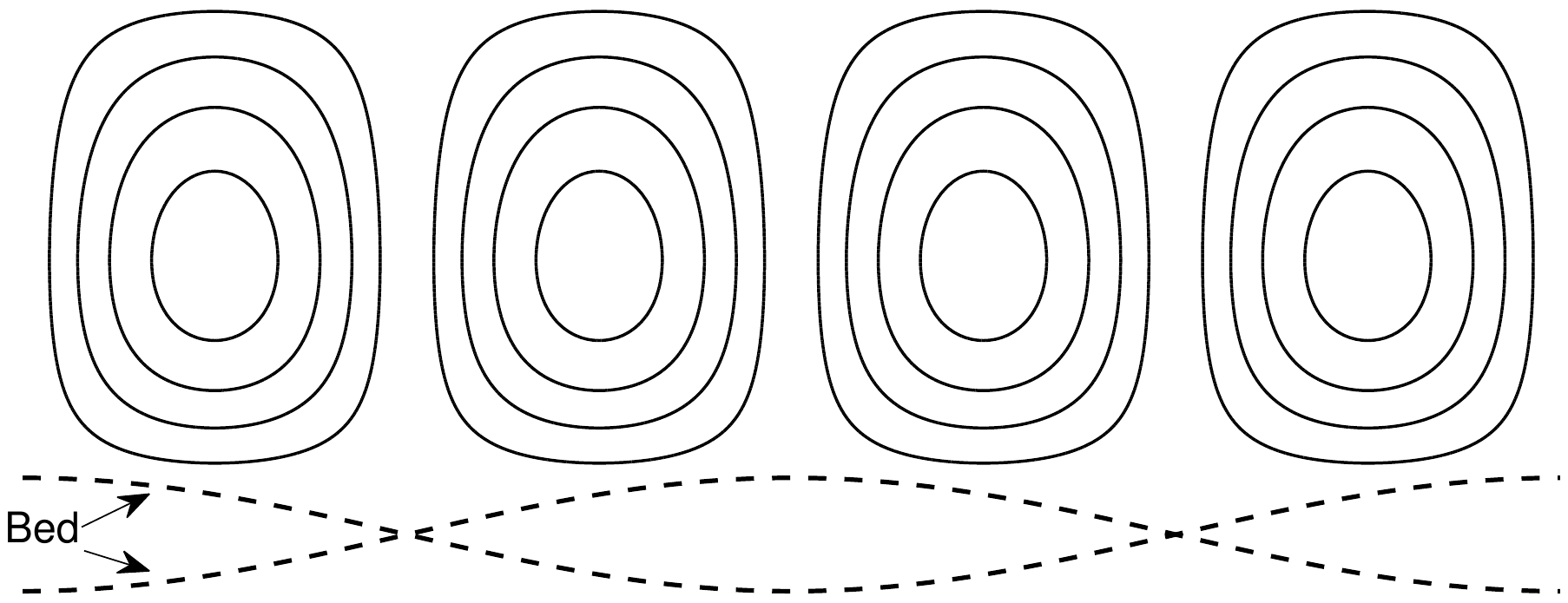}%
\caption{
Boundary topography and the orientation of vortices relative to it.
Left: bottom topography with wave-vectors $(\kxnil,\pm \kynil)$ indicated. Right: schematic illustration of swirling streamlines in the cross-flow plane.
}%
\label{fig:vortex_orientation}%
\end{figure}

Similar to \citet{Craik_1970_Langmuir_myidea}, we consider boundary undulations composed of a pair of sinusoidal waves directed symmetrically about the streamwise direction $x$:
\begin{equation}
\hat\eta = \frac{\etanil}{4}\left[\rme^{\rmi( \kxnil x + \kynil y)}+\rme^{\rmi( \kxnil x - \kynil y)} + \mathrm{c.c.} \right]
=\etanil\cos(\kxnil x)\cos(\kynil y)
\label{eq:O2_modes}
\end{equation} 
($\eta\in\{\etab,\etas\}$, $\etanil\in\{\etabnil,\etasnil\}$).
An illustration is shown in figure~\ref{fig:vortex_orientation}.
A free surface will to linear order also have the same functional form as the bathymetry, thus behaving kinematically similar to an undulated upper wall with the important difference that the the mean current and secondary flow see this as a full-slip boundary.
The first-order wave modes involved each have amplitudes $\etanil/4$ and the four wave vectors $(\pm \kxnil,\pm \kynil)$ (sign of each component is varied independently). Second order harmonics, in turn, have wave vectors $\bm\kappa$ which are sums of pairs of these. The harmonics thus come in four different types with wave vectors $\bm \kappa=\pm 2(\kxnil,\kynil)$, $\bm \kappa = 0$, $\bm \kappa=(\pm 2\kxnil,0)$ and $\bm \kappa = (0,\pm 2\kynil)$. The last type of harmonic is resonant---see \citet{Craik_1970_Langmuir_myidea} and \citet{akselsen_ellingsen_2019_ringwaves}  for further details.
The resonance will manifest in the formation of vortex pairs aligned in the current flow direction, as illustrated in figure~\ref{fig:vortex_orientation}.

From here we adopt $\bu$ and $p$ as the second-order components unless otherwise stated.
The equations of motion for the second-order components of the Stokes expansion read
\begin{subequations}
\begin{align}
(\pp_t-\Rey\inv\nabla^2)\bu 
+ \arev{\bm U'} w 
+ \nabla p &= 
-[(\bu\cdot\nabla)\bu]\crossnil  
\label{eq:EulerO2:mom}
\\
\nabla \cdot\bu&=0   \label{eq:EulerO2:cont}
\end{align}%
\label{eq:EulerO2}%
\end{subequations}%
for this particular interaction.
Left-hand variables are second-order components while the
right-hand cross terms are the second-order interactions of first-order components.
\arev{
Eliminating the horizontal velocity components and the pressure,  
the Orr--Sommerfeld equation for the second-order vertical velocity $w(z,t;\bknil)$ becomes
\begin{align}
(\pp_t-\Rey\inv\nabla^2)\nabla^2 w &= \mc R\of z; 
\label{eq:Rayleigh_O2}
\\
\mc R &=\rmi \bm \kappa \cdot \pp_z  [(\bu\cdot\nabla)\bu\_h]\crossnil + \kappa^2   [(\bu\cdot\nabla) w]\crossnil
\label{eq:R_general}
\end{align}
with $\nabla^2=\pp_z^2-\kappa^2$, $\kappa=2|\kynil|$. 
Due to symmetry, we have
$\nabla=(0,\rmi\kappa_y,\pp_z)$ and $[(\bu\cdot\nabla)u]\crossnil = 0$
which simplifies the right-hand expression \eqref{eq:R_general} significantly;
using \eqref{eq:uvw_of_p} to express the first order velocity component in terms of the first order pressure, we find the remarkably simple expression
\begin{equation}
\mc R(z) 
=  8 \frac{\kynil^2}{\kxnil^2}\frac{U'}{U^3} \sqbrac{\br{\kxnil^2-\kynil^2}\pnil^2 + (\pnil')^2}.
\label{eq:R}
\end{equation}
The first-order pressure mode $\pnil$ is the solution of \eqref{eq:Rayleigh_p}--\eqref{eq:BC_p} with $\bk=\bknil$.
}
Note that the curvature term $(U''/U)w$ is not present 
in \eqref{eq:Rayleigh_O2}
because $\kappa_x = 0$, i.e., the resonant wave is two-dimensional and orthogonal to the shear current.

For arbitrary profiles, considered in section~\ref{sec:Uzq_ODE}, the right-hand side of \eqref{eq:Rayleigh_O2}
is evaluated directly from the output of the ODE integrator. 
If the power law profile \eqref{eq:U} is adopted (section~\ref{sec:Uzq_lin_sol}) then the right-hand side can be written explicitly as
\begin{equation}
\mc R(z) = 
8 \frac{q \kynil^2 }{1+\delta} 
\sum_{s=\pm1}
\frac{\kxnil^2-s\kynil^2}{\knil^2}
\bigg\{\sum_{\pm} \cnil^{\pm} I_{\pm(q+s\frac12)}
\big[\knil(z+\delta)\big]
\bigg\}^2,
\label{eq:Rq}
\end{equation}
$\cnil^\pm$ and its velocity components are as given in \eqref{eq:velocity_field_O1}--\eqref{eq:coeffs_O1} with $\bk=\bknil$.

The resonant second-order harmonic is uniform in $x$ which allows for a stream function
\begin{equation}
\psi=\rmi w/\kappa_y
\label{eq:psi}
\end{equation}
to be expressed in the $yz$-plane.
An expression for the pressure component is given in 
the appendix.

The resonant wave interactions will come with both signs $\bm\kappa=(0,\kappa_y)=(0,\pm 2\kynil)$, with the corresponding modes being complex conjugates of each other.
It is therefore sufficient to consider only 
\arev{the positive spanwise wavenumber $\kappa_y=\kappa=2|\kynil|$}
and write, for any flow variable $\p\phi$,
\begin{equation}
\p\phi =  2 \Real(\phi)\cos(\arev\kappa y)  - 2\Imag(\phi) \sin(\arev\kappa y).
\label{eq:ReIm}
\end{equation}
The stream function $\psi$ is purely imaginary when $\etab$ and $\etas$ are real.
Contour plots of $- 2 \,\Imag\,\psi$ 
are therefore presented among the results of section~\ref{sec:results}.
\arev{
When $-\Imag\, \psi>0$,
the vertical flow is directed in towards lines aligned with peaks and troughs in the wall topography, with 
the horizontal velocity perturbation $\p u$ being strongest along these lines.
This serves to increasingly tilt vortex particle trajectories in the $xy$-plane with time. 
Vortex rotation is in the opposite direction when $-\Imag\, \psi<0$.
}
\\

In what follows, two limits of \eqref{eq:Rayleigh_O2} shall be investigated, namely the inviscid transient problem $ \Rey\inv = 0$ and the viscous steady state problem $\pp_t\,\cdot \rightarrow 0$.

\subsection{Inviscid transient problem}
\label{sec:Langmuir:inviscid}

We now set $\Rey\inv = 0$.
Because the harmonic in question is resonant it will
not reach a time-independent state without the intervention of viscosity or nonlinear dynamics.
The solution of \eqref{eq:Rayleigh_O2} can be obtained using the  method of variation of parameters which results in
\begin{align}
w &= \sum_\pm d^\pm \rme^{\pm \kappa z} + w\cross\of z;
&
w\cross\of z &= \frac{t}{\kappa} \int_0^z \!\dd \zz\, \mc R\of {\zz}  \sinh \kappa (z-\zz).
\label{eq:wO2_sol}
\end{align}
We have here imposed circulation quiescence at $t=0$.
$w\cross(z)$ is evaluated at all $z$ with simple numerical integration.

The second-order boundary conditions for 
fixed wall boundaries, Taylor expanded about $z=0$ and $z=1$,
reduce to 
\begin{equation}
w(0,t) = w(1,t) = 0.
\label{eq:BC_O2}
\end{equation}
A little more caution is required for a free surface boundary; 
kinematic and dynamic conditions then reduce to
\begin{align*}
\pp_t \etas-w =0;
\quad
\Fr^2 p -\etab = \Fr^4 \pnil \pnil',
\qquad [z=1].
\end{align*}
Combining these with the $y$-component of the second-order momentum equation \eqref{eq:EulerO2:mom} and the continuity equation \eqref{eq:EulerO2:cont} yield
\begin{equation}
\pp_t^2\pp_z w +\kappa_y^2 \Fr^{-2}w  = 0, \qquad [z=1].
\end{equation}
Vertical velocity $w$ is only linear in time so that the 
boundary conditions become \eqref{eq:BC_O2} also at a free surface.
Fixing the integration constants of \eqref{eq:wO2_sol}, our inviscid solution reads
\begin{equation}
w = w\cross\of z - w\cross(1)\frac{\sinh\kappa z}{\sinh \kappa}.
\label{eq:w_inviscid}
\end{equation}
The streamwise velocity component $u$ is retrieved directly from 
inserting \eqref{eq:w_inviscid} into
the momentum equation \eqref{eq:EulerO2:mom}. The interaction term disappears due to symmetry and we obtain
\begin{equation}
u(z,t)=-\frac{t}{2}U'(z) w(z,t),
\label{eq:u_inviscid}
\end{equation}
which is proportional to $t^2$.

\subsection{Viscous steady state problem}
\label{sec:Langmuir:viscous}
We now turn to the viscous problem assuming a steady state has been reached.
It should here be noted that the second-order solution constructed here is driven by the interaction kinematics of first-order waves which are themselves assumed to be inviscid; 
we assume that
the periodic dynamics of the first-order waves are affected little by viscosity at moderate to large Reynolds numbers, and that the role of viscosity at that order is in all essentials accounted for through the prescribed shear of $U(z)$. 
The resonant mode, on the other hand,
continues to grow in time until it becomes strong enough for viscosity to become significant, and the flow reaches a steady state.
The solution to \eqref{eq:Rayleigh_O2} when denying any time dependency can be found by applying the solution formula of the previous problem twice. 
It is
\begin{align}
w &= \sum_\pm (d_0^\pm + z d_1^\pm) \rme^{\pm \kappa z} + w\cross\of z;
\label{eq:wO2_sol_viscous}
\\
w\cross\of z &= \frac{\Rey}{2\kappa^3 } \int_0^z \!\dd \zz\, \mc R\of {\zz} G[ \kappa (z-\zz)];
& G(Z)&=\sinh(Z)-Z\cosh(Z).
\label{eq:wcO2_sol_viscous}
\end{align}
A full-slip condition is assumed at the upper boundary if considering a free surface, as is often reasonable for an atmospheric interface between water and air.
We find that all first-order cross terms cancel also for these stress conditions, even after Taylor expansion about $z=1$ (see also \citet{Craik_1970_Langmuir_myidea}).
Appropriate full-slip stress conditions are
$
v'+ \rmi \kappa_y w =  0
$
about the reference plane.
The other boundary conditions discussed in Sec~\ref{sec:Langmuir:inviscid} still hold; employing the continuity equation \eqref{eq:Euler:cont} yields
\begin{equation}
w=w''=0 \qquad[\text{full-slip}]
\end{equation}
at $z=0$ or $1$ if considering the upper or lower boundary, respectively.
We'll also consider no-slip boundaries,
which with \eqref{eq:EulerO2:cont} implies
\begin{equation}
w=w'= 0\qquad[\text{no-slip}]
\end{equation}
at $z=0$ or $1$.
With these conditions we have derived the 
integration constants $d_0^+$, $d_0^-$, $d_1^+$ and $d_1^-$ in \eqref{eq:wO2_sol_viscous} for the three cases full-slip/full-slip, no-slip/no-slip and full-slip/no-slip at the upper/lower boundary.
Explicit solutions for $w(z)$ are presented in Appendix~\ref{sec:AB}.

The horizontal velocity component of the vortex motion is again obtained by integrating the $x$-component of the linearised momentum equation 
\eqref{eq:EulerO2:mom}. 
One finds
\begin{equation}
u(z) = \sum_\pm d_u^\pm  \rme^{\pm\kappa z}
+ \frac{\Rey}{\kappa} \int_0^z
\!\dd\xi \, U'(\xi) w(\xi)\sinh\kappa (z- \xi).
\label{eq:u_O2}
\end{equation}
Appropriate no-slip and full-slip boundary conditions are $u=0$ and $u'=0$, respectively, evaluated at the appurtenant reference planes. 
(Cross terms merely represent the second-order correction of a first-order motion which itself does not support stress conditions.)
Coefficients $d_u^\pm$ are given in Appendix~\ref{sec:AB}.

\subsection{Summary of computations procedure}
\label{sec:Langmuir:summary}
In summary, we compute a solution for a general shear profile by 
first computing the linear solution for the first-order pressure $\pnil(z)$ through numerically integrating \eqref{eq:Rayleigh_p} subject to \eqref{eq:BC_p}.
From this, we compute from \eqref{eq:R} the right-hand term $\mc R(z)$ of the Rayleigh/Orr--Sommerfeld equation.
If considering a free surface flow with a power law profile, $\mc R(z)$ can instead be computed directly from \eqref{eq:Rq}.
The transient, inviscid circulation solution is then obtained from \eqref{eq:w_inviscid} after numerically integrating \eqref{eq:wO2_sol}.
This yields $w\cross$, or the stream function $\psi$ by virtue of \eqref{eq:psi}, whose contours in real space \eqref{eq:ReIm} represent streamlines. 
Spanwise velocity is the $z$-derivative of the stream function while the horizontal component is given by \eqref{eq:u_inviscid}.
Pressure can be computed from \eqref{eq:p_O2}.
In a viscous, steady state solution, equations \eqref{eq:wO2_sol_viscous} and \eqref{eq:u_O2} replace \eqref{eq:w_inviscid} and \eqref{eq:u_inviscid}, respectively. The full form of the resulting expressions are given in Appendix~\ref{sec:AB}.

\section{
\arev{
Comparison to strong-shear CL2-instability and 
the theory of the generalised Lagrangian mean 
}
}
\label{sec:GLM}

\arev{
We will start this section by relating our findings to GLM theory before considering features of strong shear CL1 instability compared to strong shear CL2 instability.

\citeauthor{andrews1978_GLM}'s (\citeyear{andrews1978_GLM}) theory of the generalised Lagrangian mean (GLM) has in recent years dominated in the study of CL2-type Langmuir circulation \citep{leibovich_1980_CL_from_GLM,craik_1982_O1,craik_1982_GLM_O0_current,phillips_1998_viscous_CLg,phillips_2005_Langmuir_spacing}.
This theory is commonly used to derive the eigenvalue problem that governs the stability of unidirectional waves to flows of longitudinal vortex form.
In order to connect our work with the GLM-based literature, we shall briefly discuss how key observations from  the GLM theory relates to the CL1-type circulation considered in the present paper.

The GLM Navier--Stokes equations, averaged in $x$, read
\begin{equation}
\olL D (\olL {\p u}_i - \p\sau_i) + \olL {\p u}_{j,i}(\olL {\p u}_j- \p\sau_j)+\p\pi_{,i} 
= \Rey\inv \sqbrac{ \olL{( \p\nabla^2\p u_i)}  + \ol{\p\xi_{j,i}(\p\nabla^2\p u_j)^l}}.
\label{eq:NS_GLM}
\end{equation}
Except for the symbol for the pseudomomentum 
\begin{equation}
\p\sau_i\equiv -\ol{\p\xi_{j,i} \p u_{j}^l}
\label{eq:pseudomomentum}
\end{equation}
and the use of hats to denote physical space variables, 
we have here adopted the common notation.
The operator $\olL{(\cdot)}$ is the average along a Lagrangian trajectory described by the displacement $\bm \xi$ of fluid particles. 
The fluctuating part of the Lagrangian velocity is denoted 
$\p\bu^l=\p \bu (\p{\bm x} + \p{\bm\xi},t)-\olL{\p\bu}$.
Gradient terms are lumped into $\p\pi$ which may be thought of as an effective pressure. 
Overlines denote the streamwise average, indices $\{1,2,3\}$ refer respectively to the spatial dimensions $\{x,y,z\}$, repeated indices are summed and comma denotes differentiation. 
We refer the reader to the above references, in particular \citet{andrews1978_GLM} and \citet{craik_1982_O1,craik_1982_GLM_O0_current}, for a more complete description.

Envisage a wave field of $O(\eps)$ generated by the wall topography. 
Only $O(\eps^2)$ terms can survive the streamwise averaging. 
To $O(\eps^2)$, \eqref{eq:NS_GLM} reduces to 
\begin{equation}
\pp_t (\olL{\p\bu}  - \p{\bm \sau})
+\p\nabla[U \olL{\p u}_1  + \p\pi-\tfrac12 U^2]
-\p\sau_1 \p\nabla U
+ \olL{\p u}_3 \bm U_{,3}  
- \Rey\inv\p\nabla^2\ol{\p\bu}
=0.
\label{eq:bmu0}
\end{equation}
The streamwise averaged Eulerian velocity, pseudomomentum and differential drift (Stokes drift) $\ol {\p\bu}{}\^S=\olL{\p\bu}-\ol{\p\bu}$ will at $O(\eps^2)$ have modes 
\[
[\bu(z,t),\bm\sau(z,t),\bu\^S(z,t)]\rme^{\rmi \kappa_y y}.
\]
The streamwise component of $\bu$ obeys, by \eqref{eq:bmu0},
\begin{equation}
(\pp_t - \Rey\inv \nabla^2 )u_{1} 
= \pp_t ( \sau_1 - u_1\^S) -  (u_3+u\^S_3) U_{,3}.
\end{equation}
The other components of \eqref{eq:bmu0},
together with continuity 
$\nabla\!\cdot\!\ol {\bm u}=0$,
can at $O(\eps^4)$ be reduced into an Orr--Sommerfeld type equation 
\begin{equation}
(\pp_t-\Rey\inv\nabla ^2)\nabla ^2  u_3
= - \kappa^2 \sau_1 U_{,3} - \pp_{t}[ \rmi \kappa_y (\sau_2-u_2\^S)_{,3} + \kappa^2   (\sau_3-u_3\^S)],
\label{eq:R_GLM}
\end{equation}
$ \nabla ^2=\pp_{zz}-\kappa^2$.
The second right-hand term vanishes when we next consider a time-independent primary flow.

For evaluating \eqref{eq:R_GLM} one must first find the linear displacement $\bm \xi$.
Following \citet{craik_1982_O1} we have derived the particle trajectories for time independent $U$, adopting Craik's method for matching trajectories to the appropriate streamlines.
Writing  $\p {\bm \xi}$ as a superposition of modes ${\bm \xi}\rme^{\rmi(k_x x + k_y y)}$ 
and using \eqref{eq:uvw_of_p} to express the $O(\eps)$ wave filed,
we eventually find the simple relation
\begin{equation}
\bm\xi= \frac{\nabla  p}{k_x^2 U^2}+O(\eps^2),
\label{eq:xi}
\end{equation}
$\nabla = (\rmi k_x,\rmi k_y,\pp_z)$.
Expression \eqref{eq:xi} implicitly assumes the absence of critical layers where $U=0$, as is common. 
Adopting this to compute the pseudomomentum \eqref{eq:pseudomomentum}, we get
\begin{subequations}
\begin{align}
\p\sau_1 &= -\frac{4}{k_x^2 U^3} \sqbrac{\br{k_x^2+k_y^2} p^2 + ( p')^2} -\frac{4\cos(2k_y y)}{k_x^2 U^3} \sqbrac{\br{k_x^2-k_y^2} p^2 + ( p')^2} + O(\eps^3),
\\
\p\sau_2 &= \p\sau_3 = 0.
\end{align}%
\label{eq:pseudomomentum_O1}%
\end{subequations}%
Inserting \eqref{eq:pseudomomentum_O1} into \eqref{eq:R_GLM} and matching it to \eqref{eq:Rayleigh_O2} successfully recovers \eqref{eq:R}.
GLM theory thus accords with the more primitive mode coupling theory utilized in section~\ref{sec:Langmuir}, 
although the benefits of GML theory are not as patent as in studies of CL2 instability.
The horizontally uniform part of the pseudomomentum can for $k_y\to\infty$ be compared to \citeauthor{craik_1982_O1}'s (\citeyear{craik_1982_O1}) equation (3.4a) 
with his $\phi =-4p'/\eps k_x^2 U$, whence we find that the spanwise periodic topography generates, for the same amplitude $a$,
half the pseudomomentum of the spanwise uniform  topography, in addition to the spanwise periodic part.
\\

When the shear intensity is of $O(1)$, the study of CL2-type instabilities
turns more complicated than its weak shear counterpart due to the influence of wave distortion \citep{craik_1982_GLM_O0_current,phillips_2005_Langmuir_spacing}. 
Wave distortion comes about as the pseudomomentum generated by the spanwise-periodic wave perturbation (here called the Langmuir wave) reaches a magnitude great enough to re-enter the 
eigenvalue problem governing the stability of the spanwise-periodic wave. 
The pseudomomentum generated by the spanwise-periodic wave itself must then be estimated in a separate analysis, first performed by \citet{craik_1982_GLM_O0_current}, yielding a coupled pair of differential equations governing the stability to longitudinal vortex form. This equation set has since been termed the `generalized Craik--Leibovich equations' (CLg), with the range of validity extending beyond instability from wind-driven surface waves and into boundary layer flows.

Wave distortion enters the CL1 problem less directly due to three prominent differences between the two mechanisms.
First, where the CL2 mechanism primarily consists of a distortion of the streamwise flow generating a spanwise-periodic wave,
the CL1 mechanism works by directly imposing a spanwise-periodic wave of prescribed wavelength.
Second, 
CL2 wave distortion is present immediately from the scaling of the initial perturbation of the primary flow, while the instability of CL1 is algebraic, growing linearly from out of a completely unperturbed state.
If not sufficiently suppressed by viscosity, the wave may eventually grow until wave distortion takes effect or other longitudinal vortex instabilities occur.
Third, where the spanwise-periodic wave in the CL2 mechanism generates pseudomomentum at the same wavelength, the CL1 equivalent is an interaction between two spanwise periodic wave fields, generating higher and lower spanwise harmonics. 

Wave distortion will eventually arise in CL1 if the viscosity is insufficient to restrain the circulation to a moderate intensity. 
This then occurs via pseudomomentum generated by the Langmuir wave. 
Only streamwise-periodic particle displacements and velocity perturbations can combine in \eqref{eq:pseudomomentum} to produce non-zero pseudomomentum (by construction, $\ol{\p{\bm\xi}}=0$).
The next order of pseudomomentum relevant in terms of wave distortion is then generated by \textit{two} primary $O(\eps)$ harmonics interacting with the Langmuir wave. 
The Langmuir wave must therefore be of $O(1)$ (and not $O(\eps)$) for this distortion to scale with the principal pseudomomentum \eqref{eq:pseudomomentum_O1}.
Such wave distortions will then be spanwise-periodic with wavenumbers $\kynil$ and $3\kynil$.
The streamwise component of the Langmuir wave is, similar to CL2,  the first to become significantly large as this is proportional to $\Rey^2$. 
Flow simulatinos are presented later in section~\ref{sec:results:LBM}.
Examining velocity fields from these simulations
(examining $\ol{\p u}$ minus the its spanwise mean) reveals that the streamwise component of the Langmuir wave is about $5\%$ the magnitude of the principal flow at $\Rey_\tau=30$, $\etanil=0.0635$, and significantly less for smaller Reynolds numbers and amplitudes. 
It seems quite possible that 
wave distortion generated by the Langmuir wave plays a role in the unstable behaviour and emergence of higher wavenumbers observed at larger Reynolds numbers.
The theoretical model accords with the observed magnitude of the streamwise Langmuir wave component and indicates that the no-slip boundary condition for the Langmuir wave is active in reducing its magnitude below $O(\eps^2 \Rey^2)$.

}

\section{Results}
\label{sec:results}

\subsection{Free surface flow}
\label{sec:results:free_surface}

In this section we consider free surface flows with the power law profile \eqref{eq:U}, i.e., adopting the free surface boundary condition in \eqref{eq:BC_p}/\eqref{eq:coeffs_O1} (using the `mixed' condition in \eqref{eq:coeff_wO2:mixed} and \eqref{eq:coeff_uO2} when discussing no-slip flows).
Even though we have presented an efficient numerical procedure for arbitrary shear currents,
the the power law \eqref{eq:U}, as adopted by \citet{phillips_1996_Langmuir_qoverz_profile},  is a suitable choice for exploring a variety of curvatures.

\begin{figure}%
\centering
\includegraphics[width=.4\columnwidth]{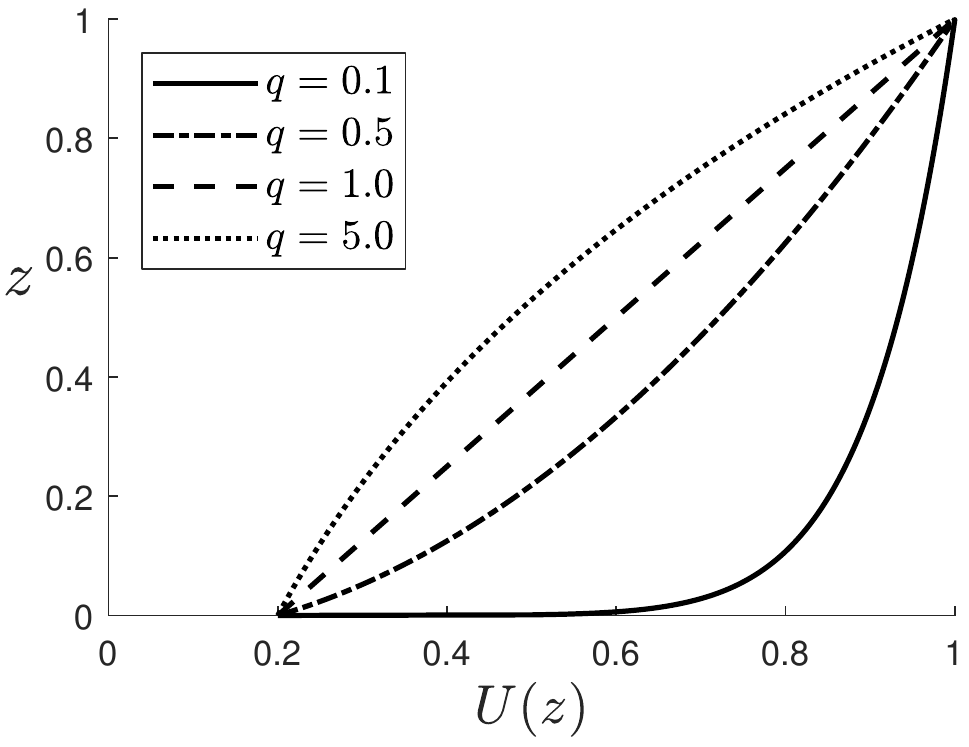}%
\caption{Primary flow profiles $U(z)$ from \eqref{eq:U}
with $\delta$ values such that $U(0) = 0.20$.}%
\label{fig:free_surf:U}%
\end{figure}

\begin{figure}%
\centering
\subfigure[Inviscid]{
\begin{minipage}{.33\columnwidth}
\includegraphics[width=\columnwidth]{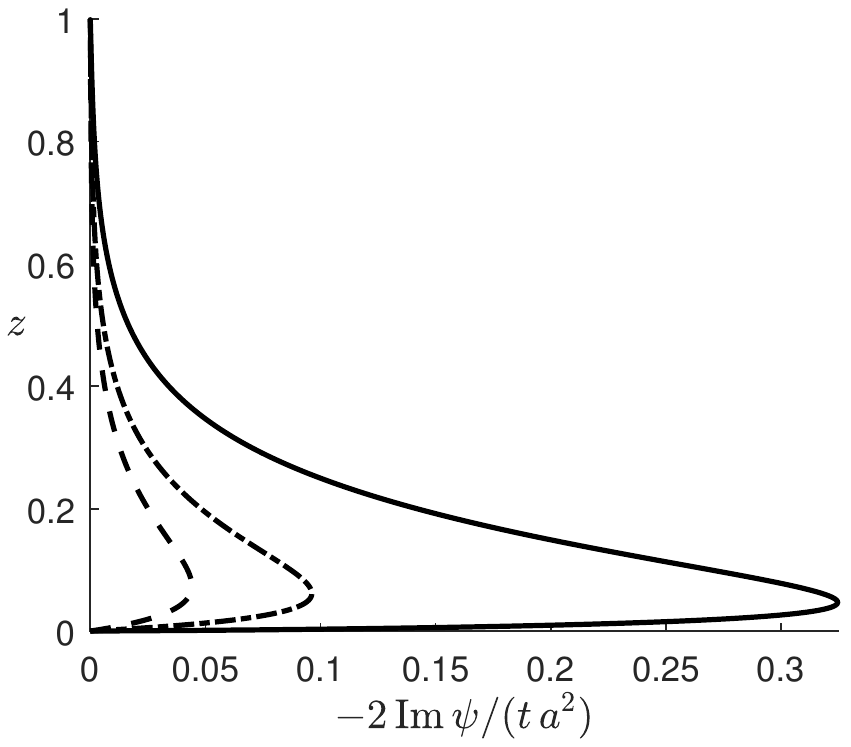}%
\\
\includegraphics[width=\columnwidth]{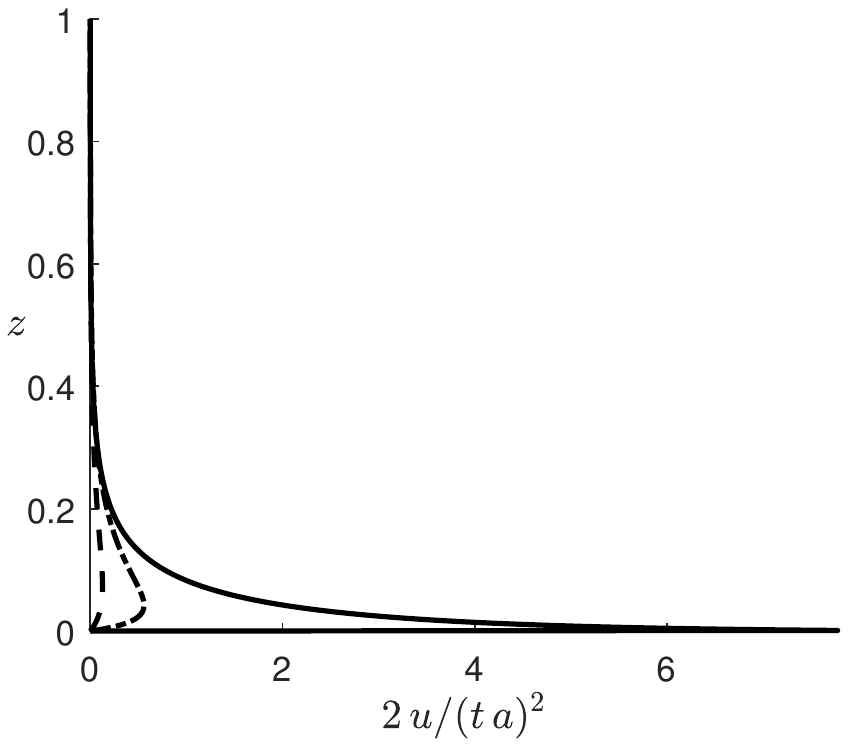}%
\end{minipage}%
\label{fig:free_surf:psi_u:k_3pi:inviscid}%
}%
\subfigure[Viscous, no-slip]{
\begin{minipage}{.33\columnwidth}
\includegraphics[width=\columnwidth]{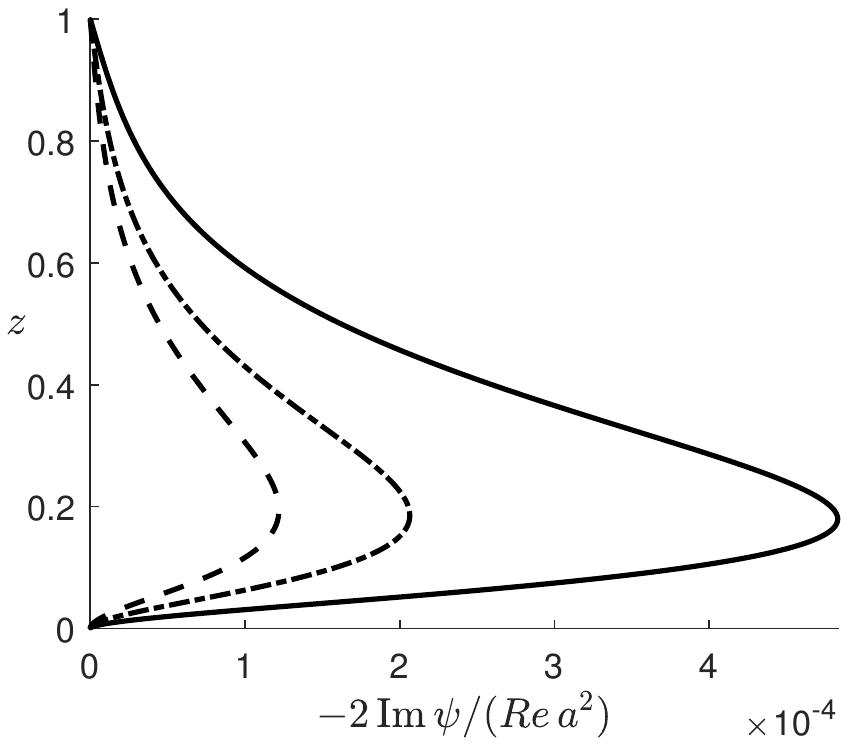}%
\\
\includegraphics[width=\columnwidth]{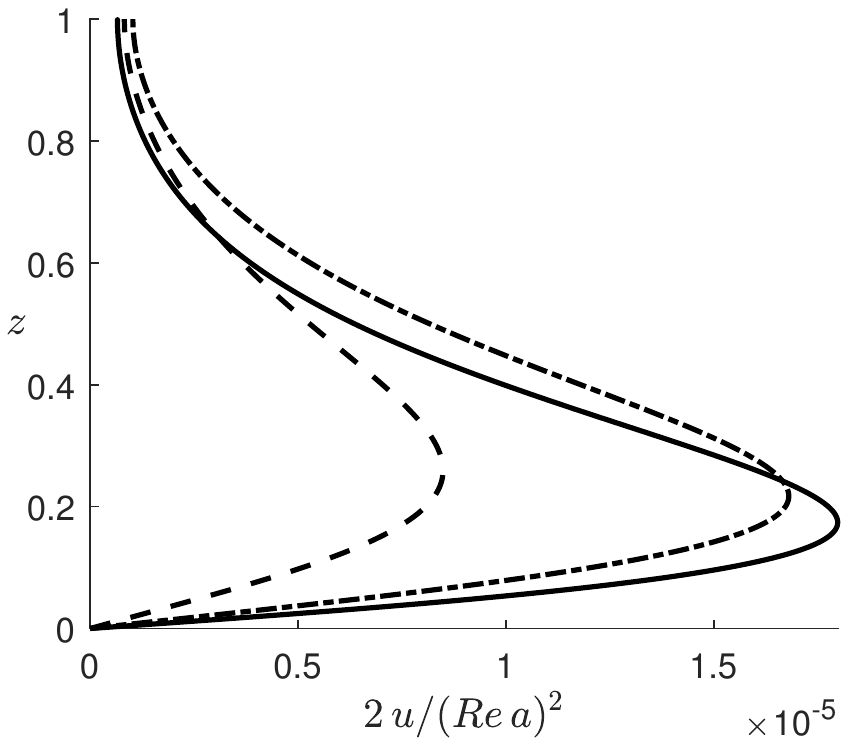}%
\end{minipage}%
\label{fig:free_surf:psi_u:k_3pi:noslip}%
}%
\subfigure[Viscous, full-slip]{
\begin{minipage}{.33\columnwidth}
\includegraphics[width=\columnwidth]{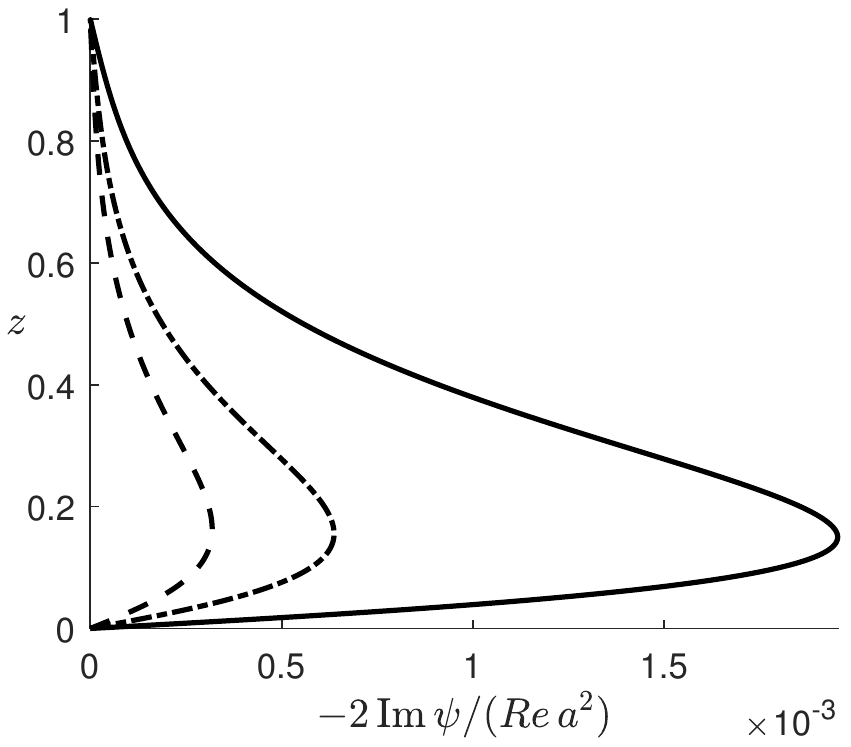}%
\\
\includegraphics[width=\columnwidth]{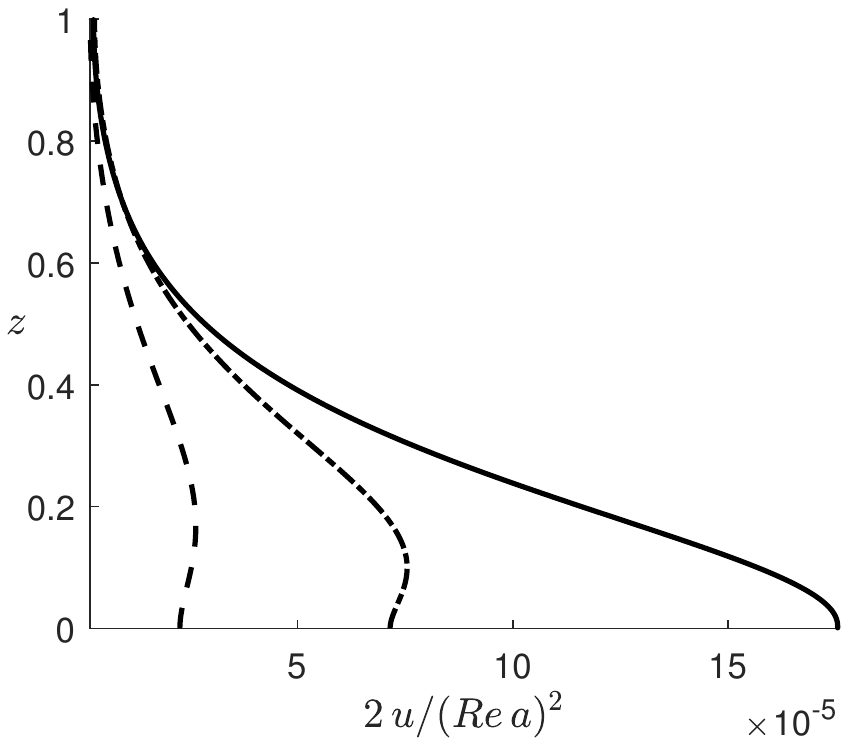}%
\end{minipage}%
\label{fig:free_surf:psi_u:k_3pi:fullslip}%
}%
\caption{
\arev{Free surface flow.}
\arev{Normalised}
stream function (top row) and \arev{normalised} streamwise velocity perturbation (bottom row) \arev{as function of vertical position}.
$\knil=3\pi$, $\arctan(\kynil/\kxnil)=\pi/8$.
Power law \eqref{eq:U} primary flow profile; 
solid, dot-dashed, dashed and stippled dotted lines respectively show $q=0.1$  , $0.5$, $1.0$ and $5.0$ with
$\delta$ values such that $U(0) = 0.20$ (cf.\ figure~\ref{fig:free_surf:U}).
}%
\label{fig:free_surf:psi_u:k_3pi}%
\end{figure}

\begin{figure}%
\centering
\subfigure[Inviscid]{
\begin{minipage}{.33\columnwidth}
\includegraphics[width=\columnwidth]{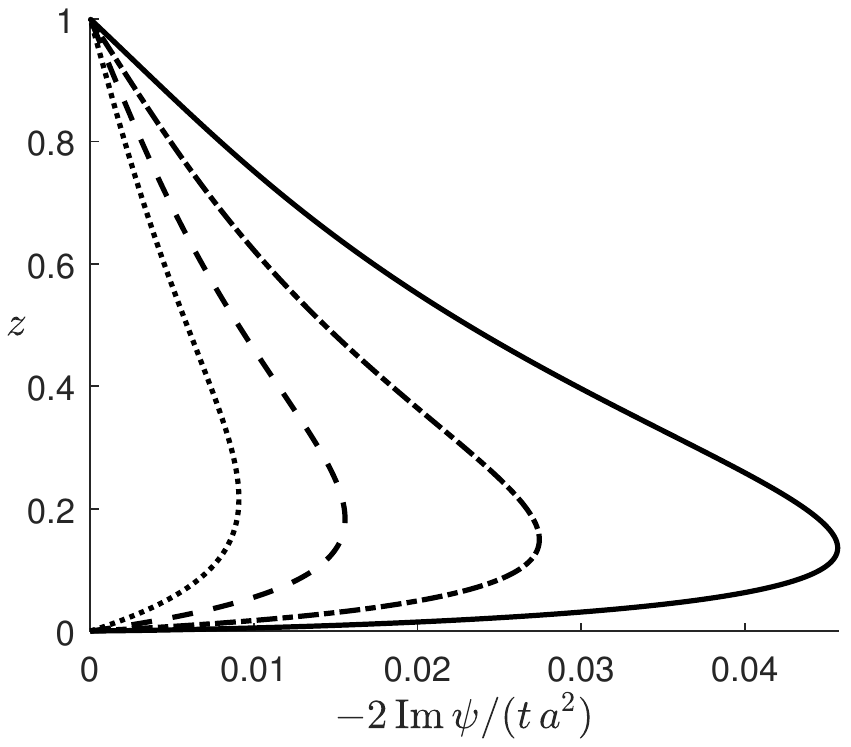}%
\\
\includegraphics[width=\columnwidth]{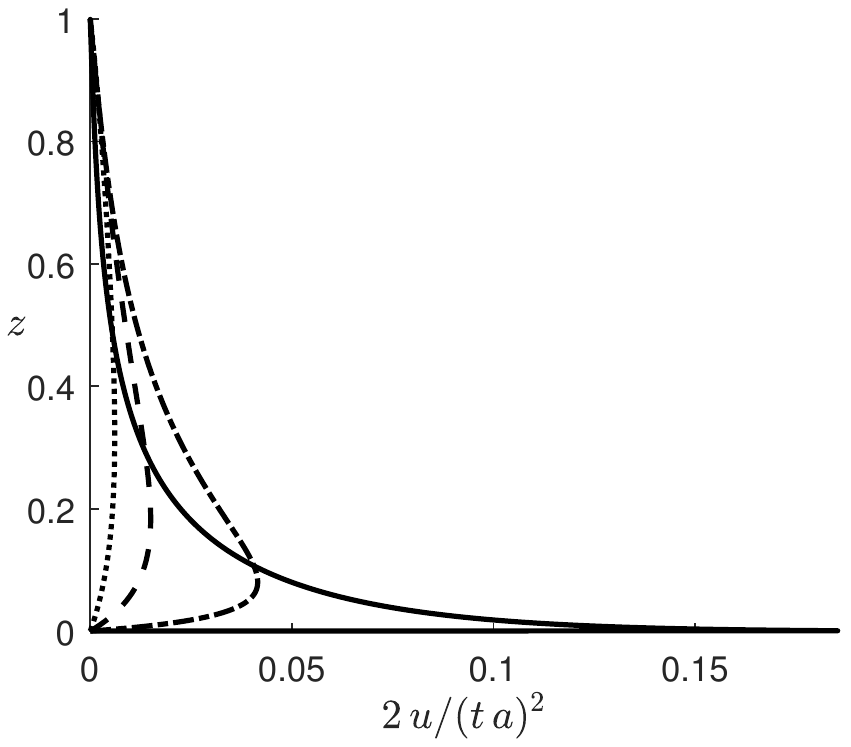}%
\end{minipage}%
\label{fig:free_surf:psi_u:k_pi:inviscid}%
}%
\subfigure[Viscous, no-slip]{
\begin{minipage}{.33\columnwidth}
\includegraphics[width=\columnwidth]{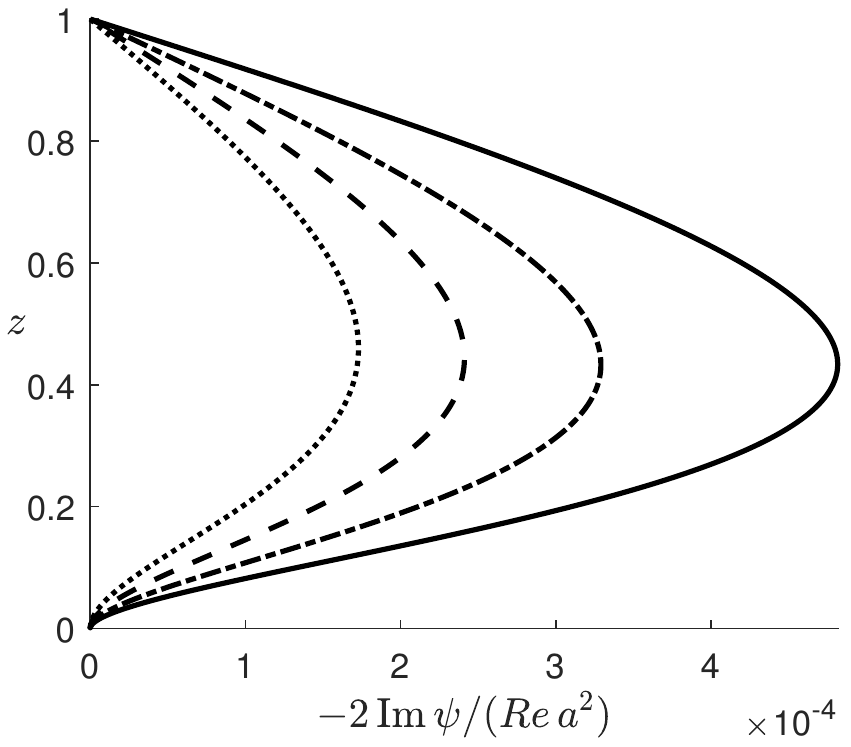}%
\\
\includegraphics[width=\columnwidth]{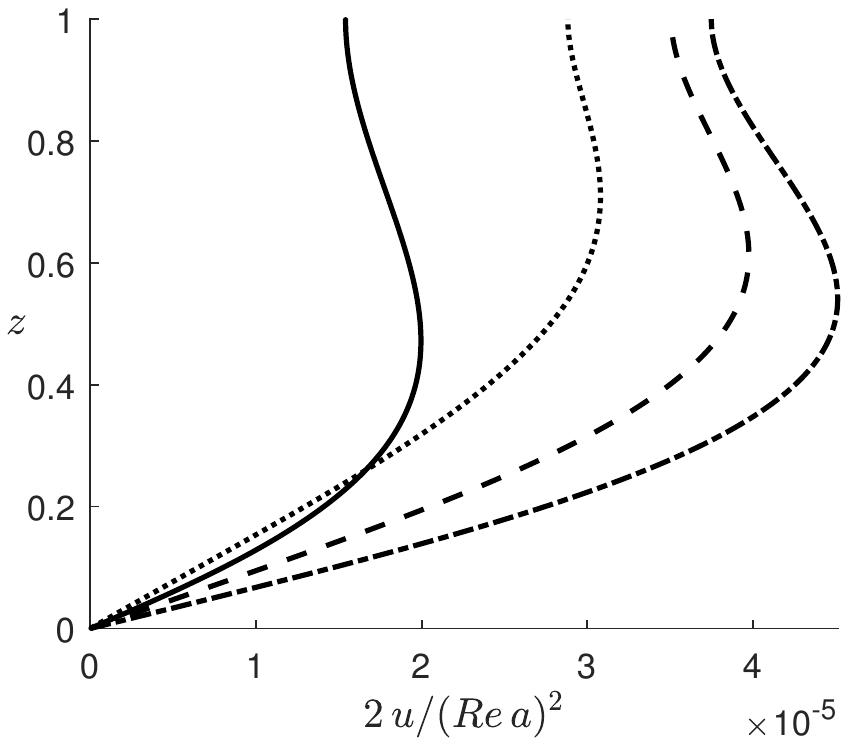}%
\end{minipage}%
\label{fig:free_surf:psi_u:k_pi:noslip}%
}%
\subfigure[Viscous, full-slip]{
\begin{minipage}{.33\columnwidth}
\includegraphics[width=\columnwidth]{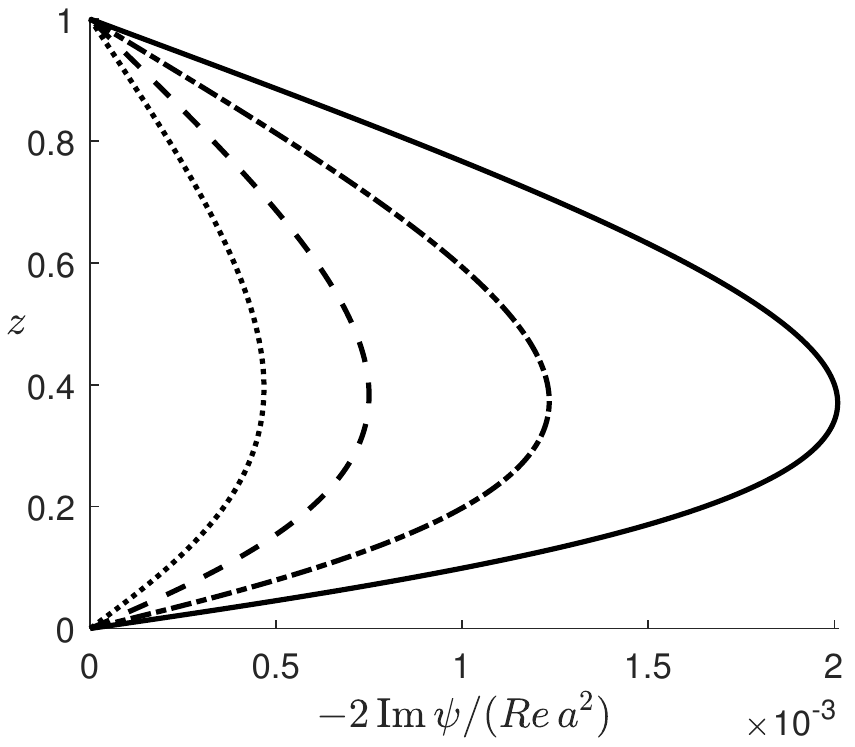}%
\\
\includegraphics[width=\columnwidth]{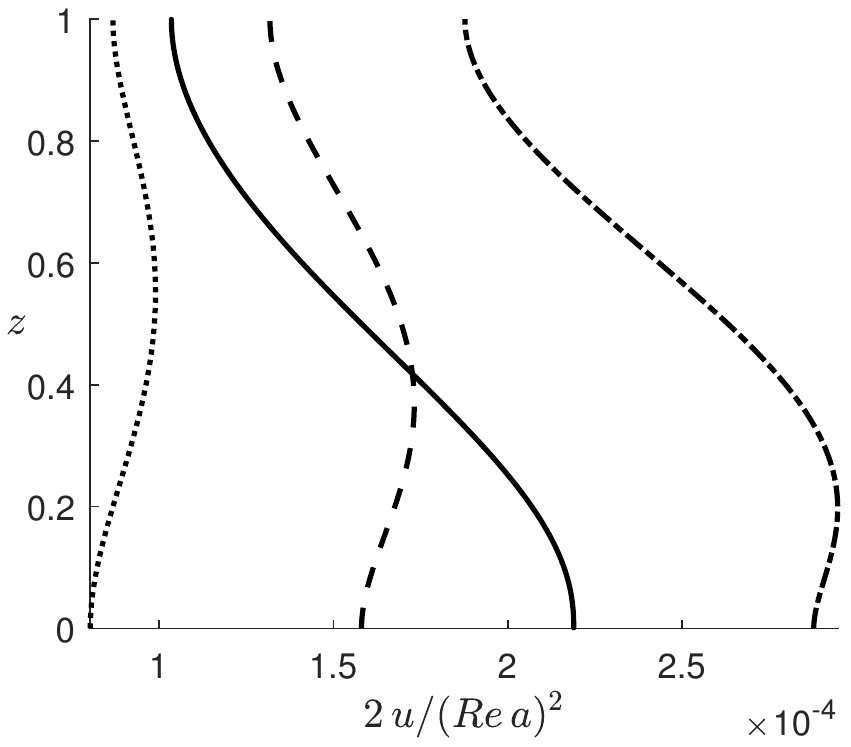}%
\end{minipage}%
\label{fig:free_surf:psi_u:k_pi:fullslip}%
}%
\caption{
Similar to figure~\ref{fig:free_surf:psi_u:k_pi}, but with increased wavelength (reduced depth) $\knil = \pi$.
}%
\label{fig:free_surf:psi_u:k_pi}%
\end{figure}

\begin{figure}%
\centering
\subfigure[Inviscid]{
\begin{minipage}{.33\columnwidth}
\includegraphics[width=\columnwidth]{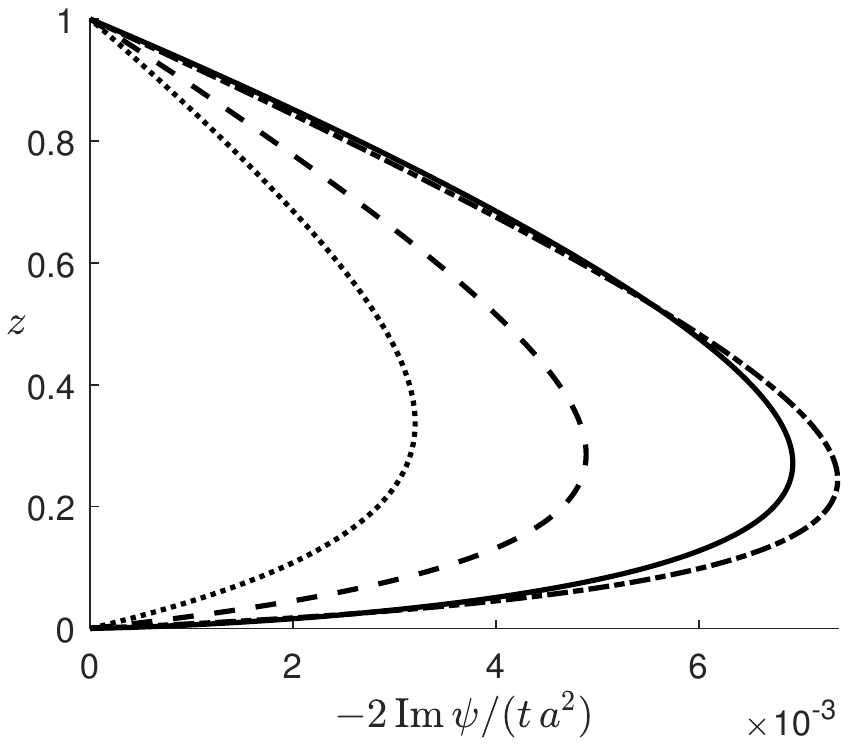}%
\\
\includegraphics[width=\columnwidth]{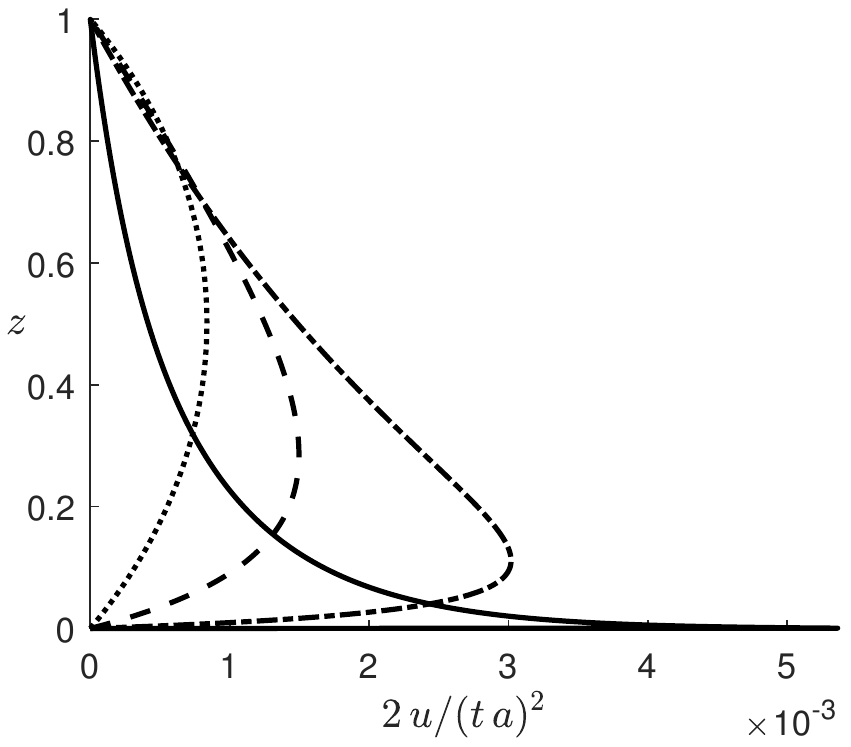}%
\end{minipage}%
\label{fig:free_surf:psi_u:k_1:inviscid}%
}%
\subfigure[Viscous, no-slip]{
\begin{minipage}{.33\columnwidth}
\includegraphics[width=\columnwidth]{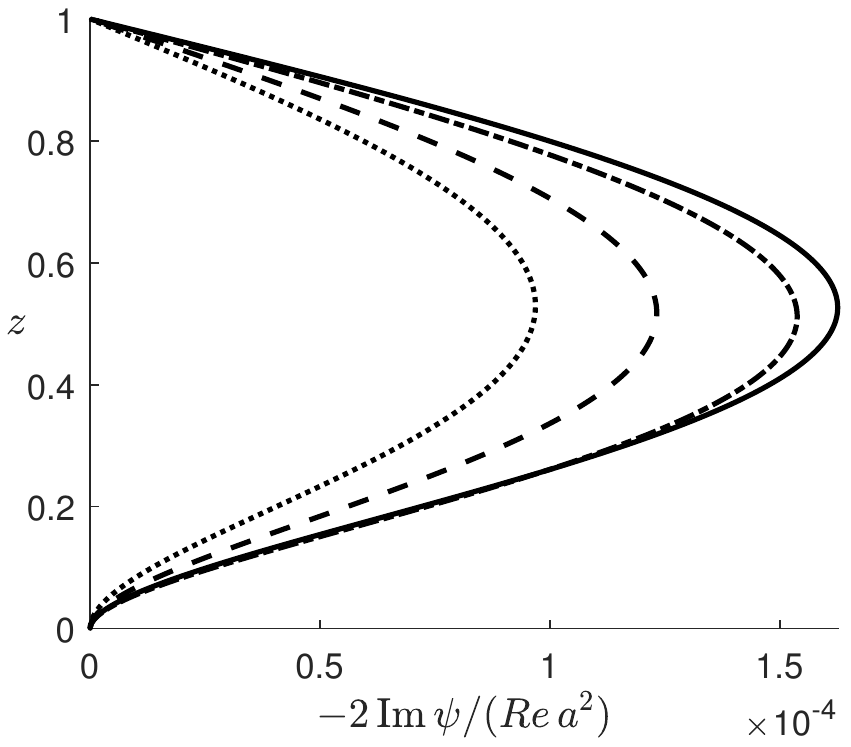}%
\\
\includegraphics[width=\columnwidth]{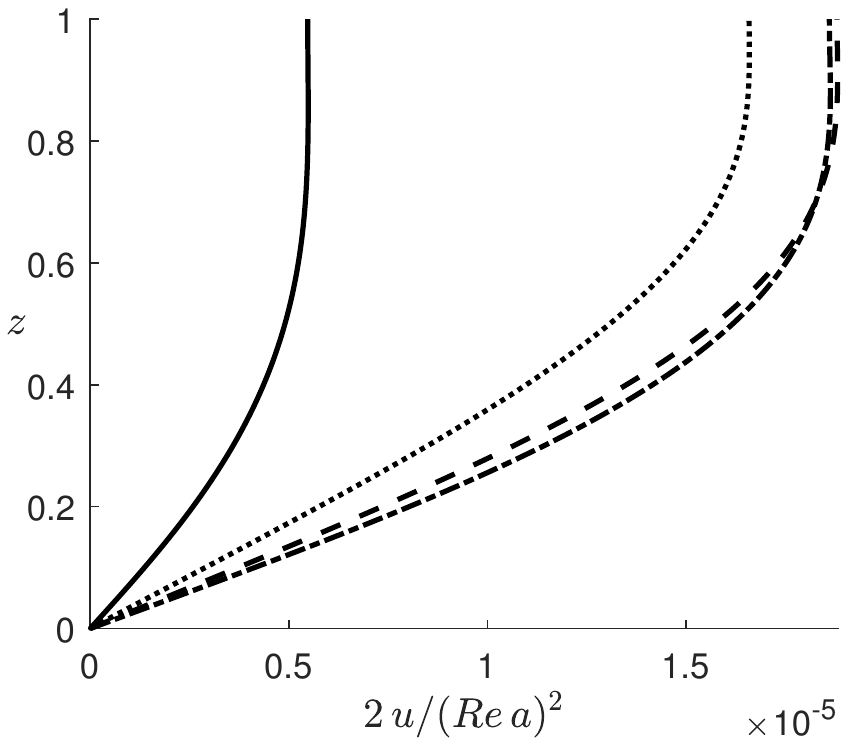}%
\end{minipage}%
\label{fig:free_surf:psi_u:k_1:noslip}%
}%
\subfigure[Viscous, full-slip]{
\begin{minipage}{.33\columnwidth}
\includegraphics[width=\columnwidth]{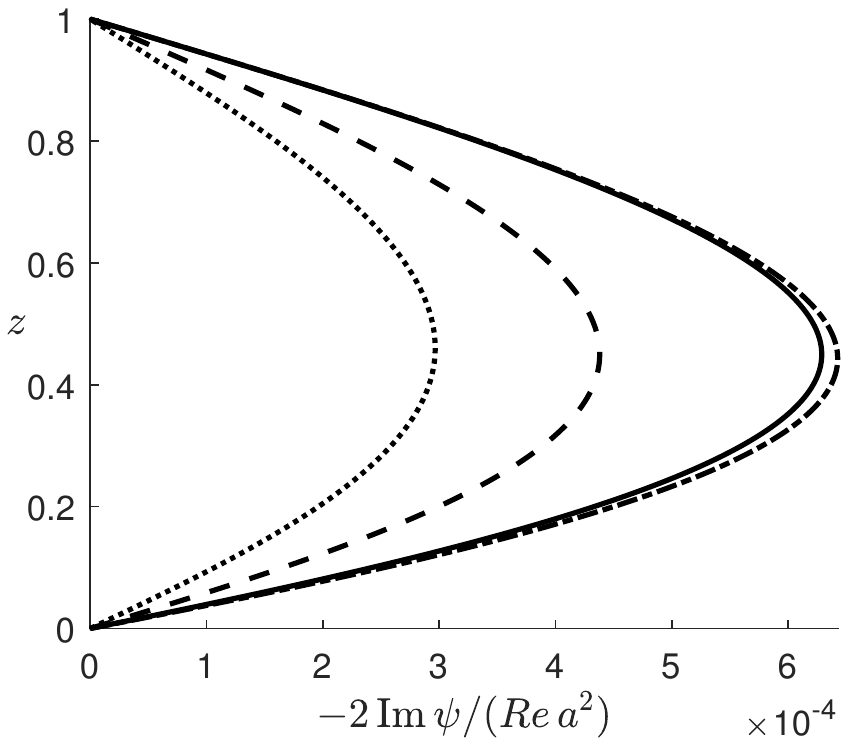}%
\\
\includegraphics[width=\columnwidth]{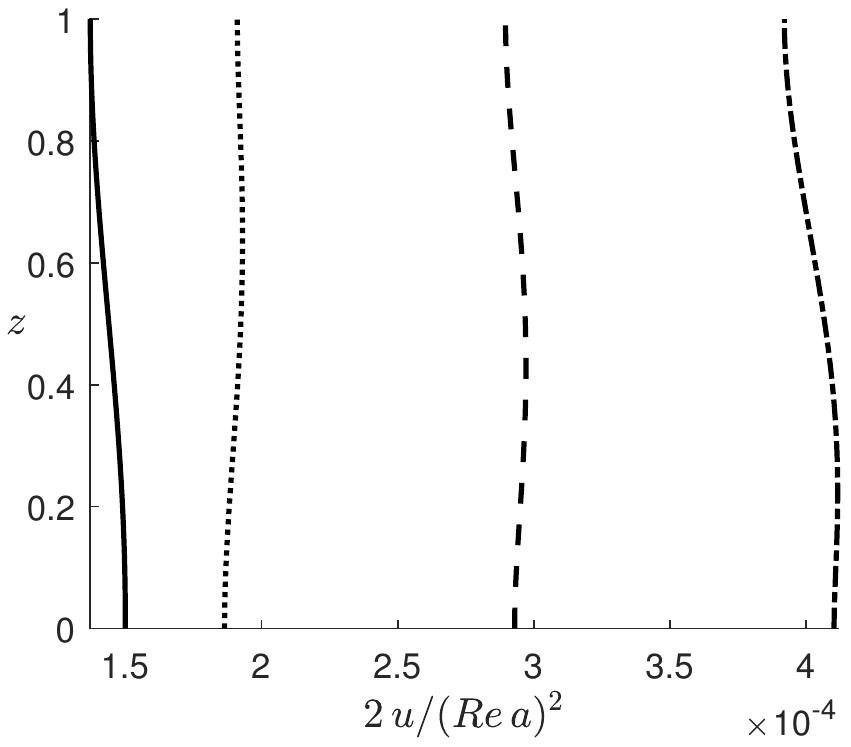}%
\end{minipage}%
\label{fig:free_surf:psi_u:k_1:fullslip}%
}%
\caption{
Similar to figure~\ref{fig:free_surf:psi_u:k_pi}, but with increased wavelength (reduced depth) $\knil = 1.0$.
}%
\label{fig:free_surf:psi_u:k_1}%
\end{figure}

\begin{figure}
\subfigure[Inviscid]{
\begin{minipage}{.33\columnwidth}
\includegraphics[width=\columnwidth]{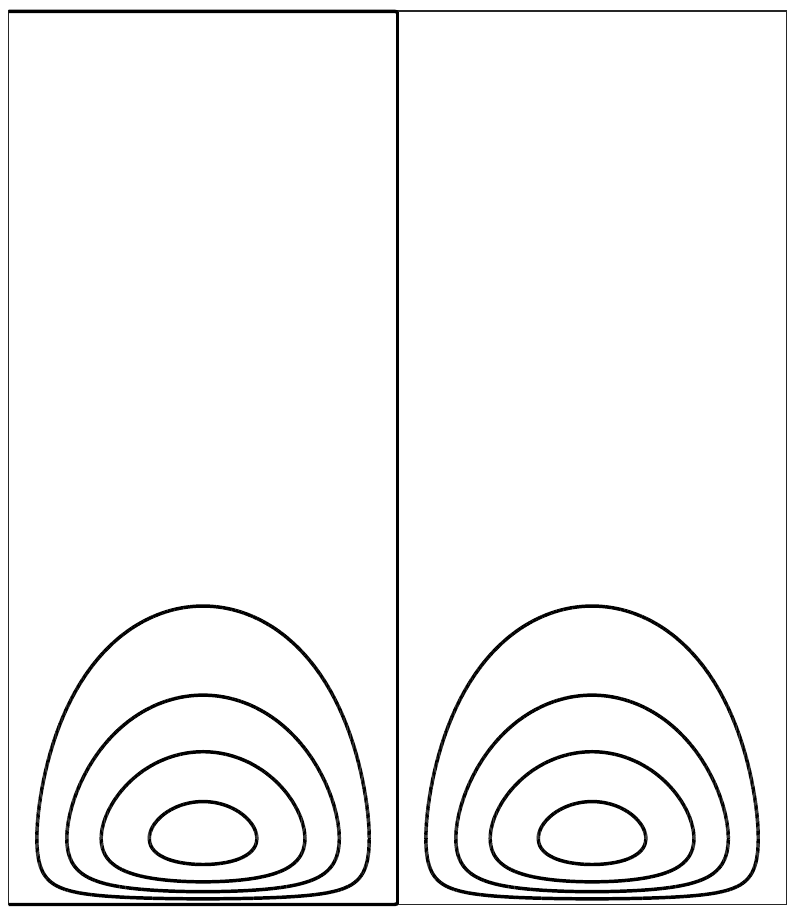}%
\\[1.5ex]
\includegraphics[width=\columnwidth]{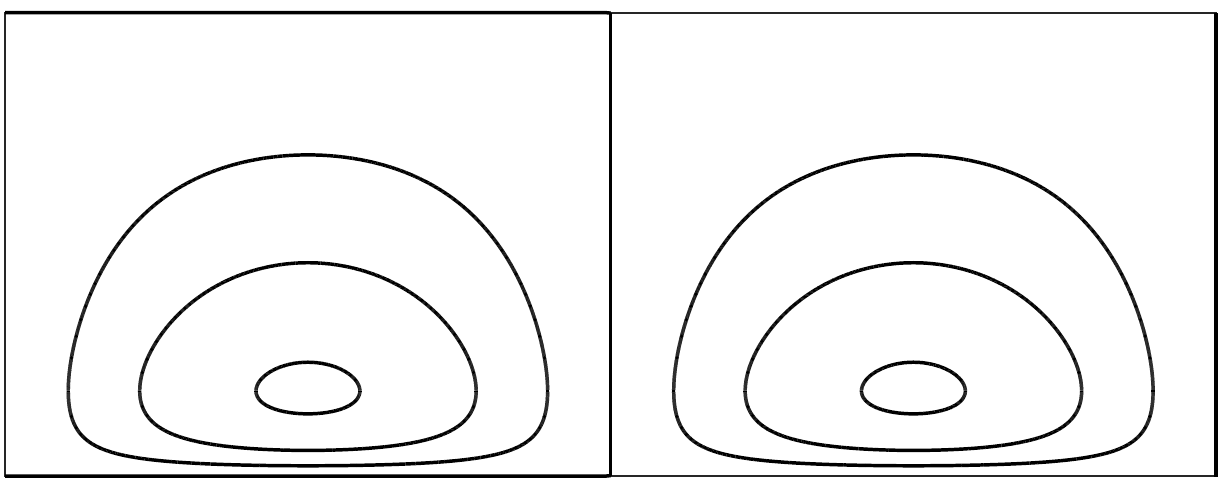}%
\\[1.5ex]
\includegraphics[width=\columnwidth]{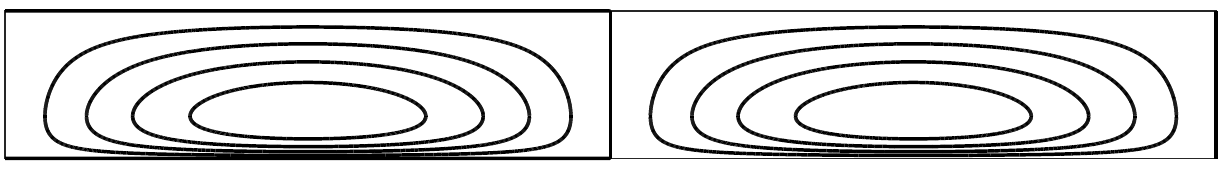}%
\end{minipage}
}%
\subfigure[Viscous, no-slip]{
\begin{minipage}{.33\columnwidth}
\includegraphics[width=\columnwidth]{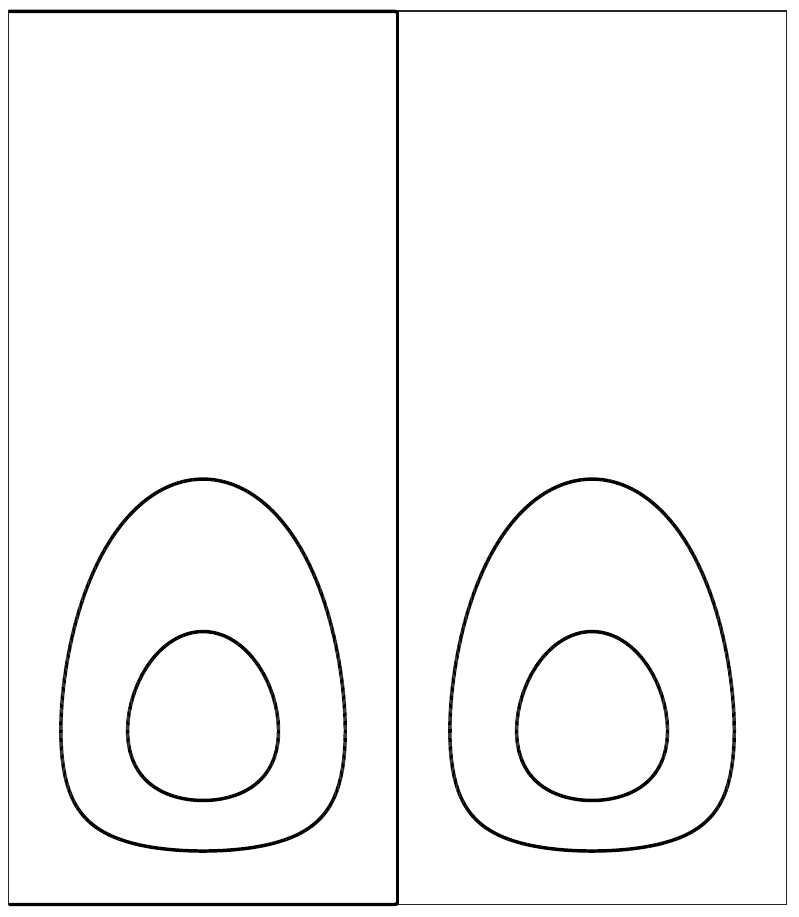}%
\\[1.5ex]
\includegraphics[width=\columnwidth]{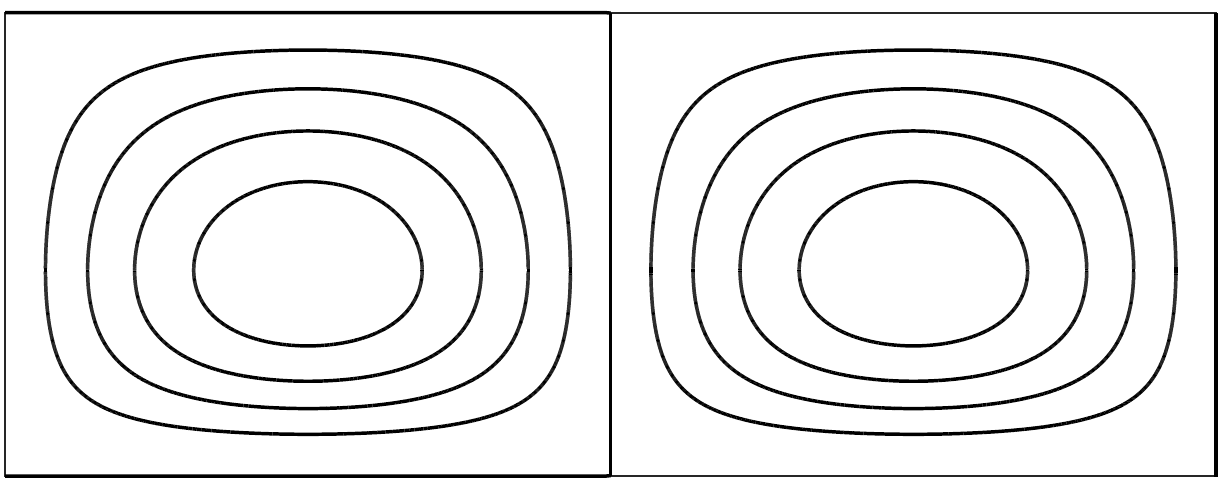}%
\\[1.5ex]
\includegraphics[width=\columnwidth]{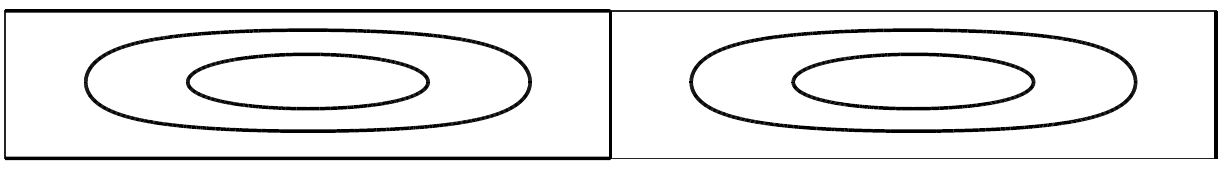}%
\end{minipage}
}%
\subfigure[Viscous, full-slip]{
\begin{minipage}{.33\columnwidth}
\includegraphics[width=\columnwidth]{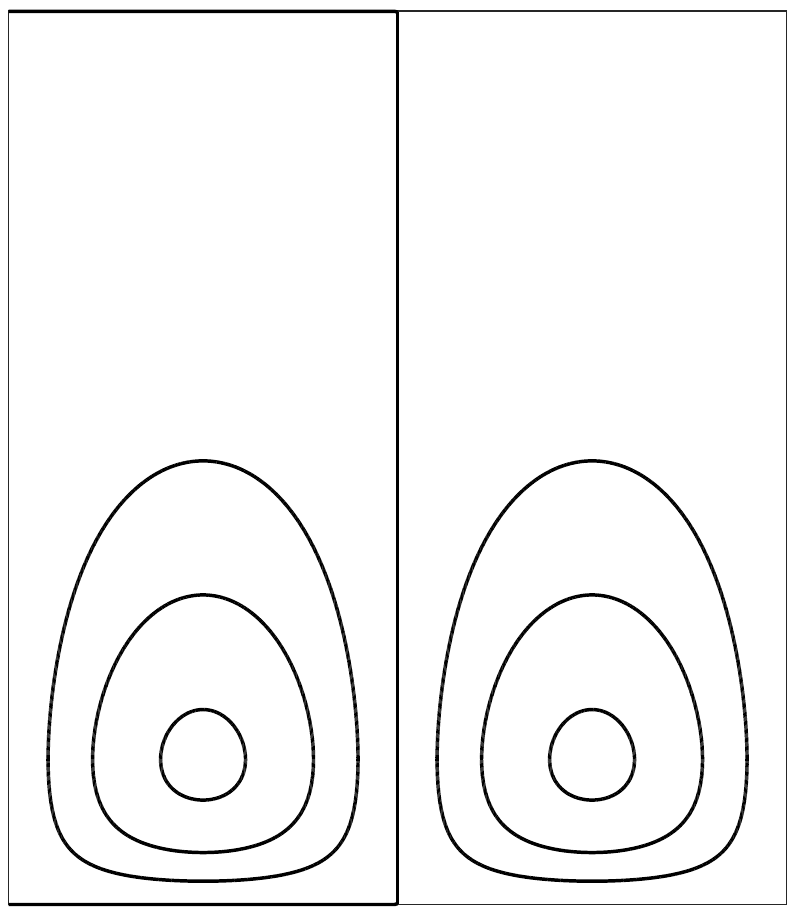}%
\\[1.5ex]
\includegraphics[width=\columnwidth]{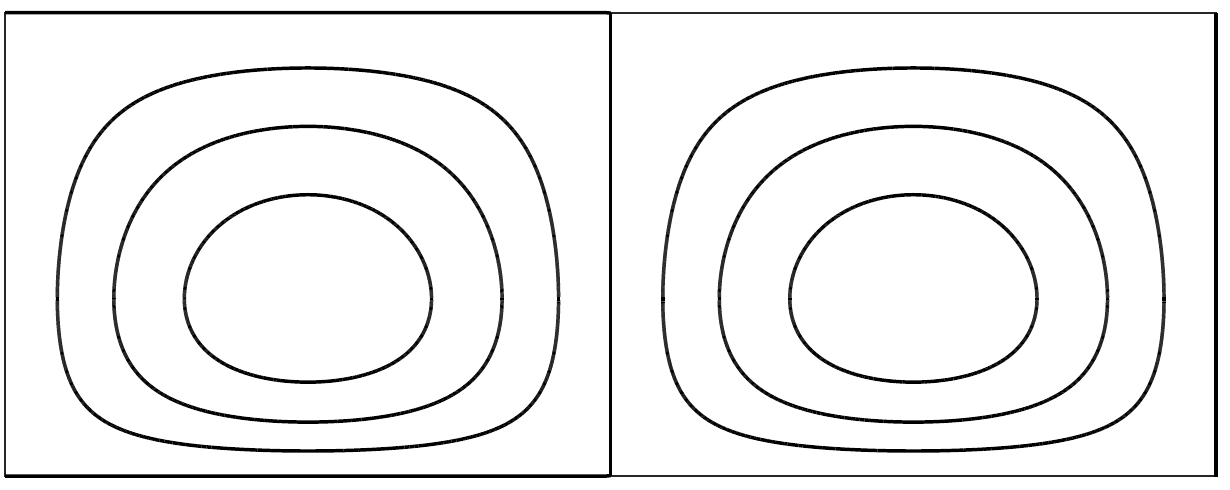}%
\\[1.5ex]
\includegraphics[width=\columnwidth]{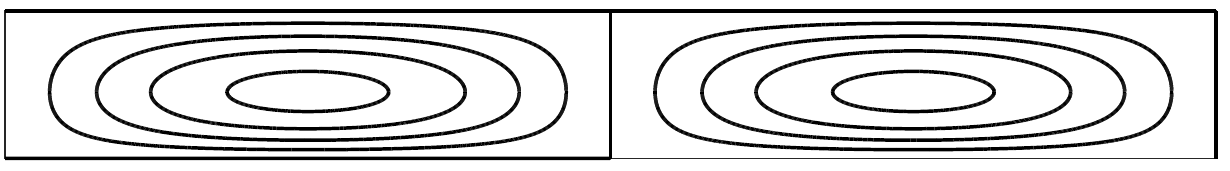}%
\end{minipage}
}%
\caption{
Streamlines $\p\psi= \mr{constant}$ \arev{in free surface flow} for the $q=1.0$ power law profile (cf. figure~\ref{fig:free_surf:U}--\ref{fig:free_surf:psi_u:k_1}. 
Top: $\knil = 3\pi$. Middle: $\knil = \pi$. Bottom: $\knil = 1.0$. $\arctan(\kynil/\kxnil)=\pi/8$.
Isoline values  of $\p \psi$ can be inferred from the graphs in figure~\ref{fig:free_surf:psi_u:k_3pi}--\ref{fig:free_surf:psi_u:k_1} \arev{as these give the stream function values in a vertical line running through the vortex centre.}
}%
\label{fig:free_surf:streamlines}%
\end{figure}

\begin{figure}%
\centering
\subfigure[Inviscid]{
\begin{minipage}{.33\columnwidth}
\includegraphics[width=\columnwidth]{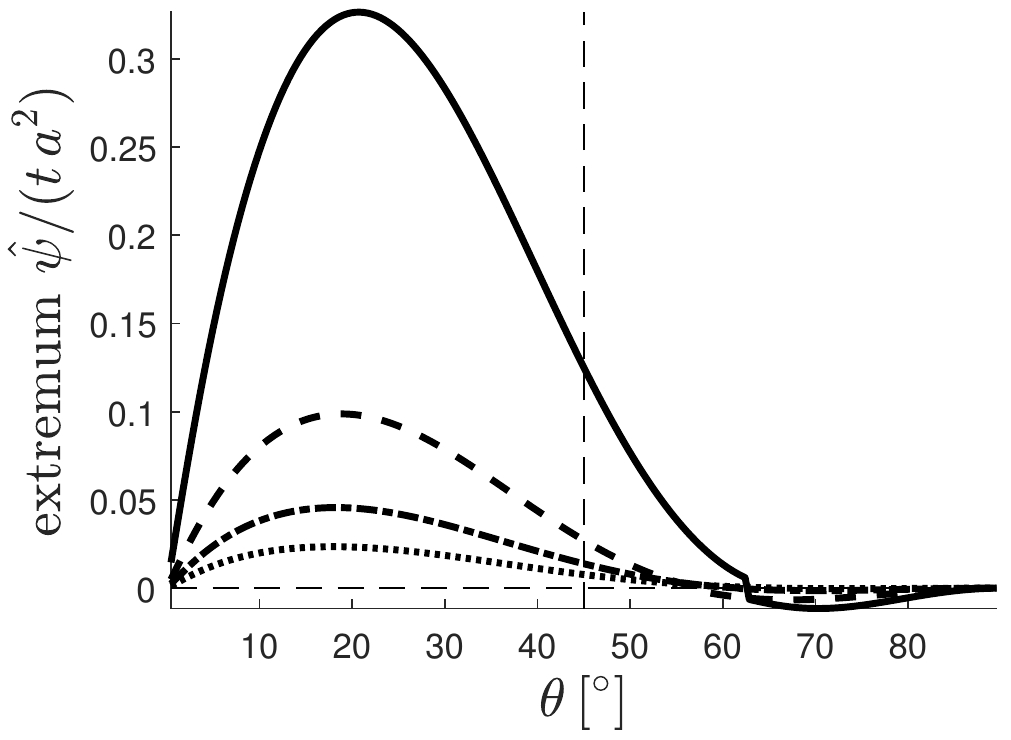}%
\\
\includegraphics[width=\columnwidth]{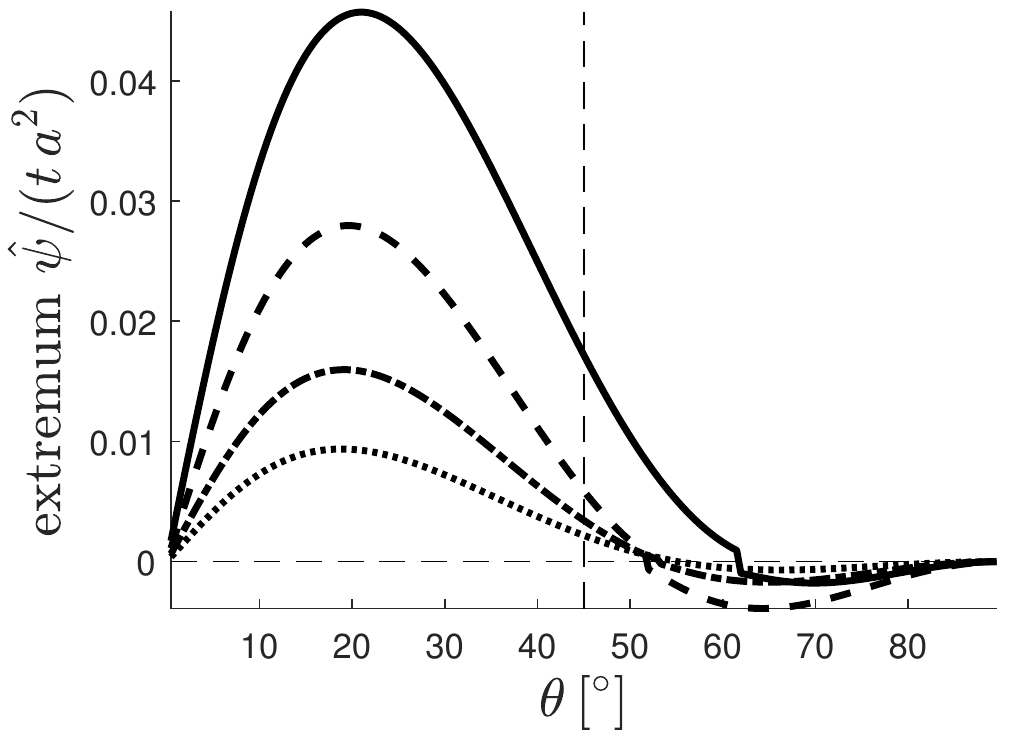}%
\\
\includegraphics[width=\columnwidth]{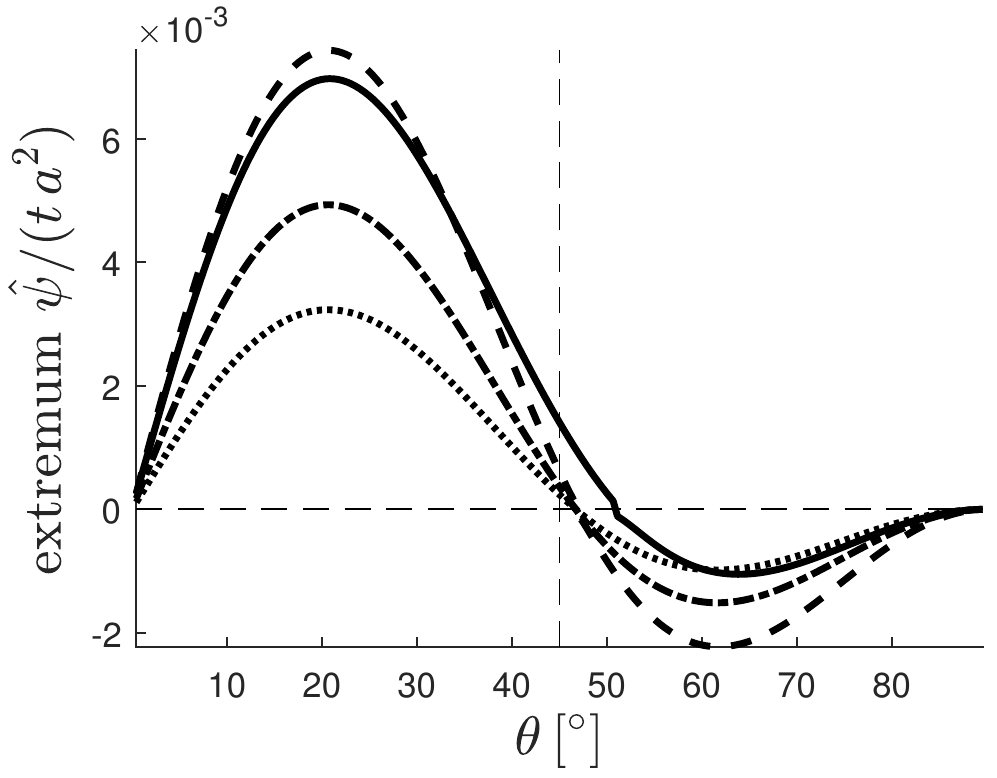}%
\end{minipage}%
\label{fig:free_surf:theta:inviscid}%
}%
\subfigure[Viscous, no-slip]{
\begin{minipage}{.33\columnwidth}
\includegraphics[width=\columnwidth]{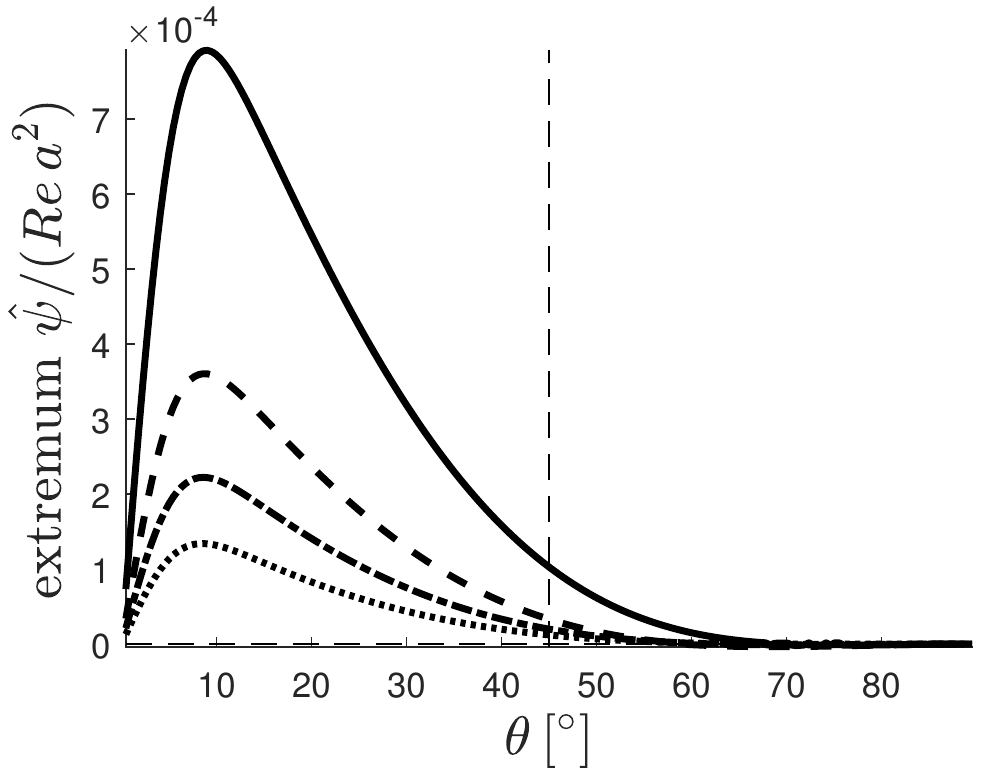}%
\\
\includegraphics[width=\columnwidth]{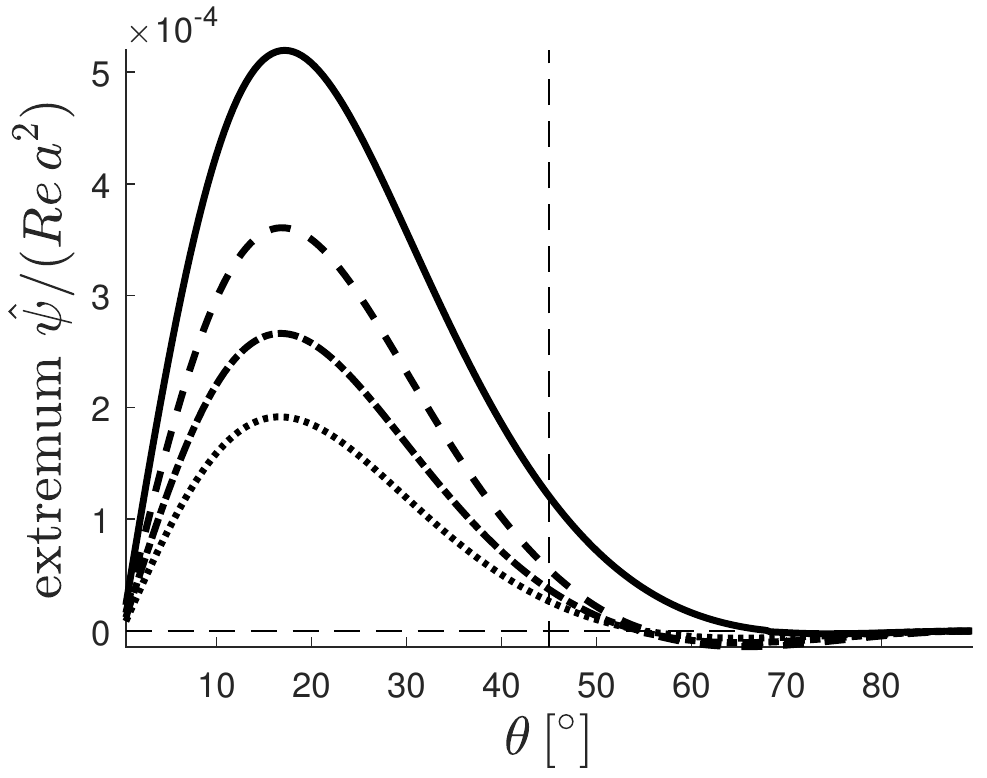}%
\\
\includegraphics[width=\columnwidth]{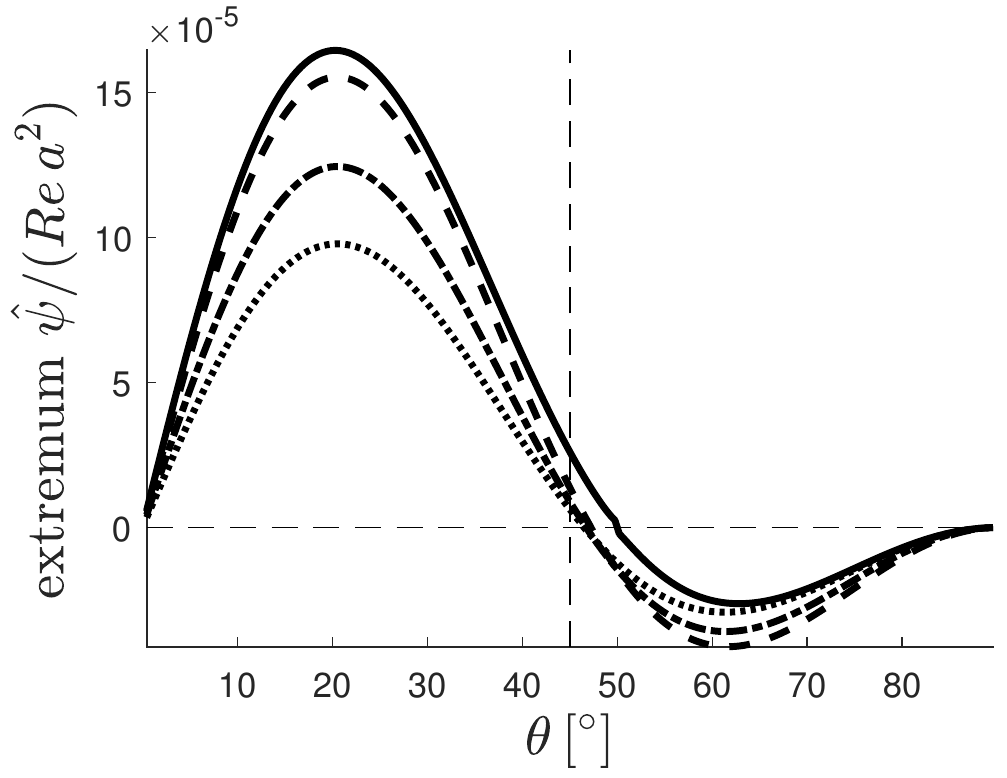}%
\end{minipage}%
\label{fig:free_surf:theta:noslip}%
}%
\subfigure[Viscous, full-slip]{
\begin{minipage}{.33\columnwidth}
\includegraphics[width=\columnwidth]{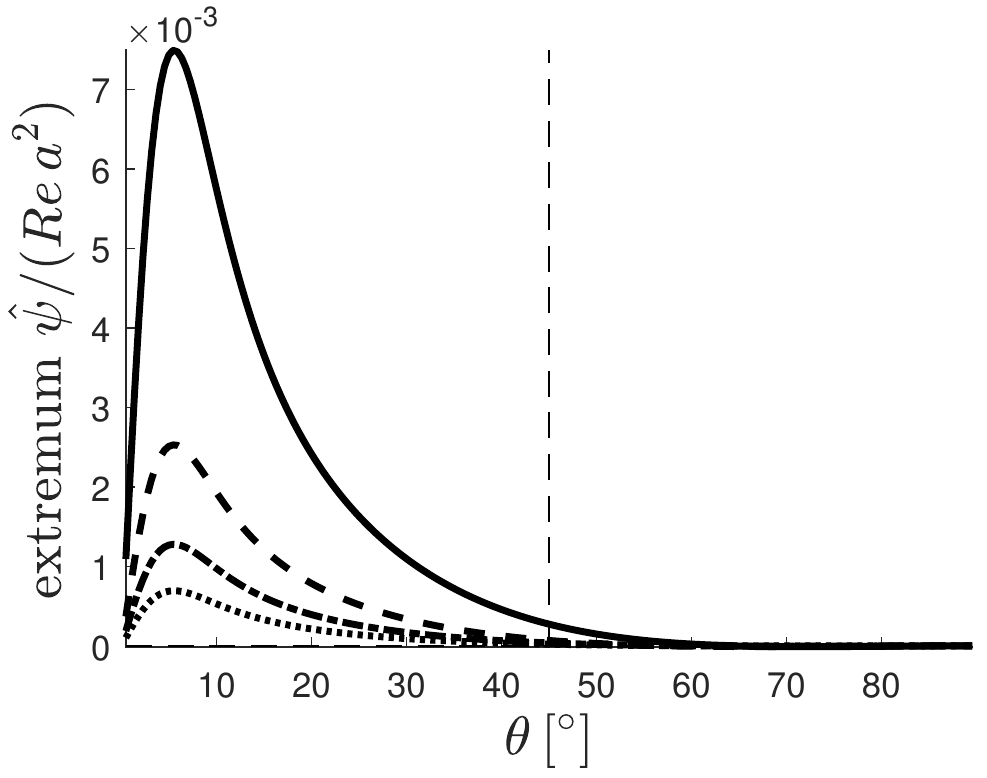}%
\\
\includegraphics[width=\columnwidth]{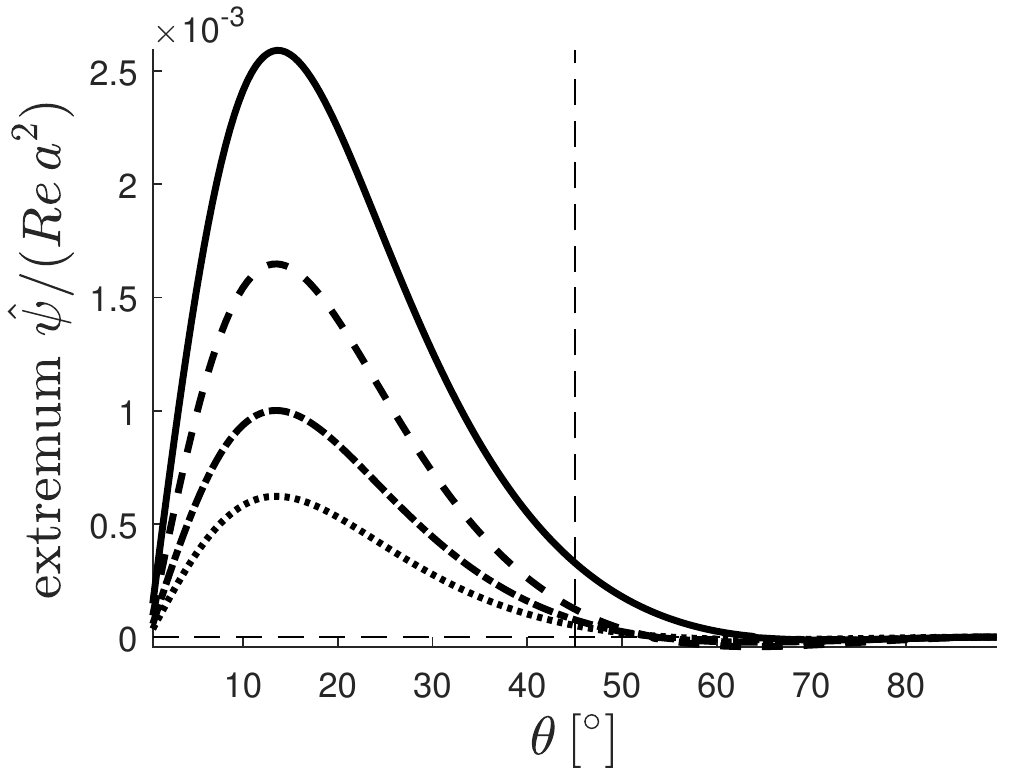}%
\\
\includegraphics[width=\columnwidth]{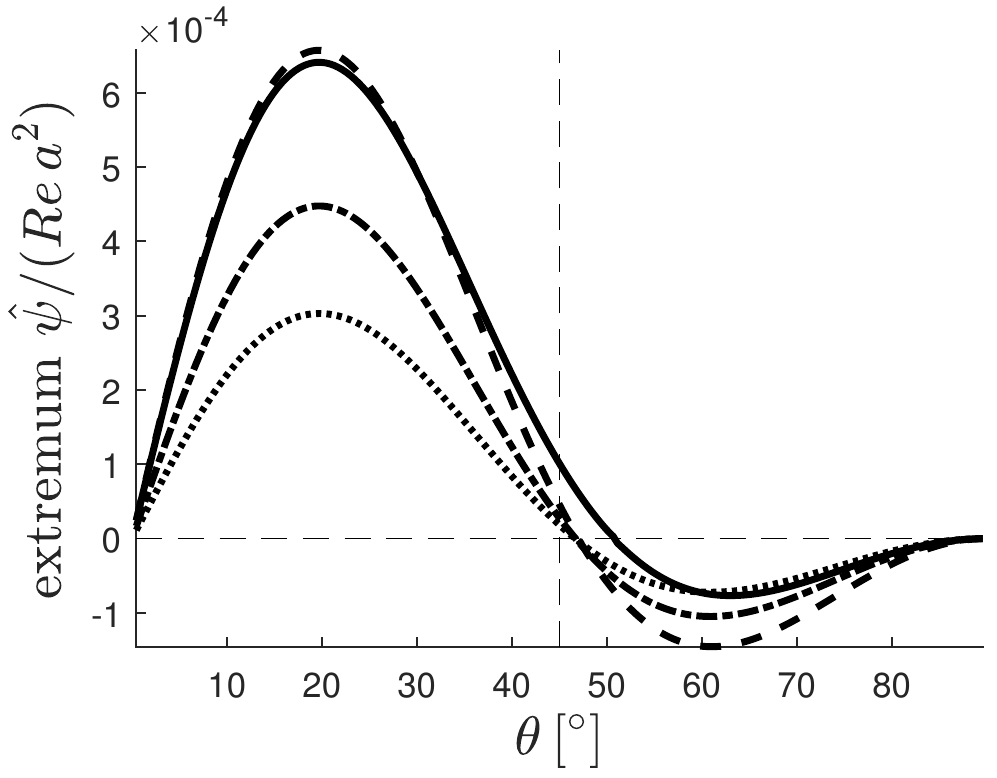}%
\end{minipage}%
\label{fig:free_surf:theta:fullslip}%
}
\caption{
Extremum of \arev{appropriate normalised stream functions} as function of the wave vector angle $\theta=\arctan(\kynil/\kxnil)$ \arev{in free surface flow}.
This value is proportional to the vortex volume flow through a cross-section from the vortex centre out. 
Top row: $k=\pi$. Bottom row: $k=1.0$.
Power law \eqref{eq:U} primary flow profile; 
solid, dot-dashed, dashed and stippled dotted lines respectively show $q=0.1$  , $0.5$, $1.0$ and $5.0$ with
$\delta$ values such that $U(0) = 0.20$ (cf.\ figure~\ref{fig:free_surf:U}).
}%
\label{fig:free_surf:theta}%
\end{figure}

The Froude number of a free surface flow affects Langmuir circulation only through the upper boundary, which is in phase with the bathymetry when the Froude number is sub-critical and in anti-phase when super-critical \arev{\citep[Art. 246]{lamb1932hydrodynamics}}.
For definiteness, we fix it at a sub-critical value $\Fr=0.5$.
The critical Froude number is unity for a uniform current and 
\begin{equation}
\Fr\_c^2 
= 
\frac{k}{k_x^2}\frac{
I_{q-\frac12}[k(1+\delta)] I_{-q+\frac12}(k\delta)
-I_{-q+\frac12}[k(1+\delta)] I_{q-\frac12}(k\delta)
}{
I_{q+\frac12}[k(1+\delta)] I_{-q+\frac12}(k\delta)
-I_{-q-\frac12}[k(1+\delta)] I_{q-\frac12}(k\delta)
},
\label{eq:Fr_crit}
\end{equation}
with the power law profile \eqref{eq:U}---see e.g. \citet[Art.~246]{lamb1932hydrodynamics} or \citet{akselsen_ellingsen_2019_river}. 
\arev{
Whether the upper boundary is a wall or a free surface affects the Langmuir circulation mainly by determining whether this motion should be subject to a full-slip condition or not.  
The Froude number enters only the boundary condition of the linear wave.
Wall undulations must generate surface undulations of comparable magnitude for the Froude number to notably affect the circulation.
This happens only when the flow itself is shallow ($k \ll 1$) or when the Froude number is  close to criticality.
Neither is the case in the presented examples so that the sensitivity to the Froude numeber is weak.
\\

We start by considering the primary current profiles drawn in figure~\ref{fig:free_surf:U} for the power law \eqref{eq:U} with varying exponent values $q=0.1$, $0.5$, $1.0$ and $5.0$.
Values of displacement $\delta$ (need not be small) are chosen to give the same wall velocity $U(0)=U\_b=0.2$, i.e., 
$\delta = U\_b^{1/q}/(1-U\_b^{1/q})$.
In figure~\ref{fig:free_surf:psi_u:k_3pi}--\ref{fig:free_surf:psi_u:k_1}, we compare the stream functions $\psi(z)$ and the streamwise velocity component $u(z)$ generated by these profiles.
Three depths, $\knil = 3\pi$, $\pi$ and $1$, are considered.
These represent deep, intermediate and shallow water, respectively,
as can be seen in the real-space streamline plots 
in figure~\ref{fig:free_surf:streamlines}.
Note that $\knil = 3\pi$ and $\pi$ can both be considered `deep' in terms of the first order Rayleigh wave, yet the Langmuir wave penetrates much deeper. 
This feature makes Langmuir circulation 
an effective mixing mechanism.

Vortex centres are vertically located at the extrema of the stream function $\psi(z)$.
An extremum of $\psi(z)$ 
serves as a measure of vortex turnover as it equals half the volume flux per unit along the $x$-axis through a cross-section passing from a vortex centre to one of the boundaries.
($\psi(z)$ is chosen zero at the boundaries.) 
Vortex solutions are directly proportional to powers of $\etanil$, $\Rey$ and, in the case of transient inviscid flow, $t$.
We therefore adopt normalised stream functions
$-2\Imag \psi/(t\,\etanil^2)$ when inviscid and $-2\Imag \psi/(\Rey\,\etanil^2)$ for stationary viscous flow, and the normalised streamwise velocity perturbations $2 u/(t\,\etanil)^2$ and $2 u/(\Rey\,\etanil)^2$ in the same flows, respectively. This renders the normalised solutions independent of  $t$, $\Rey$ and $\etanil$.

Compared to the inviscid, transient case, an effect of viscosity is to push the vortex centres out away from the boundaries. 
This is expected and was also demonstrated by \citet{Craik_1970_Langmuir_myidea}.
In addition, the no-slip boundary condition  is seen to
significantly reduce the steady state vortex intensity and also to
push the vortex slightly further away from the lower boundary. 
No-slip conditions are appropriate for rigid walls and full-slip conditions for free surfaces.

Vortex intensity is governed by the presence of the two main ingredients for Langmuir circulation---the wave-like perturbation of current streamlines, proportional to the 
vertical velocity enforced by 
the bed, and the shear
of the primary flow.
Since the current profiles have been made to have the same velocity at the bed, the profile with the strongest shear near the bed (the lowest $q$) has the greatest vortex intensity. 

The power law profile with a low exponent value $q$ has historically often been used as a model for turbulent boundary layers.
In fact, a general consensus has in recent years formed that the
power law performs better than the log law over rough boundaries or at low Reynolds numbers 
\citep{barenblatt_1993_power_law_theory,bergstrom_2001_power_law_low_Re,djenidi_1997_advantages_power_vs_log_law,George_1997_extended_power_law}. 
A range of exponent values have over the years been suggested for such flows,
ranging between $1/3$ to $1/12$ depending on roughness and Reynolds number \citep{Cheng_2007_power_laws,Chen_1991_power_laws,dolcetti_2016_channel_directional_spectrum}.
Models of varying sophistication for relating the power law exponent to a particular a boundary layer are available. 
For example, adopting the much celebrated model proposed by \citet{barenblatt_1993_power_law_theory} (his equation 16 with a hydraulic diameter assumption), we compute that the example exponent $q=0.1$ is a suitable representation of a turbulent boundary layer of
wall Reynolds number $2.8\times10^4$.
\\

The influence of the angle $\theta$ of $\bknil$, $\theta=\arctan(\kynil/\kxnil)$, is demonstrated in figure~\ref{fig:free_surf:theta}, showing for fixed moduli $\knil$ peak values of $\psi(z)$ proportional to vortex turnover.
Wavenumbers are $\knil=\pi$  and $1.0$  in the top and bottom panels, respectively.
Strongest vortex intensity is typically observed within the range $10^\circ<\theta<25^\circ$
where $- \Imag\,\psi>0$, meaning that zones of downwelling are aligned with extrema in the bathymetry 
such that vortex rolls push fluid down towards peaks and troughs in the bed topography.
The streamwise velocity component $u$ is also positive in this range so that the strongest  velocity perturbation in the downstream direction  is found along the extrema in the bathymetry while the strongest counter-current perturbations lie above the `saddle-point ridge'.
Naturally, vortex intensity approaches zero in the orthogonal orientations $\theta = 0$ and $90^\circ$. 

Circulation intensity increases with increasing 
non-dimensional wavenumbers modulus $\knil$, which constitutes increasing
the relative depth.
Since depth is used as the normalisation length scale, increasing $\knil$ amounts to changing the wall undulation wavelength while keeping its amplitude and the water depth fixed. 
We have observed with the viscous flows that 
the increase in circulation intensity flattens out around $\knil\gtrsim\pi$ beyond which point 
the vortex motion is no longer restricted
by the upper boundary. 
With the transient inviscid flows, the intensity of vortex wave growth continues to increase with higher wavenumber moduli.

Figure~\ref{fig:free_surf:theta} reveals 
that the rotational direction of vortices relative to the bathymetry can switch as a function of $\theta$, i.e. at a critical value of $\theta$ the extrema of $\psi(z)$ change sign.
From \eqref{eq:R} we see that the forcing term $\mc R(z)$ in the Rayleigh/Orr--Sommerfeld equation is always positive in the range $0^\circ<\theta < 45^\circ$ whenever $U(z)$ is monotonically increasing. 
The switch in rotational direction therefore
occurs in the range $45^\circ<\theta < 90^\circ$.
Furthermore, $\mc R(z)$ is directly proportional to $(\pnil')^2$ at $\theta = 45^\circ$, which vanishes in the shallow water limit $\knil\to0$ as the shallow flow becomes depth uniform.
The critical angle is therefore always $\theta=45^\circ$ in the shallow water limit. 
Tests with a wide range of parameters indicate that
the critical angle increases with increasing wavenumber and so does the relative difference in intensity between the two rotational orientations; 
in the deeper viscous flows ($k_1=3\pi$ and $\pi$), the `negative' circulation at high $\theta$-values is orders of magnitude weaker than at the `positive' peak in the $5^\circ$--$25^\circ$ range. 
Increasing the depth reduces the $\theta$ of strongest circulation.

The vertical location of the vortex centres is depicted in figure~\ref{fig:free_surf:zmax}, again as function of $\theta$.
Locations are not strongly dependent on $\theta$, but a sharp jump is observed where the switch in rotational direction occurs. 
This is because more than a single vortex, rotating in opposite directions, is present vertically in the region of this switch. 
These are both weak in intensity.
The plot in figure~\ref{fig:free_surf:zmax} shows the location of the vortex whose magnitude is greater, hence there is a jump in location as a new vortex starts to dominate. 
To illustrate,  
figure~\ref{fig:free_surf:rotation_switch} 
shows the stream function $\psi(z)$ at a state close to the directional switch;
two extrema are present, meaning counter-rotating vortices aligned vertically. 

}

\begin{figure}%
\centering
\subfigure[
Primary flow profiles $U(z)$ from \eqref{eq:U}.
]{
\includegraphics[width=.4\columnwidth]{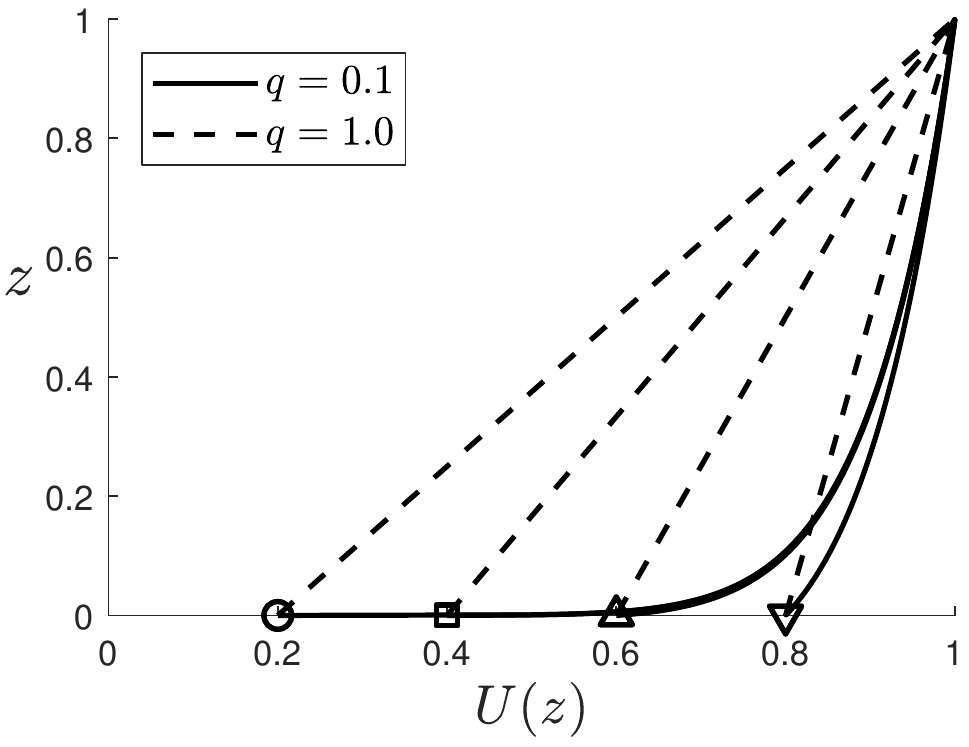}%
\label{fig:free_surf:delta_sensitivity:U}%
}
\\
\subfigure[Inviscid]{
\begin{minipage}{.33\columnwidth}
\includegraphics[width=\columnwidth]{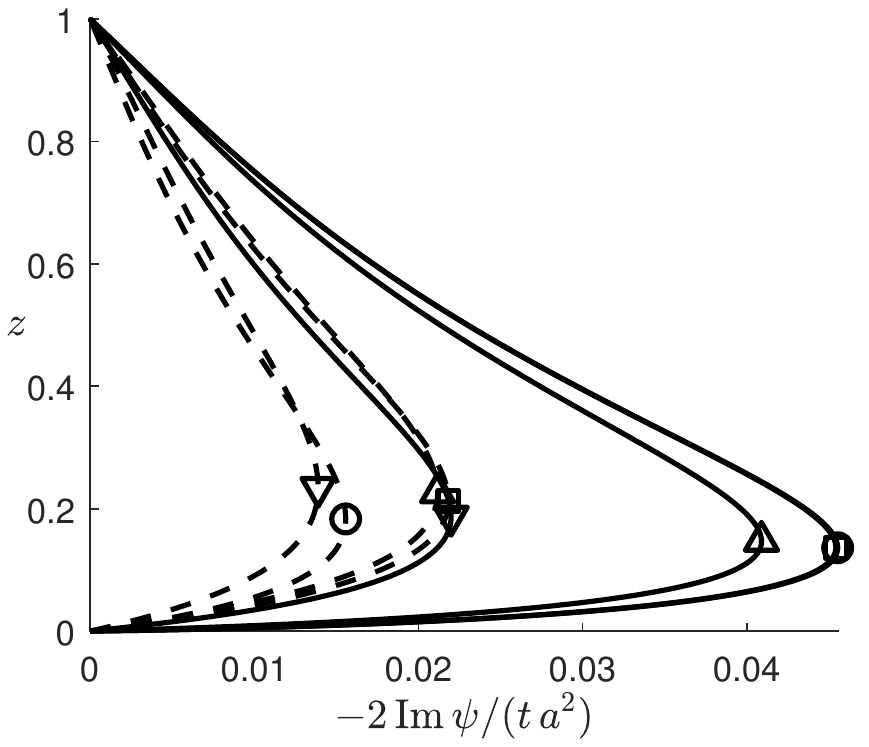}%
\\
\includegraphics[width=\columnwidth]{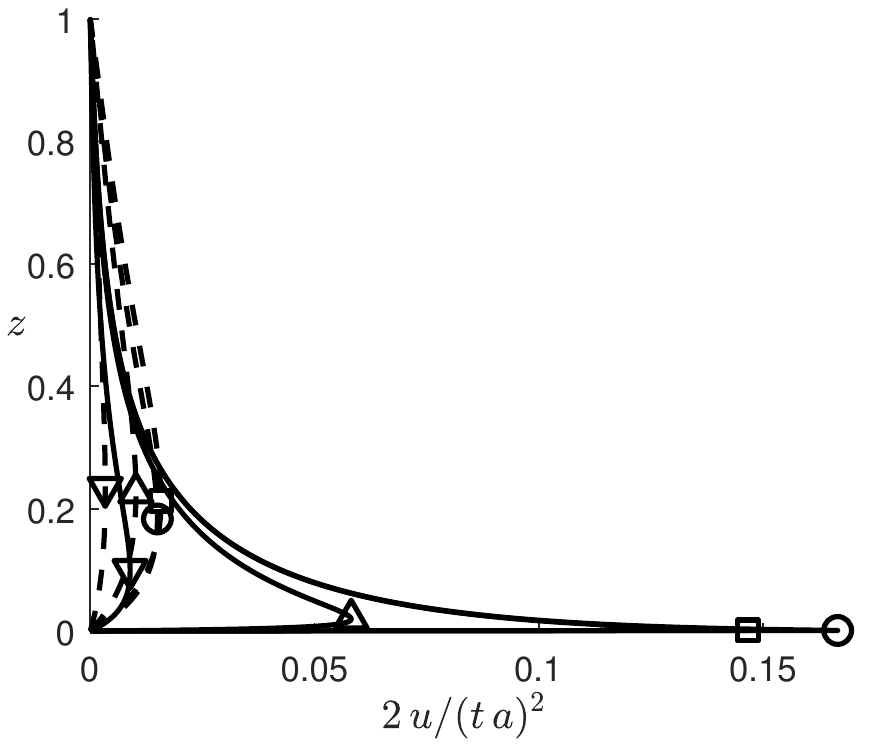}%
\end{minipage}%
\label{fig:free_surf:delta_sensitivity:inviscid}%
}%
\subfigure[Viscous, no-slip]{
\begin{minipage}{.33\columnwidth}
\includegraphics[width=\columnwidth]{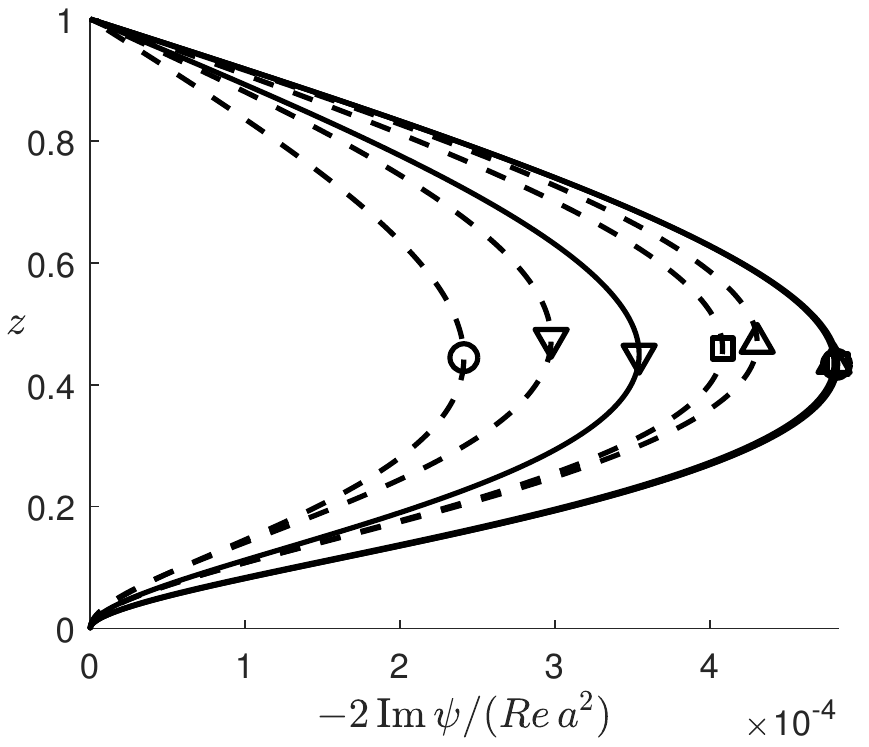}%
\\
\includegraphics[width=\columnwidth]{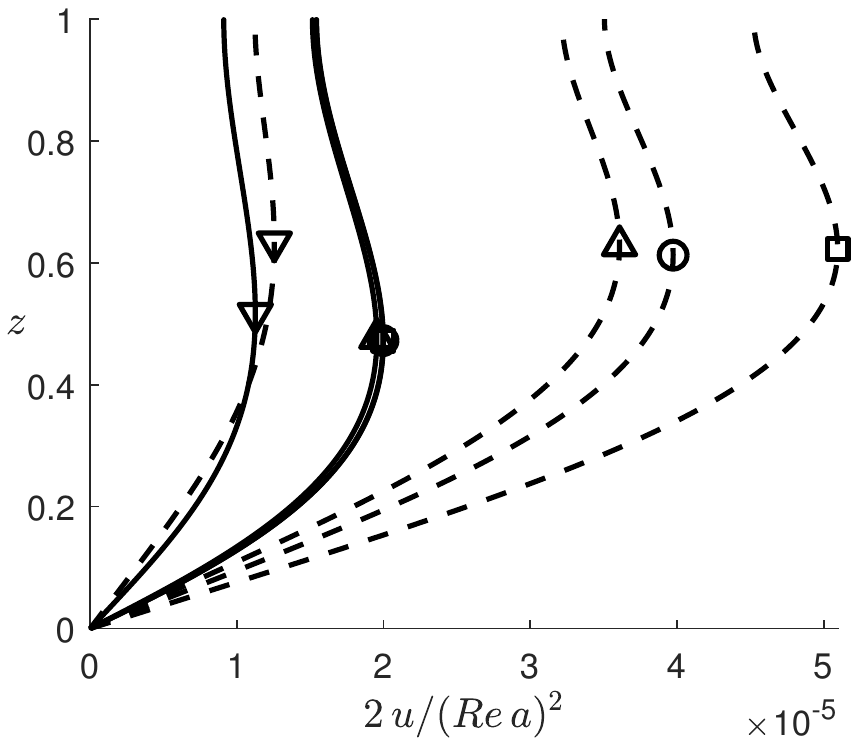}%
\end{minipage}%
\label{fig:free_surf:delta_sensitivitynoslip}%
}%
\subfigure[Viscous, full-slip]{
\begin{minipage}{.33\columnwidth}
\includegraphics[width=\columnwidth]{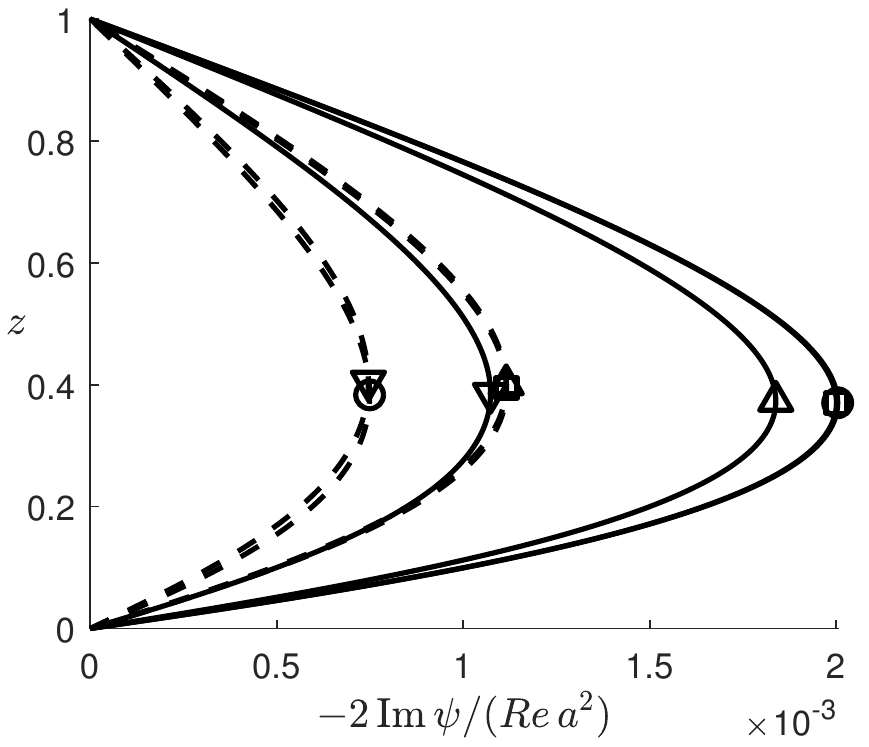}%
\\
\includegraphics[width=\columnwidth]{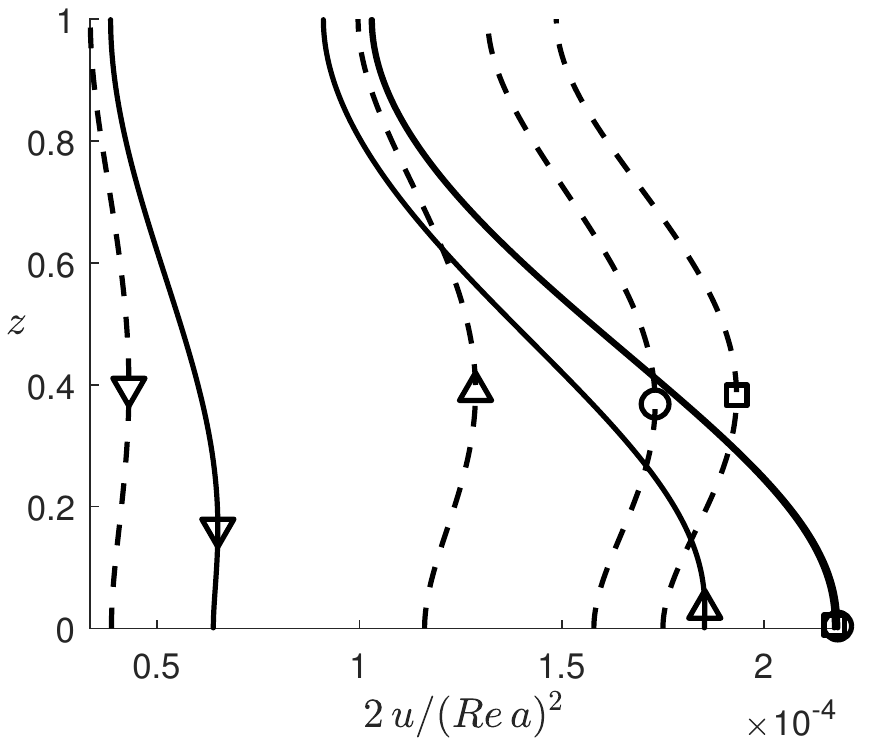}%
\end{minipage}%
\label{fig:free_surf:delta_sensitivity:fullslip}%
}%
\caption{
\arev{Free surface flow} for varying values of displacement parameter $\delta$ with
$\delta$-values such that $U(0) = 0.20$, $0.4$, $0.6$ and $0.8$, respective markers indicated in figure~\subref{fig:free_surf:delta_sensitivity:U}.
Solid: $q=0.1$, dashed: $q=1.0$.
Figure~\subref{fig:free_surf:delta_sensitivity:U}: Current profile. 
Figures~\subref{fig:free_surf:delta_sensitivity:inviscid}--\subref{fig:free_surf:delta_sensitivity:fullslip}:
\arev{Normalised}
stream function (top row) and \arev{normalised} streamwise velocity perturbation (bottom row) \arev{as function of vertical position, graphed} for various values of $\delta$ with the power law model \eqref{eq:U}.
$\knil=\pi$, $\arctan(\kynil/\kxnil)=\pi/8$.
(Curves occasionally overlap.) 
}%
\label{fig:free_surf:delta_sensitivity}%
\end{figure}

The sensitivity to the displacement parameter $\delta$ is considered next. 
Figure~\ref{fig:free_surf:delta_sensitivity:U} shows the near-logarithmic profile $q=0.1$ and the linear profile $q=1.0$ for  $\delta$'s
corresponding to $U(0) = 0.20$, $0.4$, $0.6$ and $0.8$.
The other panels of figure~\ref{fig:free_surf:delta_sensitivity} show $\psi$ and $u$ for the $\knil=\pi$ bed topography.
The strongest circulation is sometimes observed at an intermediate $U(0)$-value  since its intensity depends on both shear strength and current velocity at the wall.
Notice that the near-logarithmic profile $q=0.1$ is fairly insensitive to $\delta$ for all but the largest $\delta$-value. 
We interoperate this as a balance in logarithmic profiles between decreasing (increasing) shear and increasing (decreasing) current velocity at the wall.

\begin{figure}%
\centering
\subfigure[Inviscid]{
\begin{minipage}{.33\columnwidth}
\includegraphics[width=\columnwidth]{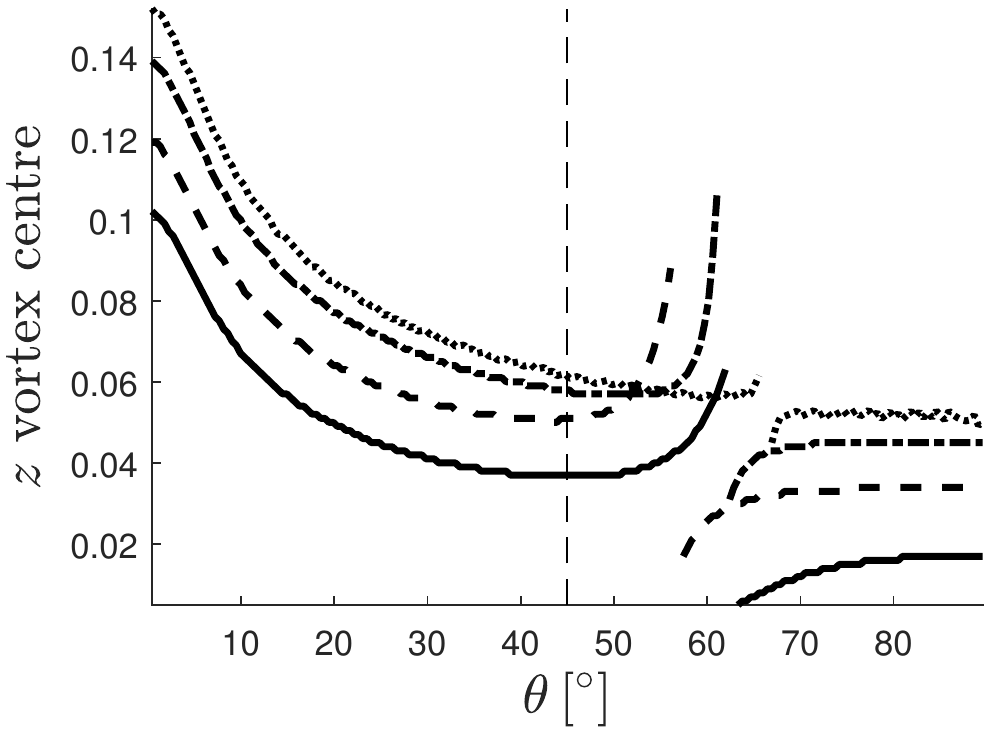}%
\\
\includegraphics[width=\columnwidth]{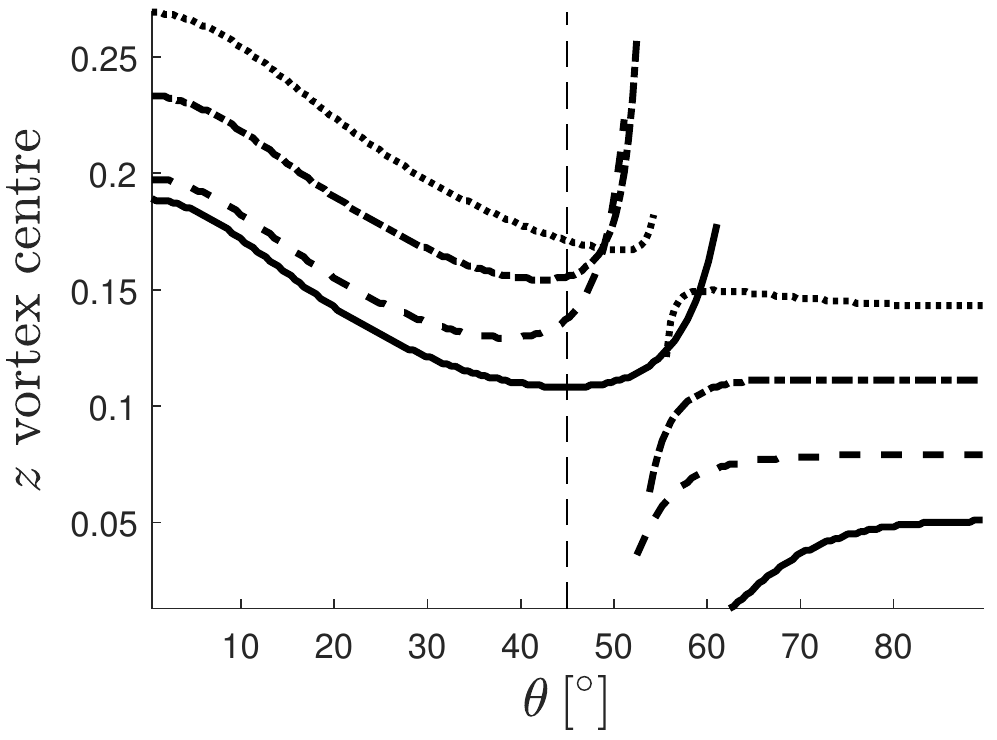}%
\\
\includegraphics[width=\columnwidth]{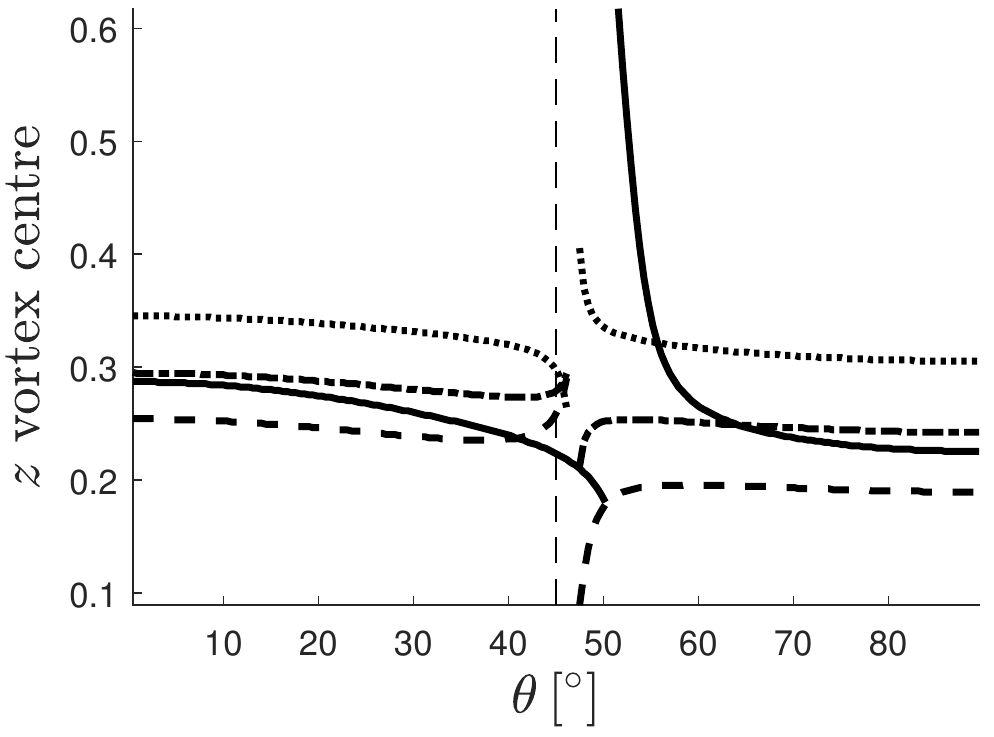}%
\end{minipage}%
\label{fig:free_surf:zmax:inviscid}%
}%
\subfigure[Viscous, no-slip]{
\begin{minipage}{.33\columnwidth}
\includegraphics[width=\columnwidth]{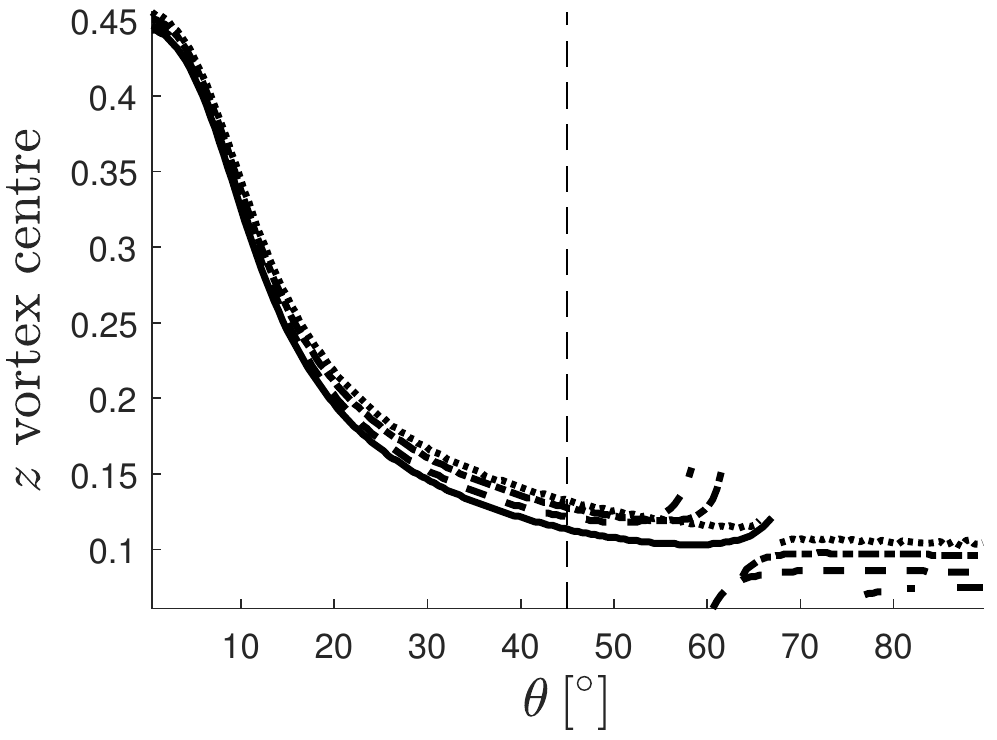}%
\\
\includegraphics[width=\columnwidth]{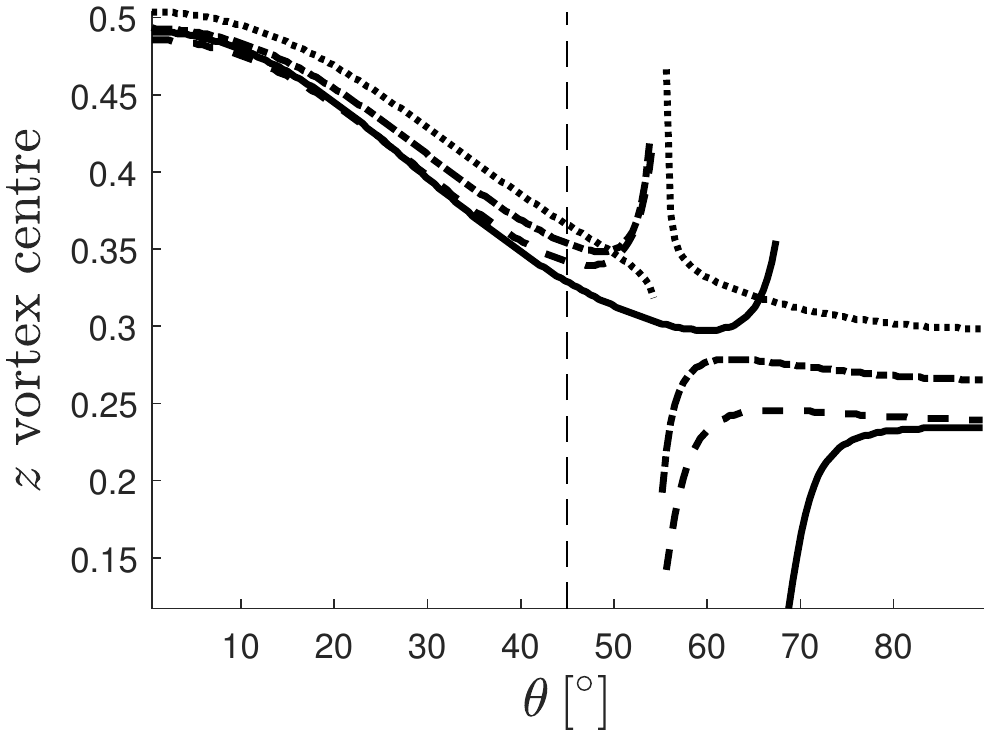}%
\\
\includegraphics[width=\columnwidth]{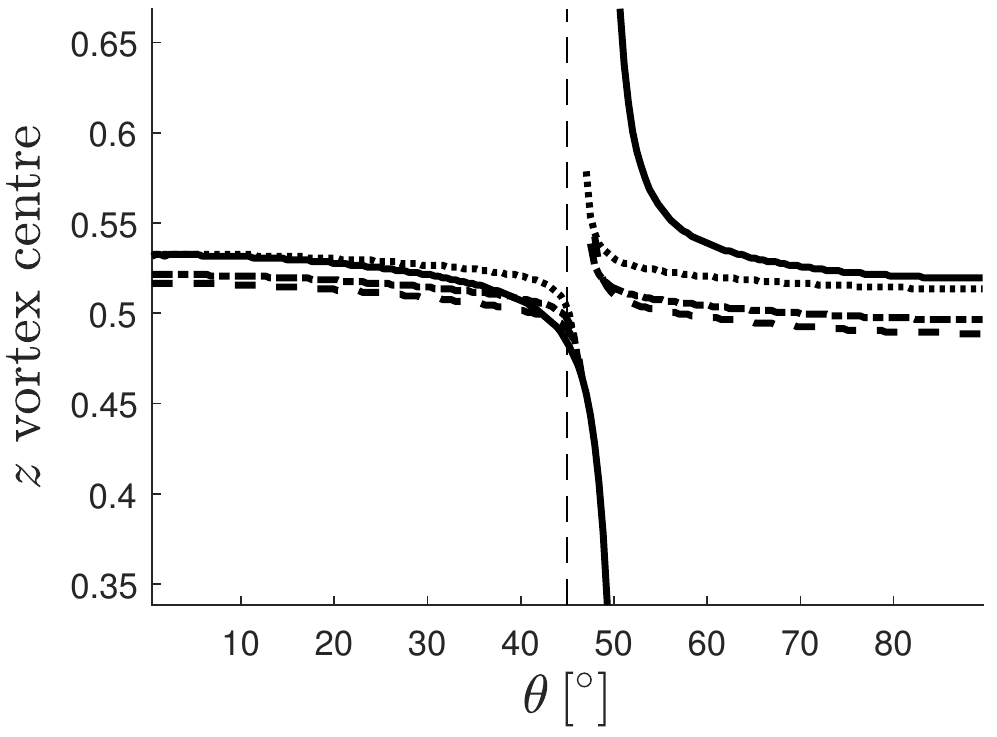}%
\end{minipage}%
\label{fig:free_surf:zmax:noslip}%
}%
\subfigure[Viscous, full-slip]{
\begin{minipage}{.33\columnwidth}
\includegraphics[width=\columnwidth]{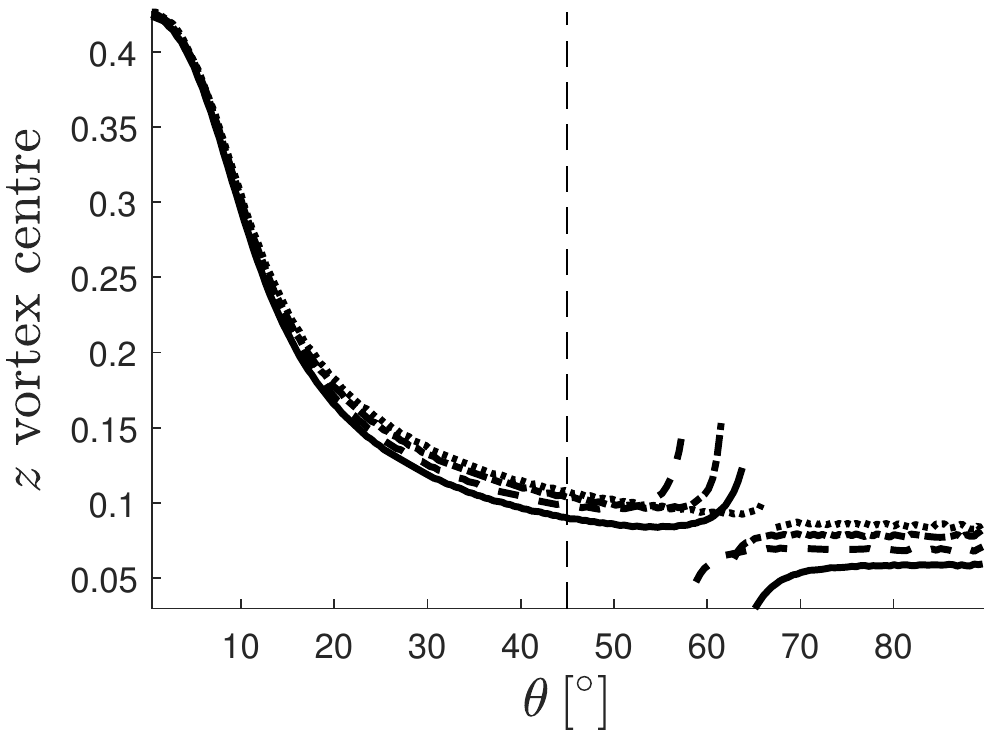}%
\\
\includegraphics[width=\columnwidth]{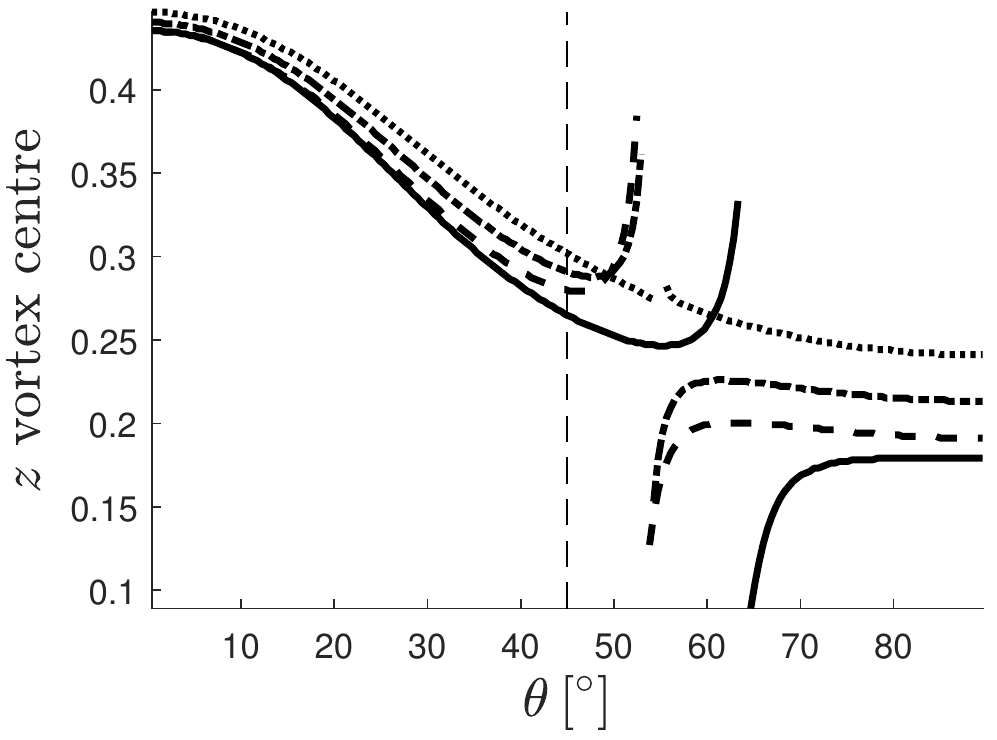}%
\\
\includegraphics[width=\columnwidth]{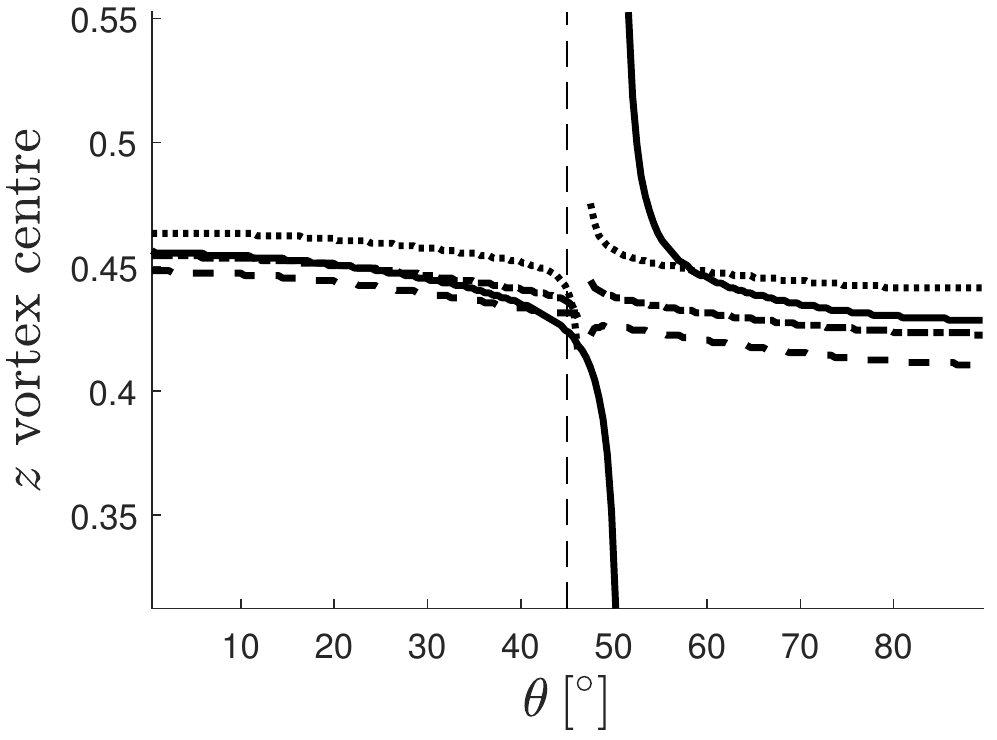}%
\end{minipage}%
\label{fig:free_surf:zmax:fullslip}%
}
\caption{
Vertical position of vortex centre as function of the wave vector angle $\theta=\arctan(\kynil/\kxnil)$ \arev{in free surface flow}.
See caption of figure~\ref{fig:free_surf:theta} for further details.
}%
\label{fig:free_surf:zmax}%
\end{figure}

\begin{figure}%
\centering
\begin{minipage}{.33\columnwidth}
\subfigure[Stream function]{
\includegraphics[width=\columnwidth]{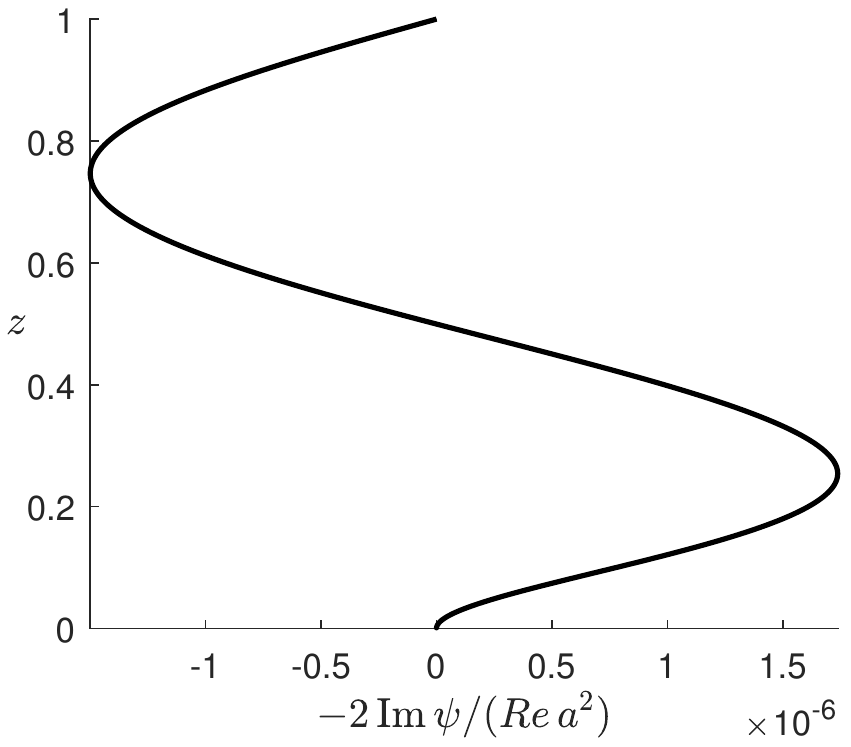}%
}
\end{minipage}
\hfill
\begin{minipage}{.6\columnwidth}
\subfigure[Streamlines]{
\includegraphics[width=\columnwidth]{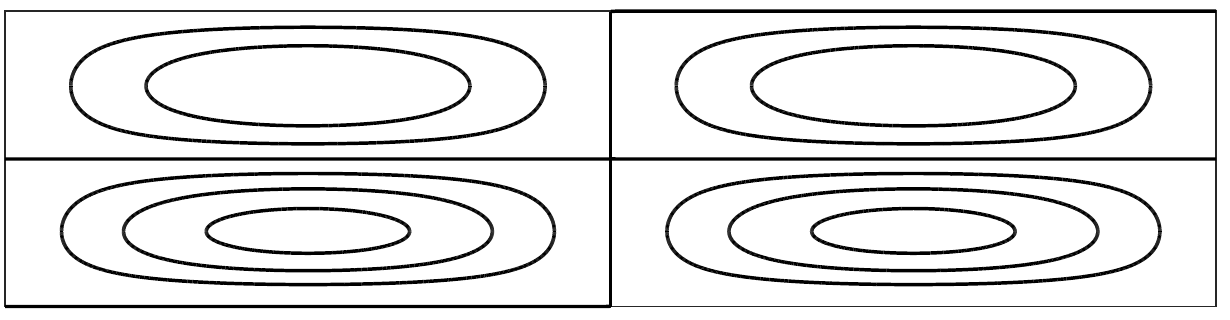}%
}
\end{minipage}
\caption{\arev{Normalised} stream function \arev{and streamlines} at a value of $\theta$ near the point where the rotational direction of vortices switches. 
Viscous, no-slip case with $\knil = 1.0$ $\theta=50^\circ$, $q=0.1$ and $U(0)=0.20$.
}%
\label{fig:free_surf:rotation_switch}%
\end{figure}

\subsection{Wall bounded laminar flow}
\label{sec:results:Langmuir_river}

\begin{figure}%
\centering
\includegraphics[width=.33\columnwidth]{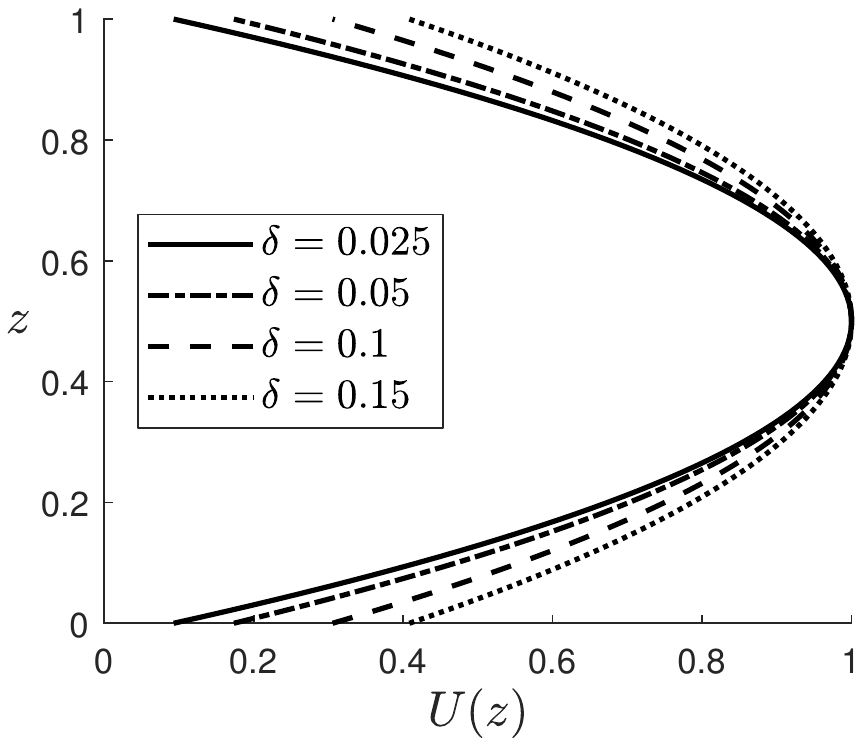}%
\caption{Parabolic current profile.}%
\label{fig:parabolic_profile}%
\end{figure}

For the flow bounded by walls both above and below we adopt a parabolic current profile
\begin{equation}
U(z) = 4 [1-\zeta(z)]\zeta(z); \qquad \zeta(z) = 
\arev{\frac{z + \delta}{1 + 2\delta}},
,
\label{eq:parabolic_profile}
\end{equation}
representing the laminar flow sketched in figure~\ref{fig:parabolic_profile}. Numerical integration
is employed for the first-order solution.

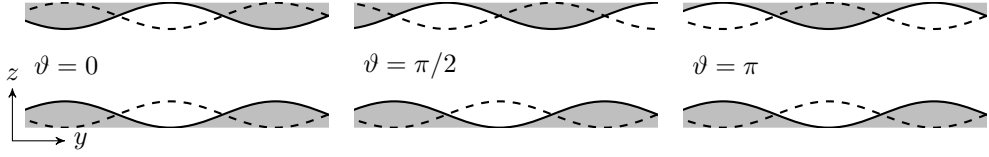
\begin{figure}%
\centering
\input{./figures/drawing_wallshift.tex}
\caption{Sketch of relative wall phase shift \arev{when looking along the streamwise direction}.}%
\label{fig:wallshift}%
\end{figure}

We can adjust the relative phase of the upper and lower boundary by allowing the mode amplitudes $\eta$ to be complex; 
letting the boundary undulation amplitude be $\etanil$ on both boundary, we 
shift the phase of the upper boundary an angle $\pi + \thy$ in the $y$-direction by letting
$\etab=\frac{\etanil}4$ and 
$\etas=-\frac{\etanil}4   \rme^{\pm \rmi\thy}$.
We then have
\begin{equation}
\petas 
= \sum_{s_x,s_y=\pm1}\br{ -\frac{\etanil}4   \rme^{\rmi s_y \thy} } 
\rme^{\rmi (s_x\kxnil x + s_y\kynil y)}
=-\etanil\cos (\kxnil x) \cos(\kynil y+\thy)
\label{eq:O2_modes_shifted}
\end{equation} 
while the lower boundary $\petab$ is still given by \eqref{eq:O2_modes}.
Relative wall positions are sketched in figure~\ref{fig:wallshift}.

\begin{figure}%
\centering
\subfigure[Inviscid]{
\begin{minipage}{.33\columnwidth}
\includegraphics[width=\columnwidth]{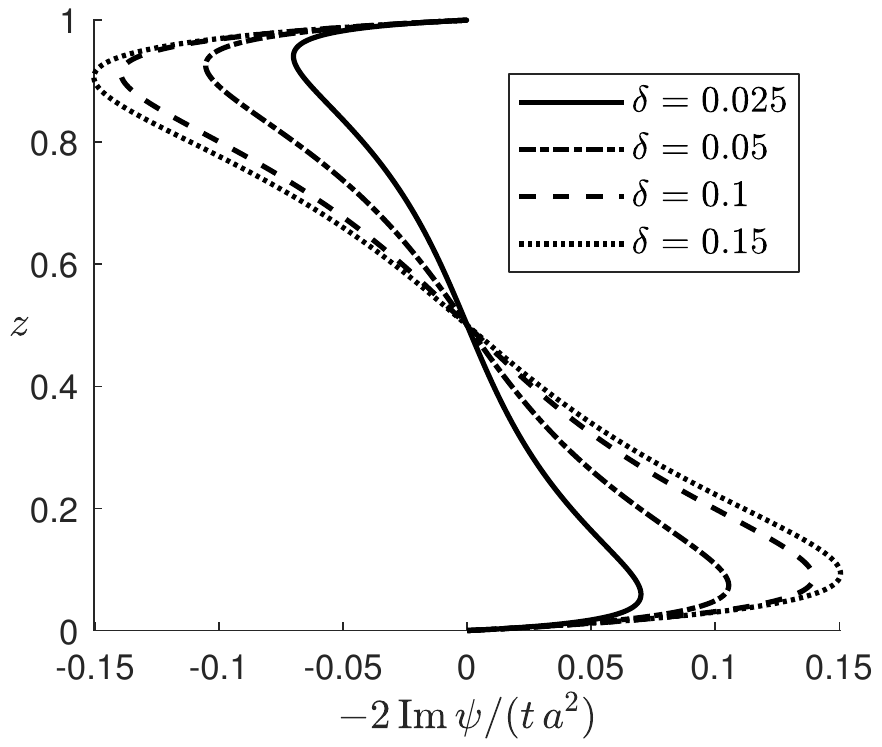}%
\\
\includegraphics[width=\columnwidth]{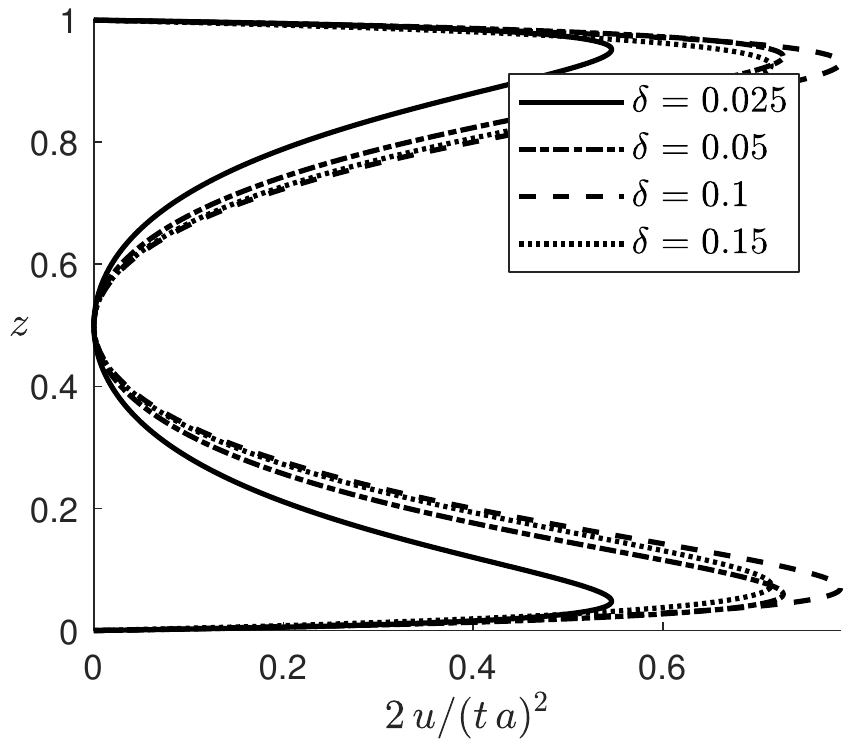}%
\end{minipage}%
\label{fig:parabolic_delta:theta0:inviscid}%
}%
\subfigure[Viscous, no-slip]{
\begin{minipage}{.33\columnwidth}
\includegraphics[width=\columnwidth]{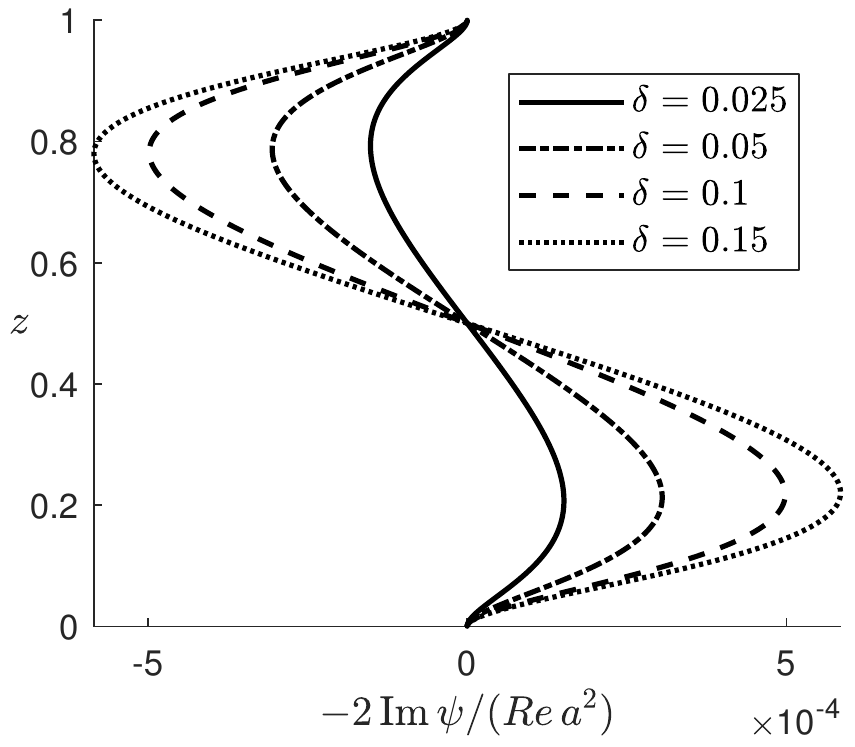}%
\\
\includegraphics[width=\columnwidth]{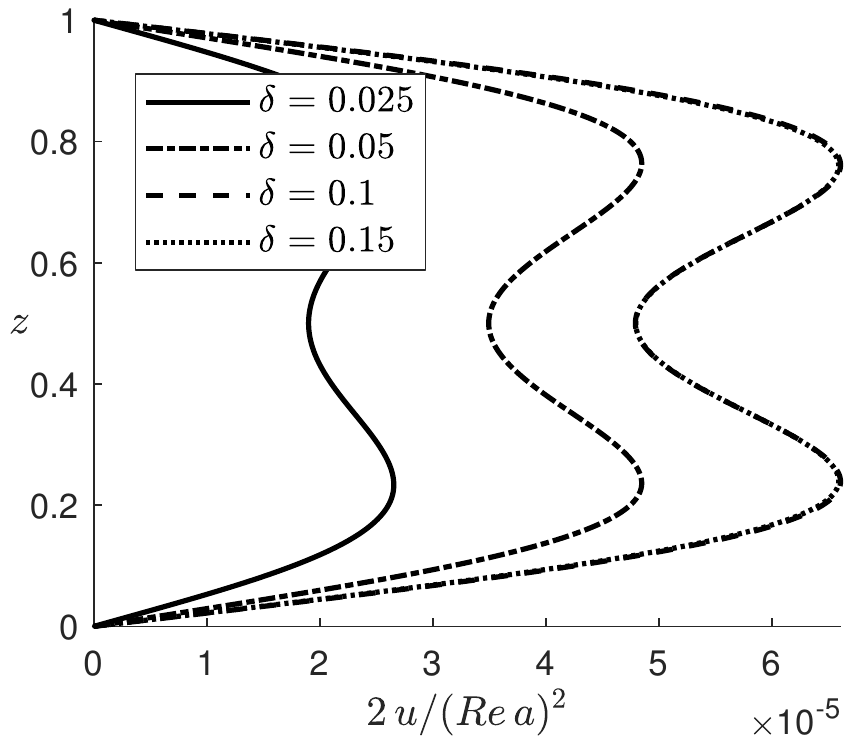}%
\end{minipage}%
\label{fig:parabolic_delta:theta0:noslip}%
}%
\subfigure[Viscous, full-slip]{
\begin{minipage}{.33\columnwidth}
\includegraphics[width=\columnwidth]{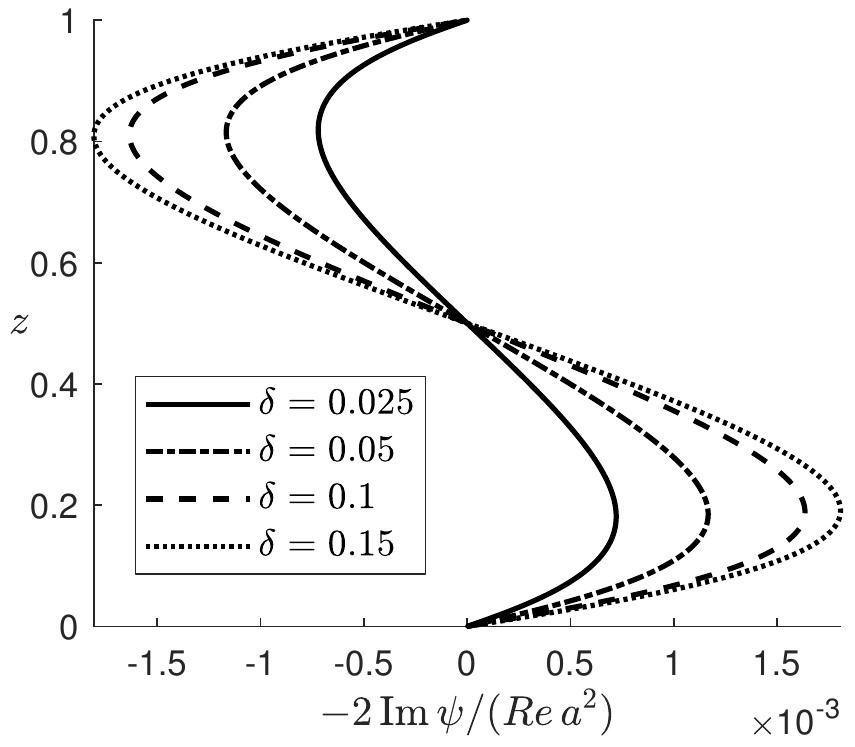}%
\\
\includegraphics[width=\columnwidth]{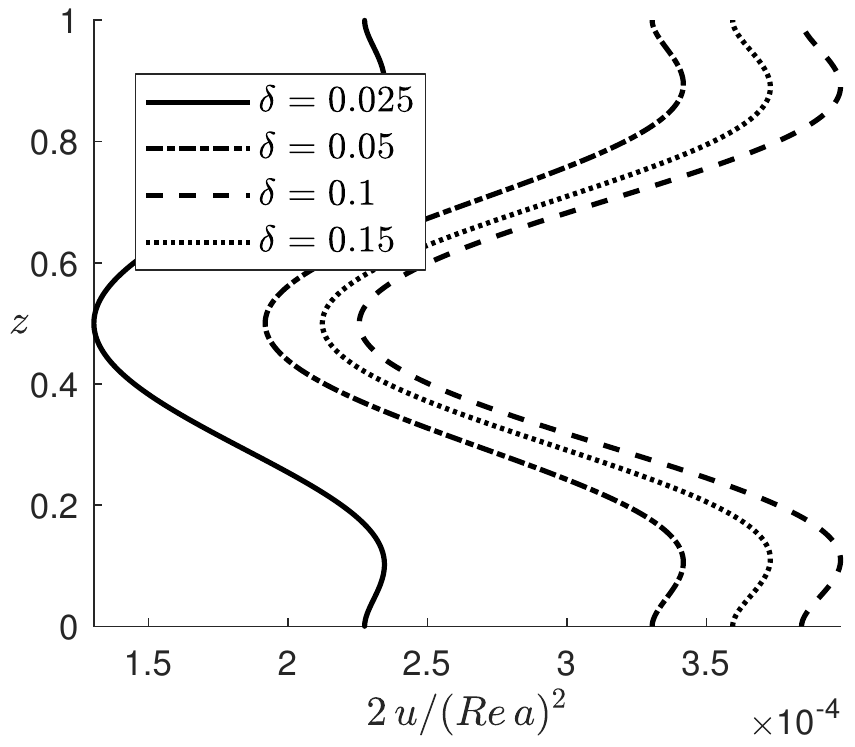}%
\end{minipage}%
\label{fig:parabolic_delta:theta0:fullslip}%
}%
\caption{
\arev{Doubly wall-bounded flow.}
\arev{Normalised} stream function and streamwise velocity perturbation for three values of $\delta$.
Parabolic current profile (figure~\ref{fig:parabolic_profile}) and a wall bounded domain with $\thy=0$ (see figure~\ref{fig:wallshift}).
$\knil = 2\pi$, $\theta=\pi/8$.
}%
\label{fig:parabolic_delta:theta0}%
\end{figure}

\begin{figure}%
\centering
\subfigure[Inviscid]{
\begin{minipage}{.33\columnwidth}
\includegraphics[width=\columnwidth]{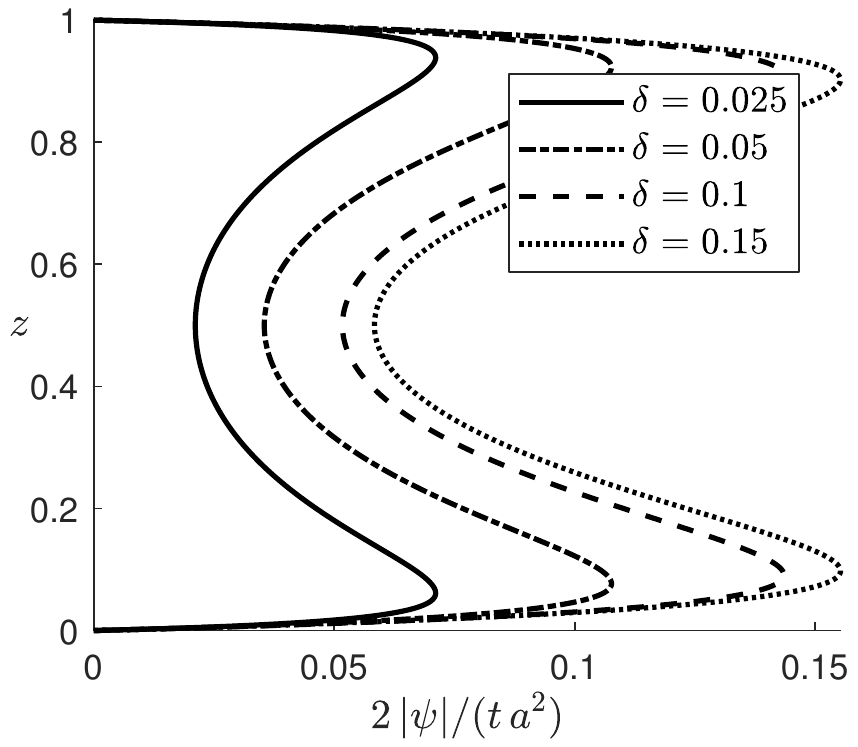}%
\\
\includegraphics[width=\columnwidth]{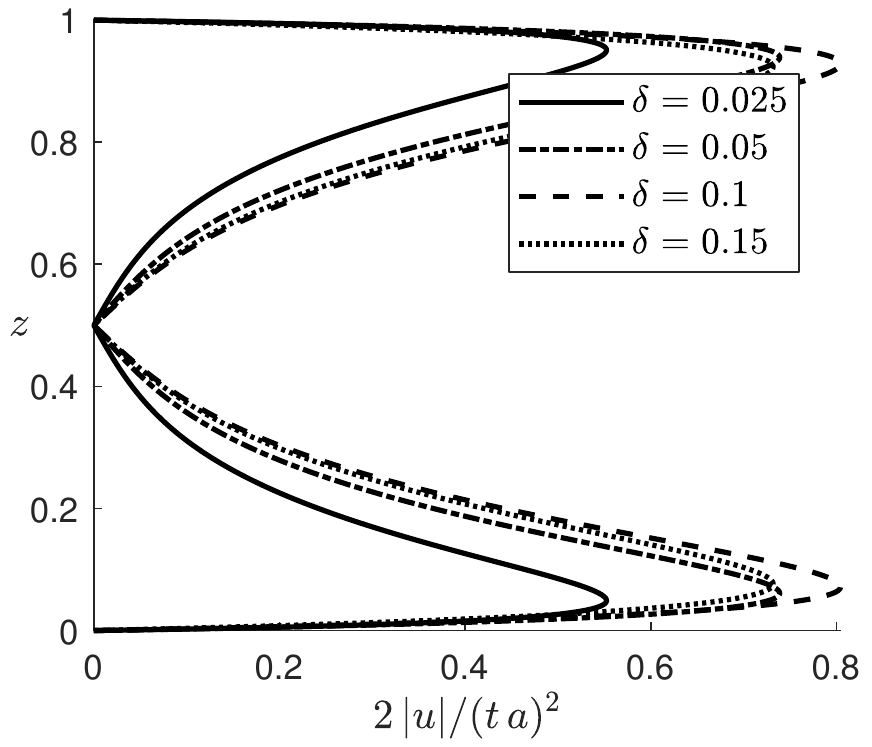}%
\end{minipage}%
\label{fig:parabolic_delta:thetapi2:inviscid}%
}%
\subfigure[Viscous, no-slip]{
\begin{minipage}{.33\columnwidth}
\includegraphics[width=\columnwidth]{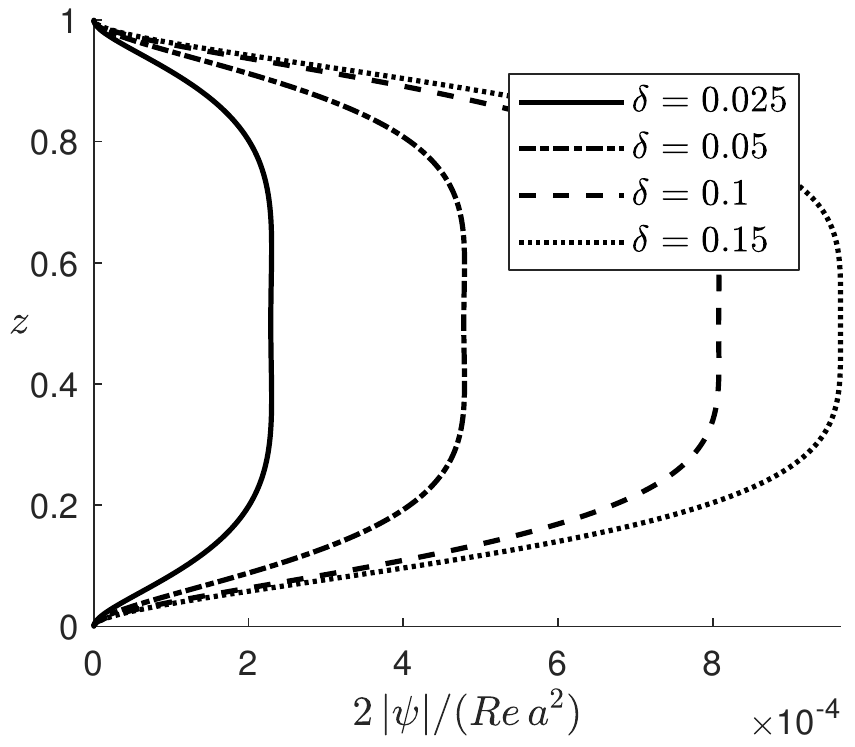}%
\\
\includegraphics[width=\columnwidth]{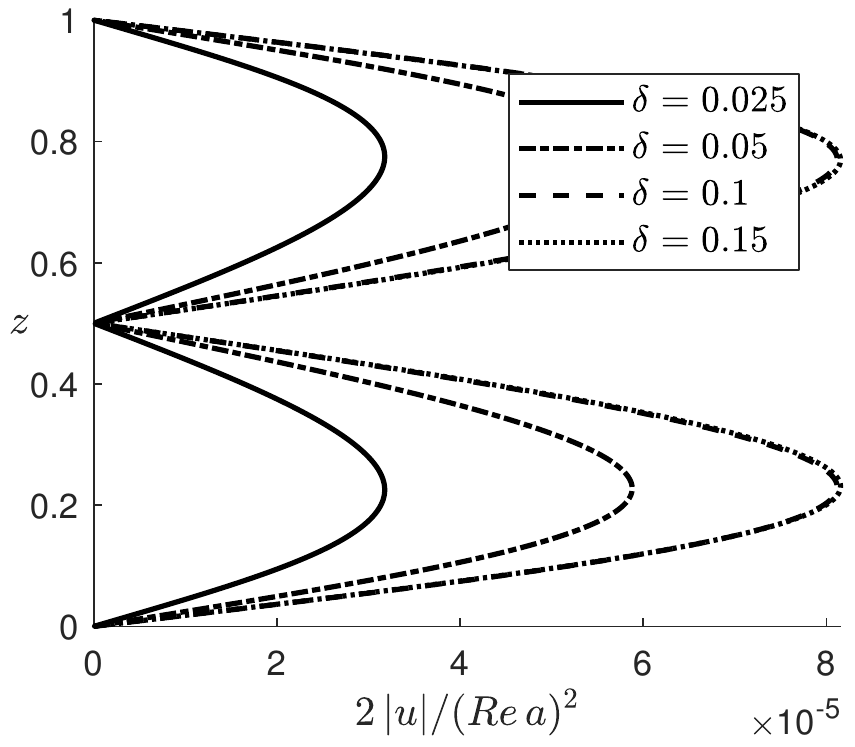}%
\end{minipage}%
\label{fig:parabolic_delta:thetapi2:noslip}%
}%
\subfigure[Viscous, full-slip]{
\begin{minipage}{.33\columnwidth}
\includegraphics[width=\columnwidth]{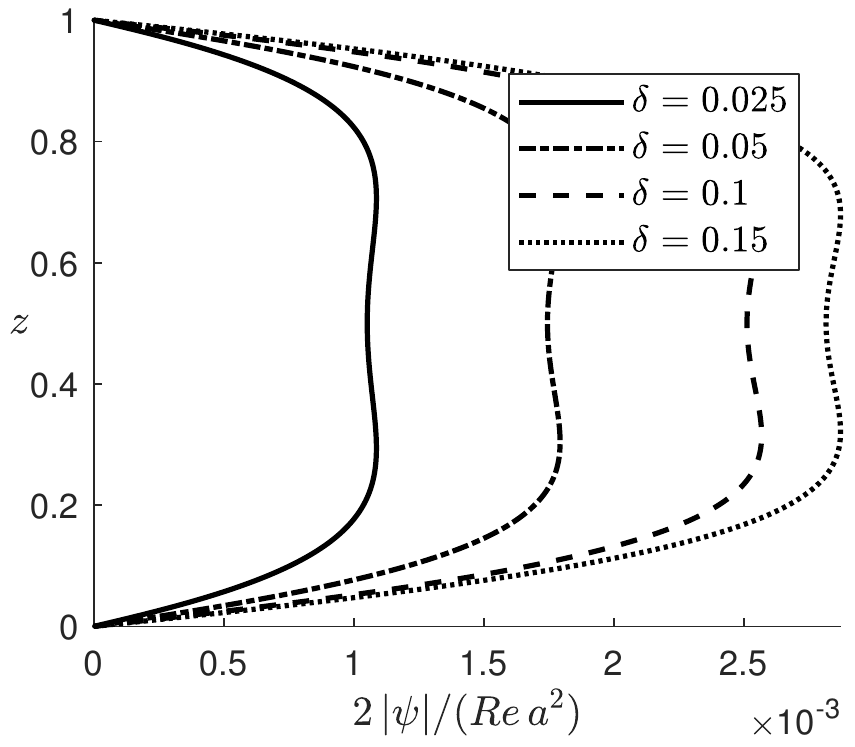}%
\\
\includegraphics[width=\columnwidth]{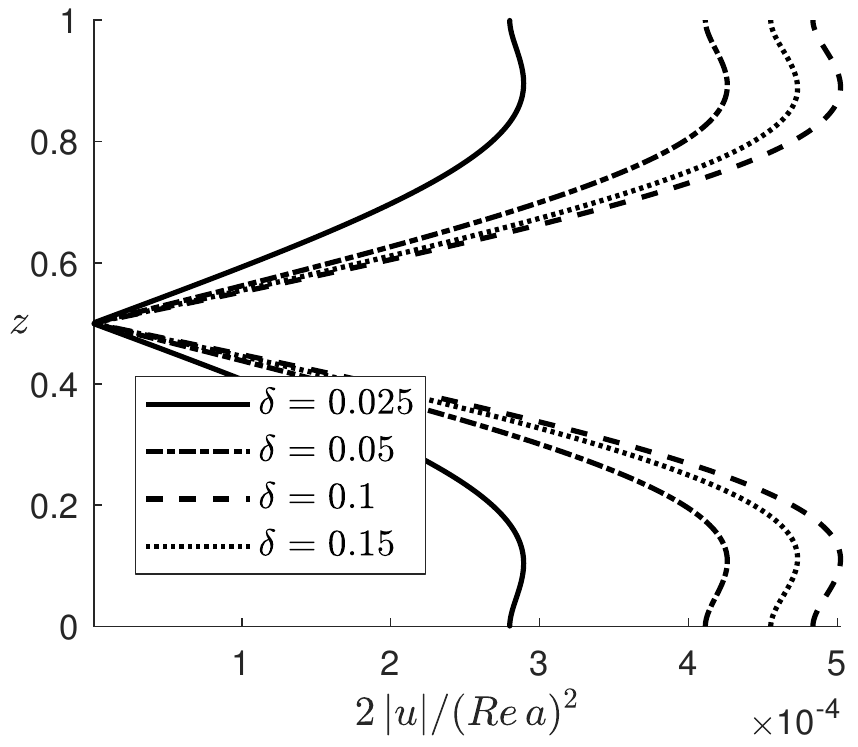}%
\end{minipage}%
\label{fig:parabolic_delta:thetapi2:fullslip}%
}%
\caption{
Same as figure~\ref{fig:parabolic_delta:theta0} but with  $\thy=\pi/2$.
}%
\label{fig:parabolic_delta:thetapi2}%
\end{figure}

\begin{figure}%
\subfigure[Inviscid]{
\begin{minipage}{.33\columnwidth}
\includegraphics[width=\columnwidth]{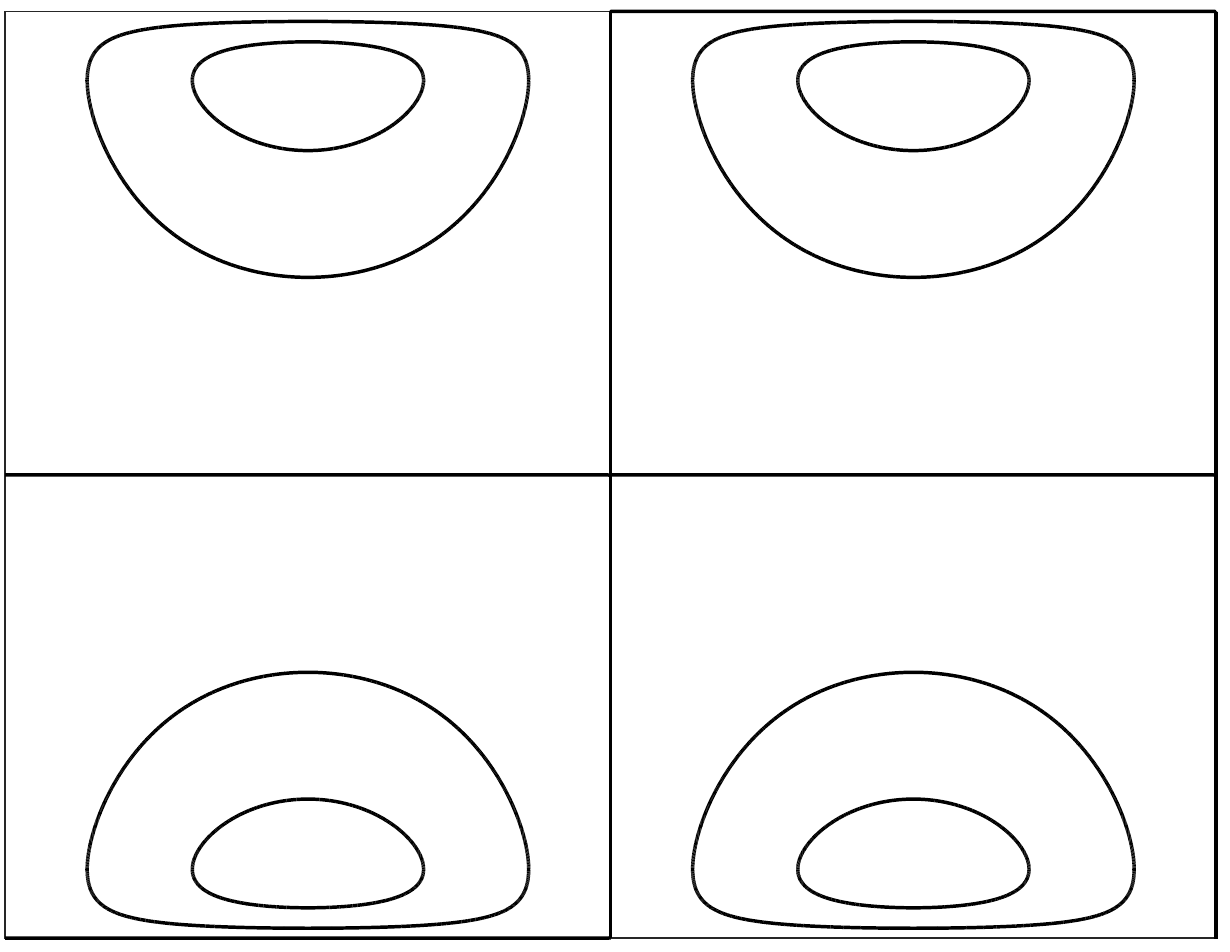}%
\\
\includegraphics[width=\columnwidth]{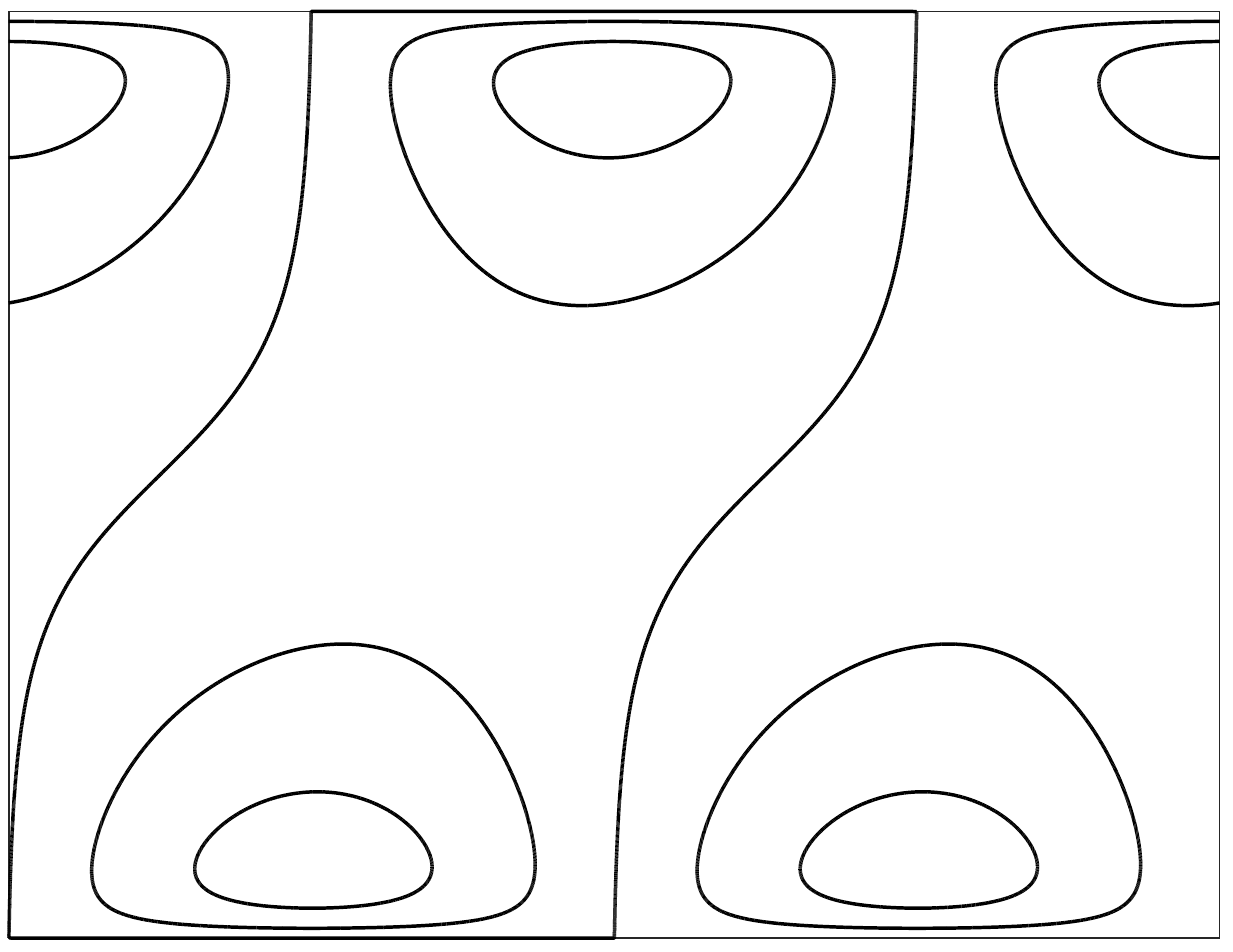}%
\\
\includegraphics[width=\columnwidth]{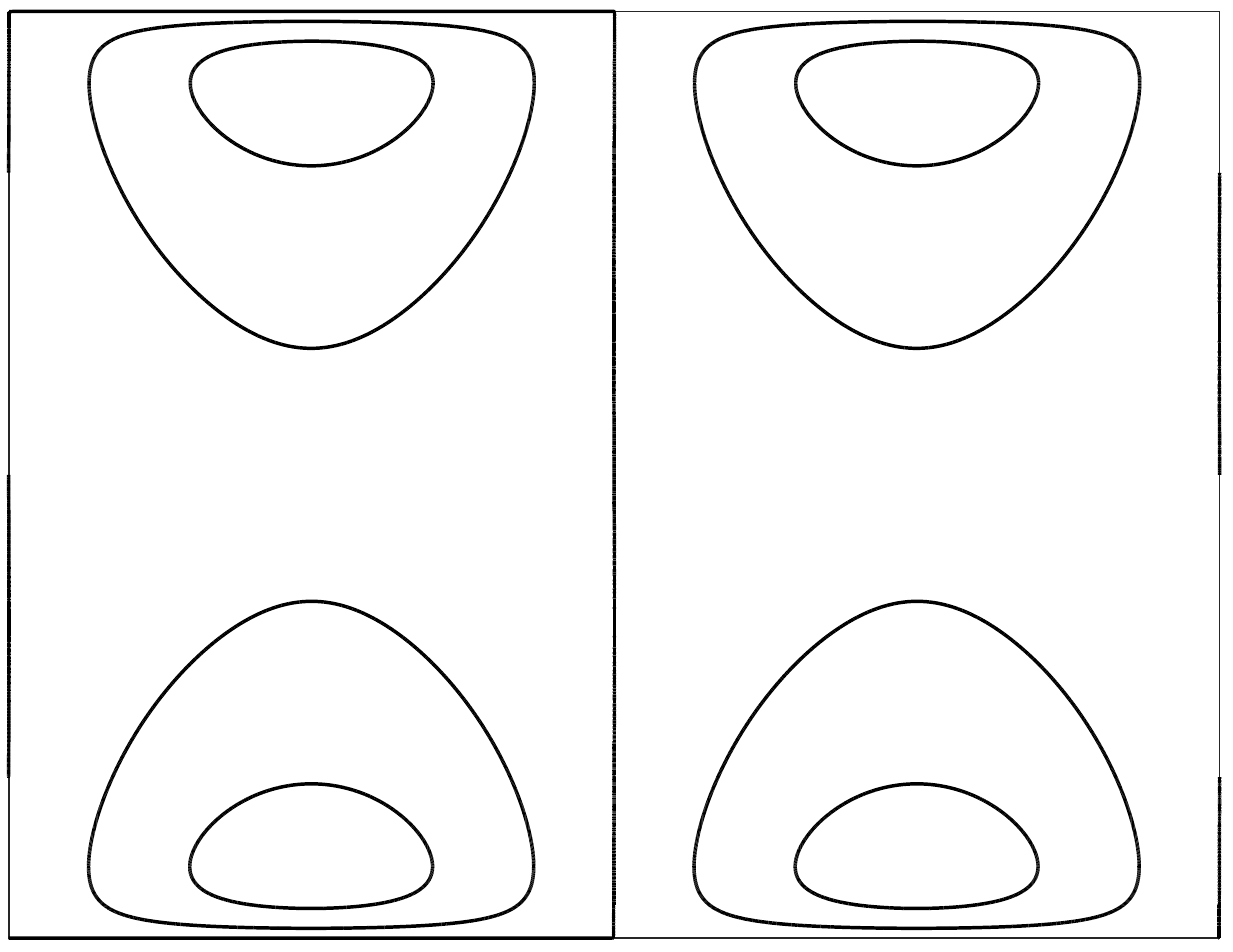}%
\end{minipage}
}%
\subfigure[Viscous, no-slip]{
\begin{minipage}{.33\columnwidth}
\includegraphics[width=\columnwidth]{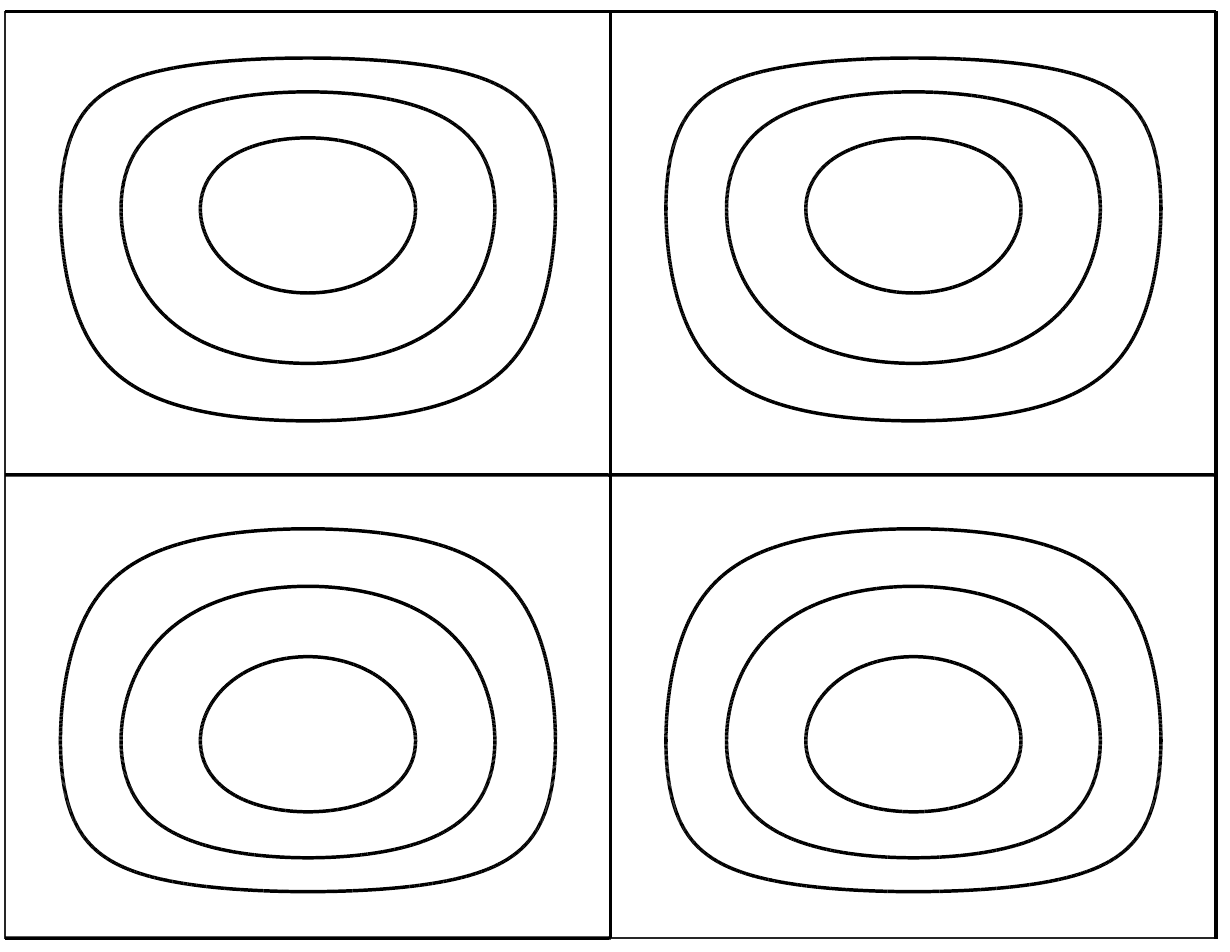}%
\\
\includegraphics[width=\columnwidth]{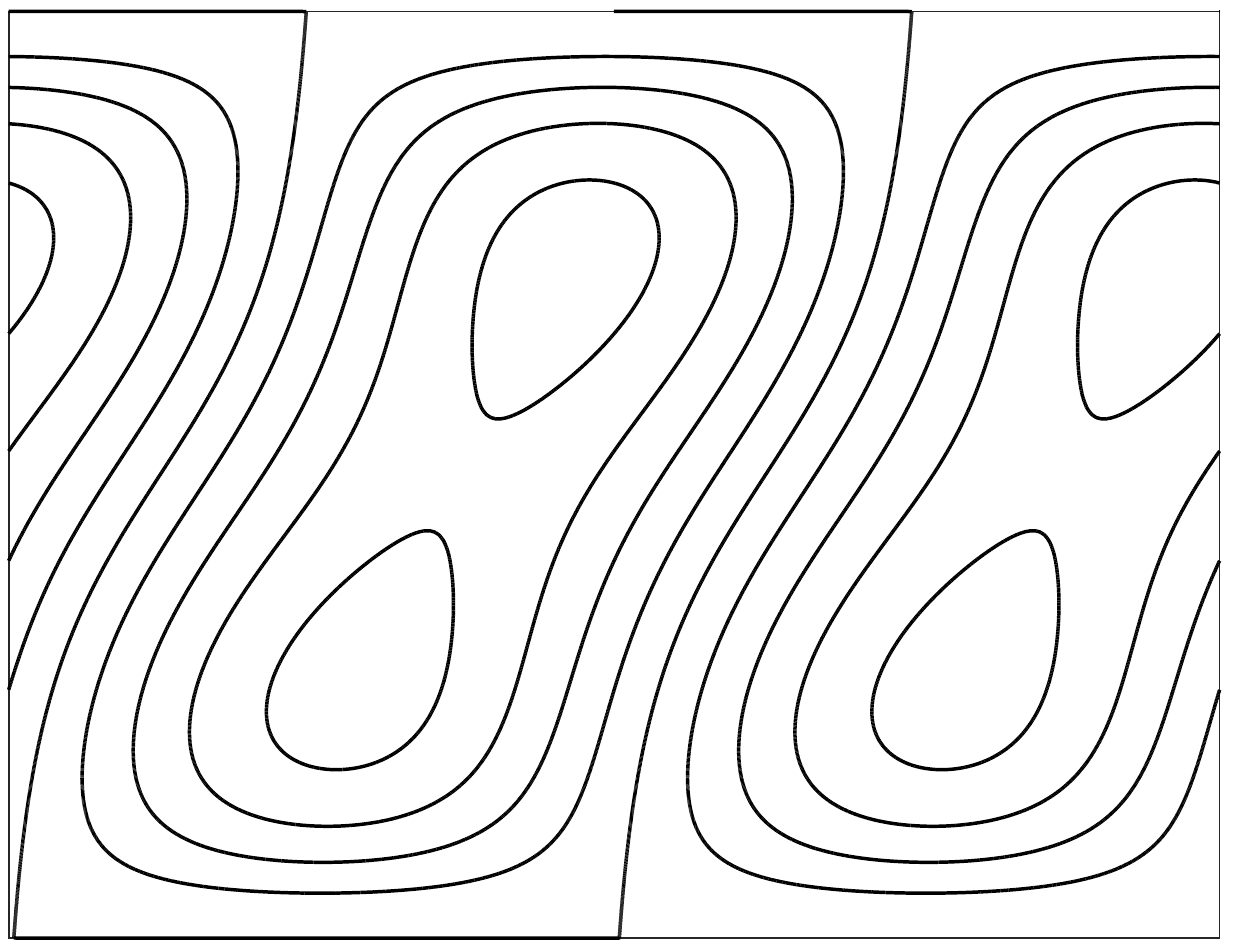}%
\\
\includegraphics[width=\columnwidth]{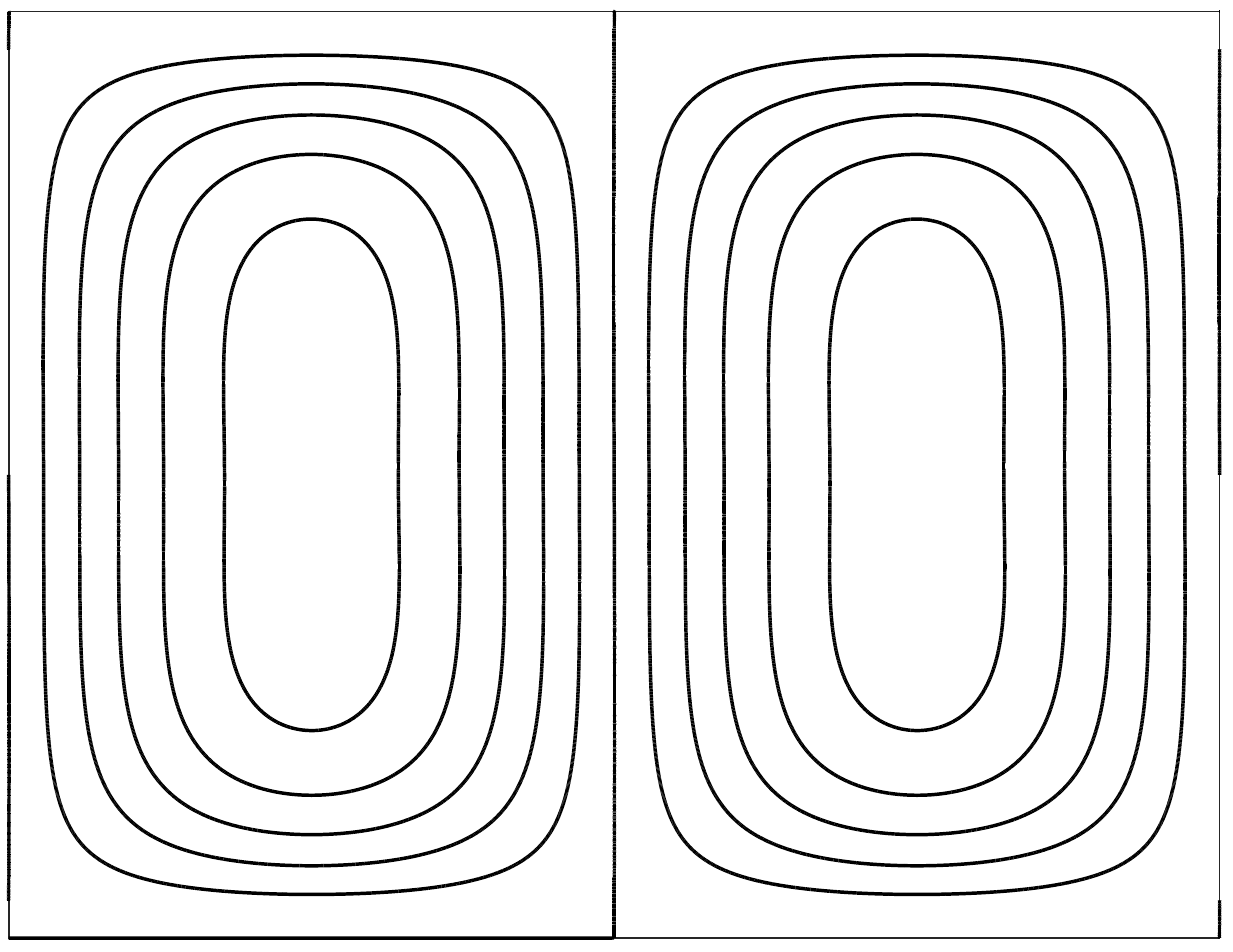}%
\end{minipage}
}%
\subfigure[Viscous, full-slip]{
\begin{minipage}{.33\columnwidth}
\includegraphics[width=\columnwidth]{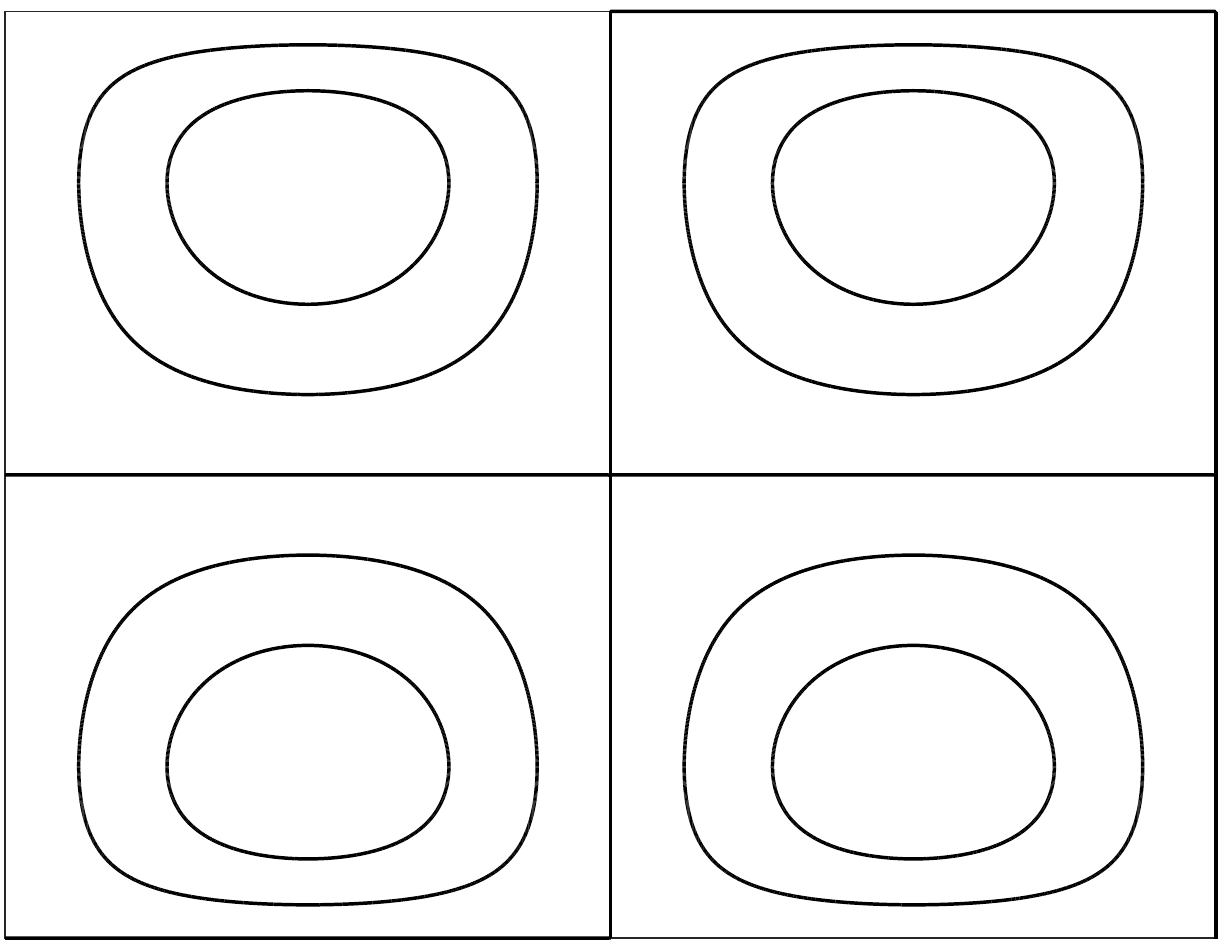}%
\\
\includegraphics[width=\columnwidth]{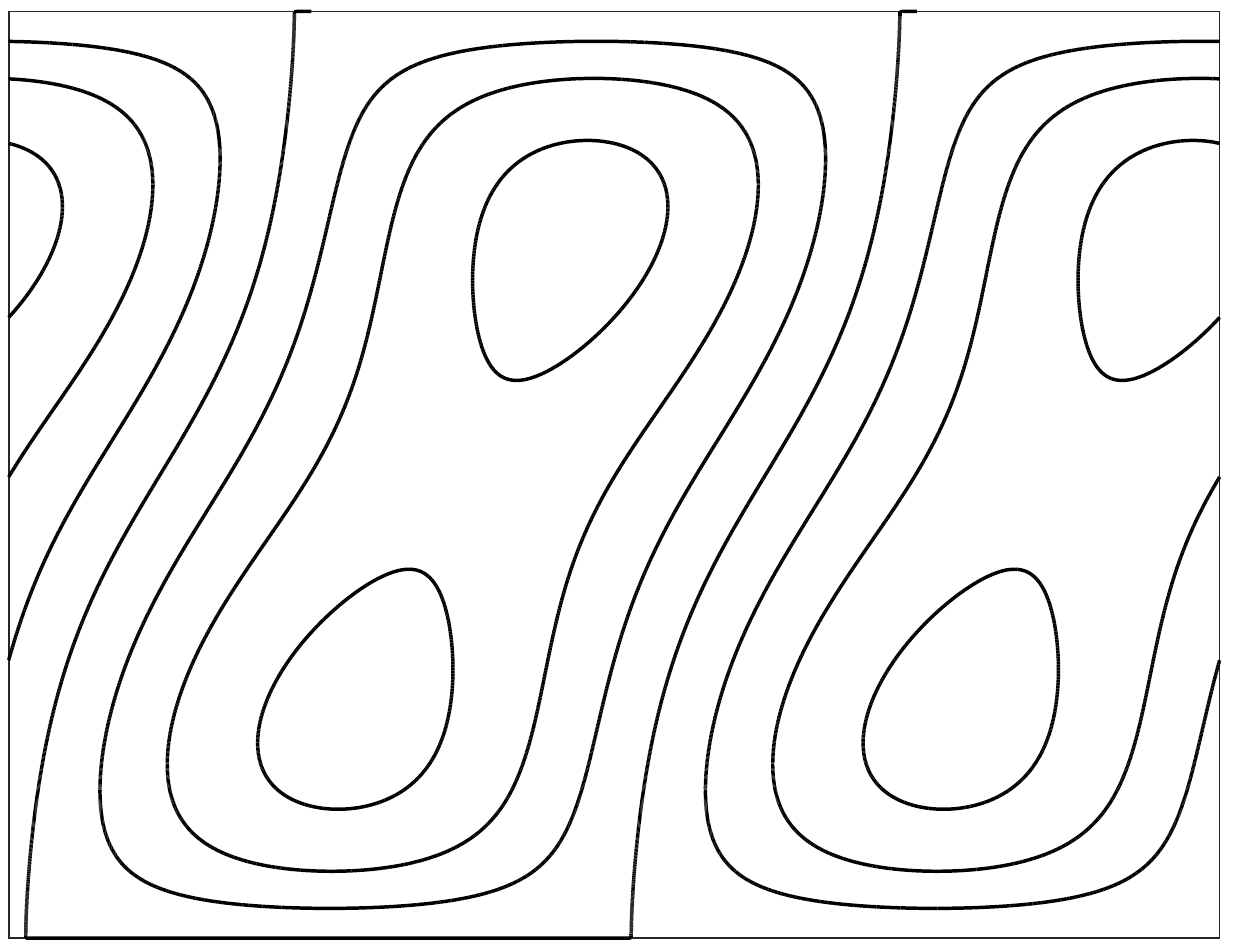}%
\\
\includegraphics[width=\columnwidth]{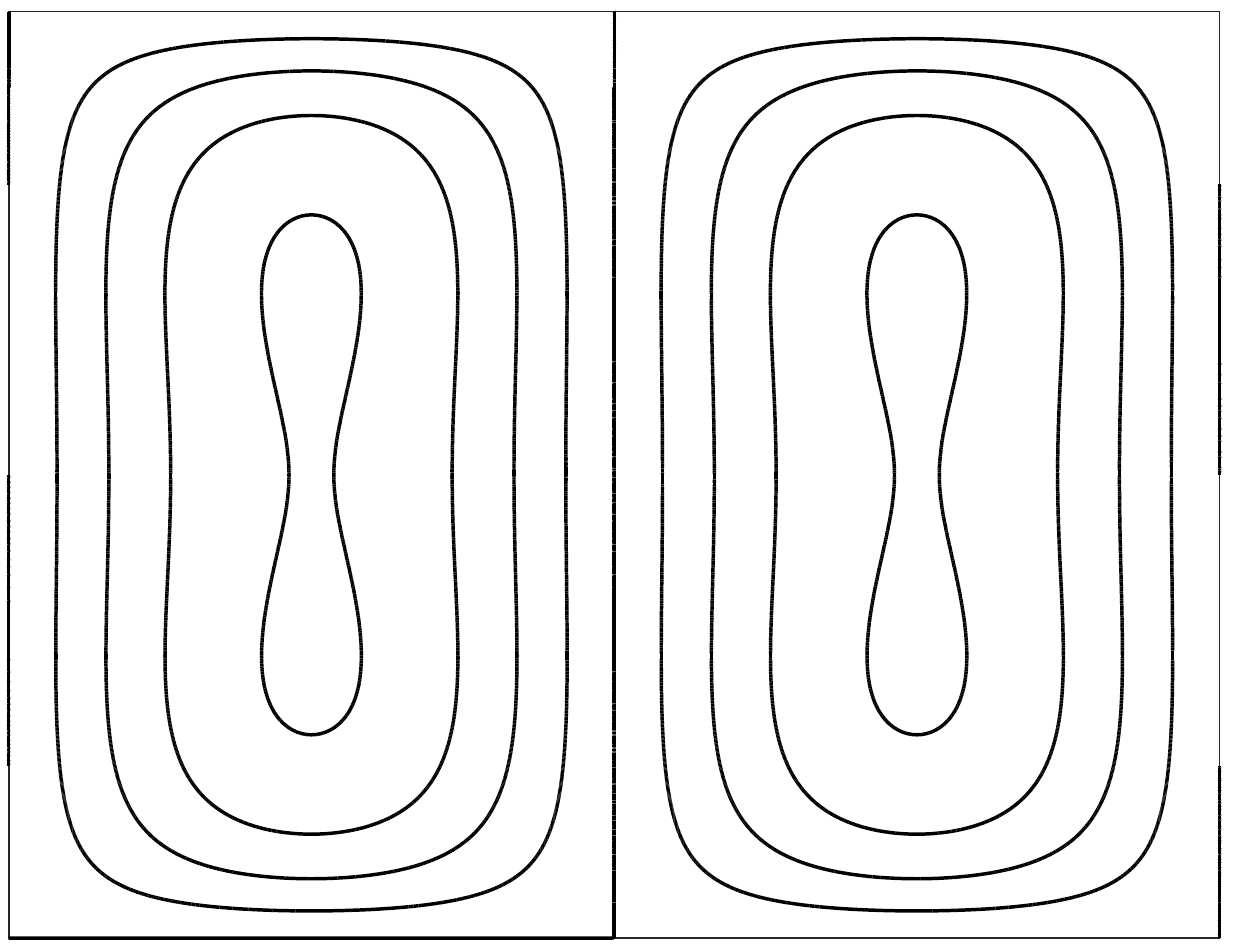}%
\end{minipage}
}%
\caption{
\arev{Wall bounded flow.}
Streamlines $\p\psi= \mr{constant}$ for $\thy=0$ (top row, cf.\ figure~\ref{fig:parabolic_delta:theta0}), $\thy=\pi/4$ (middle row) and $\thy=\pi/2$ (bottom row, cf.\ figure~\ref{fig:parabolic_delta:thetapi2}).
$\delta = 0.05$ and $\knil = 2\pi$, $\theta=\pi/8$.
}%
\label{fig:parabolic_streamlines}%
\end{figure}

Similar to the previous section, amplitudes of the stream function and streamwise velocity component are shown in figure~\ref{fig:parabolic_delta:theta0} and \ref{fig:parabolic_delta:thetapi2} as functions of $z$ and $\delta$.
Functions $\psi(z)$ and $u(z)$ each have both real and imaginary components whenever $\thy\neq0$ due to the spanwise skewness
induced (see \eqref{eq:ReIm}).
Absolute values are therefore plotted when $\thy\neq0$.
Streamlines, as they appear in physical space, are shown in figure~\ref{fig:parabolic_streamlines}.
Figure~\ref{fig:parabolic_delta:theta0}--\ref{fig:parabolic_streamlines} show that
boundaries which are in-phase ($\thy=0$) generate vertical vortex pairs which 
rotate in opposite senses, 
generating horizontal jets in the cross-flow plane along the horizontal plane 
$z=1/2$.
Conversely, boundaries which are shifted $\thy=\pi/2$ generate co-rotating vortex pairs.
If sufficiently closely situated, these will negate each other's horizontal flow along the symmetry plane  $z=1/2$ and merge into a single vortex spanning the cross-section.
Such a single vortex generally generates stronger turnover than does a vortex pair.
Also, 
we observe in numerous computations that
the closer the vortices are squeezed together (the lower the value of $\knil$) the easier they merge into a single vortex spanning the vertical domain---more so for viscous flows than inviscid flows.

\section{Simulated laminar flow}
\label{sec:results:LBM}

\begin{figure}%
\centering
\begin{tabular}{ccc}
\def\y{.15} \def\d{.1}
\input{./figures/drawing_real_life.tex}
&\tikz{\node at (0,.5){$\Longrightarrow$}; \node at (0,0){};}&
\def\y{.15} \def\d{.1}
\input{./figures/drawing_transformed_flat.tex}
\\[2ex]
\def\y{.075} \def\d{.15}
\input{./figures/drawing_real_life.tex}
&\tikz{\node at (0,.5){$\Longrightarrow$}; \node at (0,0){};}&
\def\y{.075} \def\d{.15}
\input{./figures/drawing_transformed_flat.tex}
\end{tabular}
\caption{
Conceptual representation of a boundary layer with no-slip at the walls with a slip velocity reference boundary located some distance $\delta$ within the boundary layer.
Streamlines resemble the actual boundary curvature more closely when $\delta$ is smaller than the amplitude $\etanil$ (top sketch), although this makes the perturbation amplitudes large relative to the 
assumptions of the linear theory.
The situation is the opposite if  $\delta>\etanil$ (bottom sketch).
}%
\label{fig:real_life}%
\end{figure}
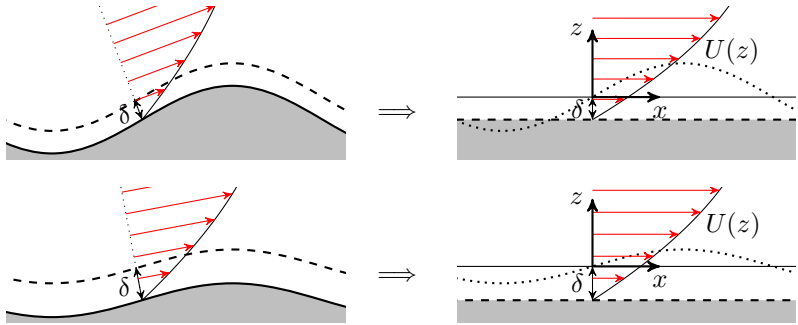

The perturbation theory presented so far sheds light on the subject of structurally generated Langmuir circulation, yet a vital question still remains to be answered: 
What happens to the boundary layer? 
The presence of Langmuir circulation beneath free surfaces is of course well established, but these boundaries allow for a slip velocity whereas flow stagnation takes place at a rigid boundary.
Will Langmuir circulation even appear near a no-slip boundary?
We have so far circumvented the issue by assuming that the bulk current velocity $U(z)$ \textit{does} have some slip velocity at the reference planes $z=0,1$.
This assumption is perhaps less alien when considering a turbulent boundary layer---assuming a slip governed by the `law of the wall' is a common approach within turbulence modelling---but is not easily justified in laminar flows. 
We must then consider wall undulation amplitudes that penetrate some depth into the flow field, and these amplitudes must in turn somehow be related to the undulation amplitude $\etanil$ and the displacement height $\delta$ introduced in the perturbation theory.
A sketch is presented in figure~\ref{fig:real_life} where
we regard a streamsurface within the flow as the `seen' surface.
The closer this surface is to the actual wall (the smaller the $\delta$) the more the boundary undulations will resemble the actual known undulation of the wall.
Further away the `seen' amplitude will be smaller and less sinusoidal as also the opposing boundary contributes to compress the streamlines.
On the other hand, 
the perturbation theory rests on the assumption that
the variation in current velocity across the perturbation is small relative to the velocity at the reference plane, which improves in validity at larger $\delta$.

We have at our disposal a code for numerical flow simulations using the lattice-Boltzmann method, with which we have generated simulation results within the laminar flow regime for the purpose of demonstrating that the 
artefacts of our perturbation theory do in fact manifest in real, nonlinear flows.
Although the lattice--Boltzmann method is usually not first choice in terms of accuracy and computational efficiency, it proved easily adaptable to the present problem 
since the undulated boundaries could be represented simply by marking nodes outside the flow domain as `solid' nodes. 
\arev{The code has been validated against two- and three-dimensional lid-driven cavity flow cases.
These benchmarks are presented in the supplementary  material.   
}

\begin{figure}%
\centering
\subfigure[$\thy=0$, $\etanil= 0.0391$]{
\includegraphics[width=.33\columnwidth]{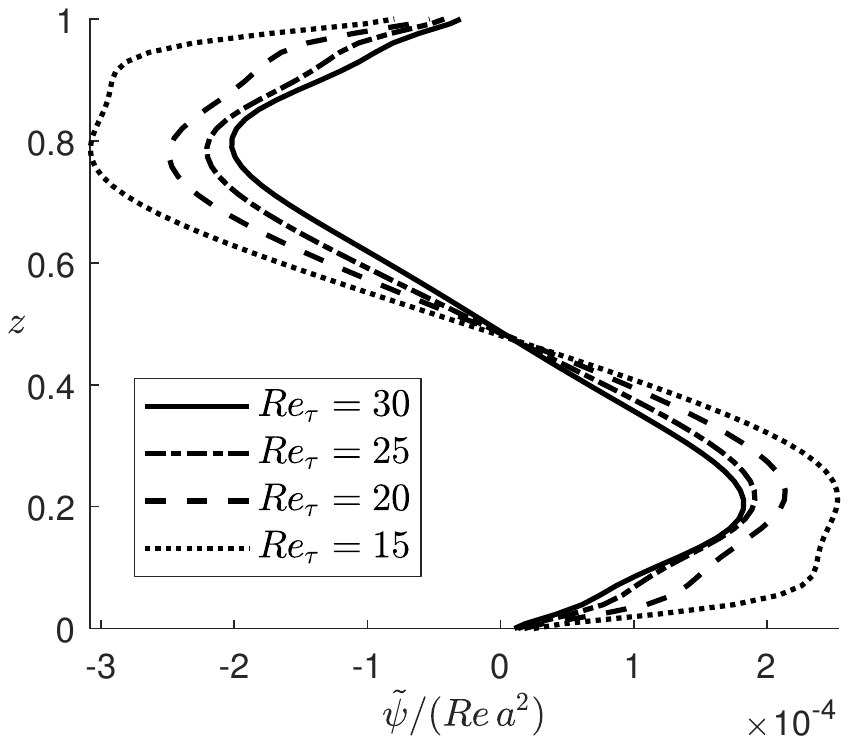}%
}\subfigure[$\thy=0$, $\etanil= 0.0625$]{
\includegraphics[width=.33\columnwidth]{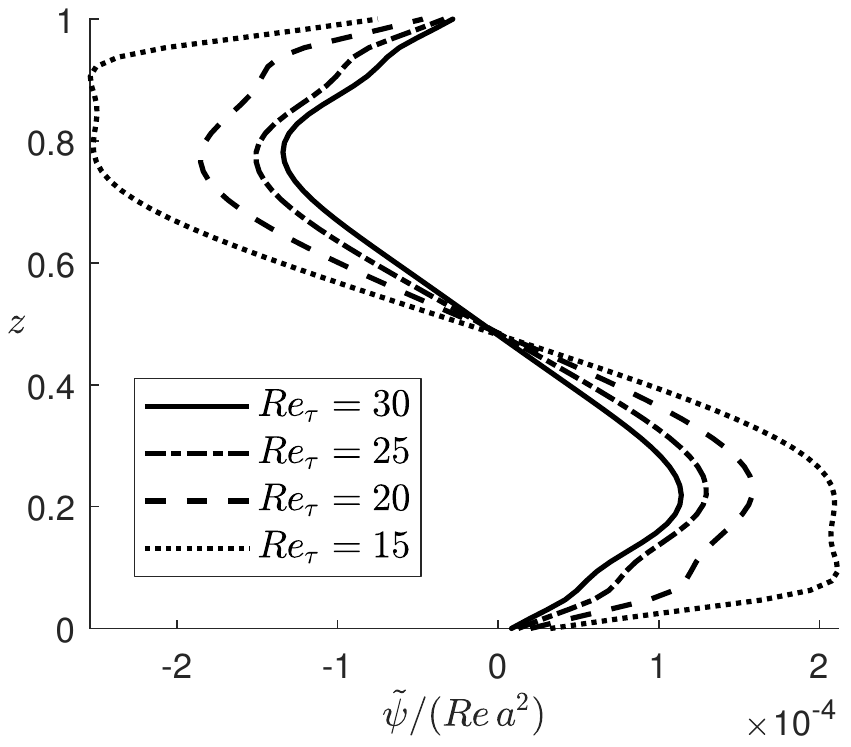}%
}\subfigure[$\thy=0$, $\etanil= 0.0938$]{
\includegraphics[width=.33\columnwidth]{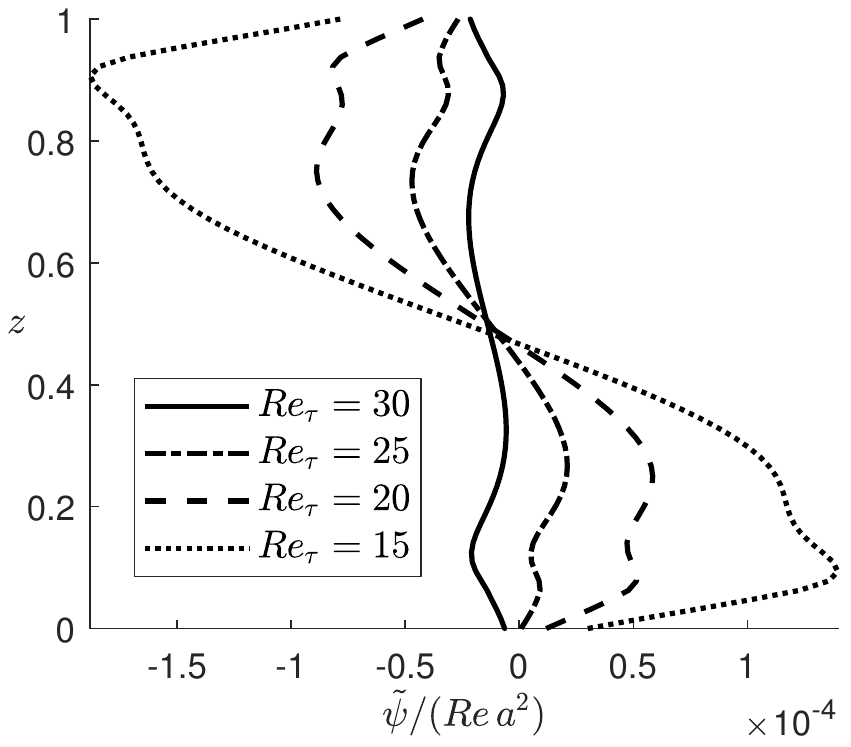}%
}
\\
\subfigure[$\thy=\pi/2$, $\etanil= 0.0391$ ]{
\includegraphics[width=.33\columnwidth]{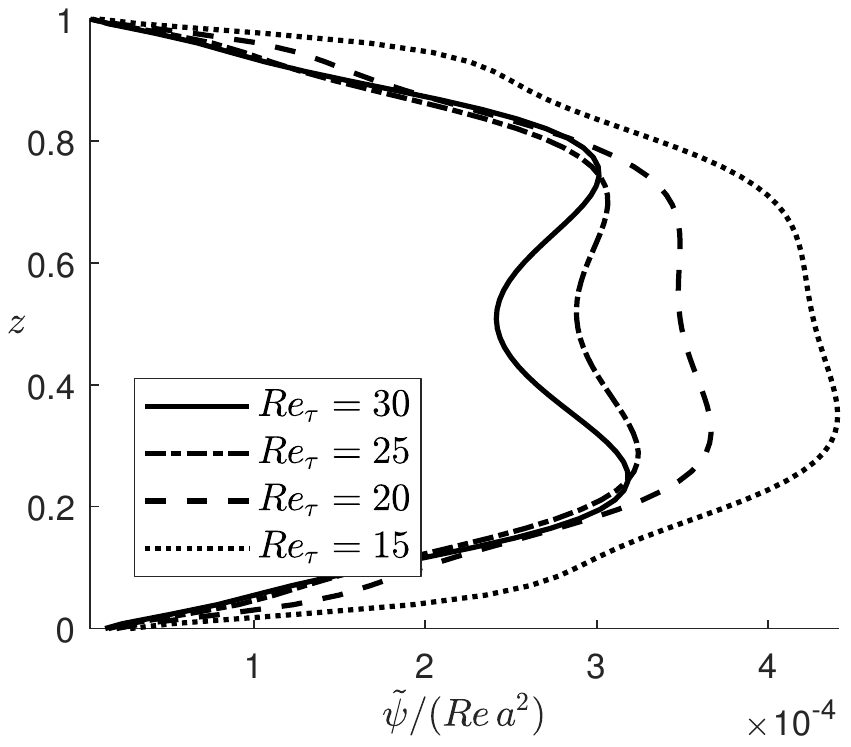}%
}\subfigure[$\thy=\pi/2$, $\etanil= 0.0625$ ]{
\includegraphics[width=.33\columnwidth]{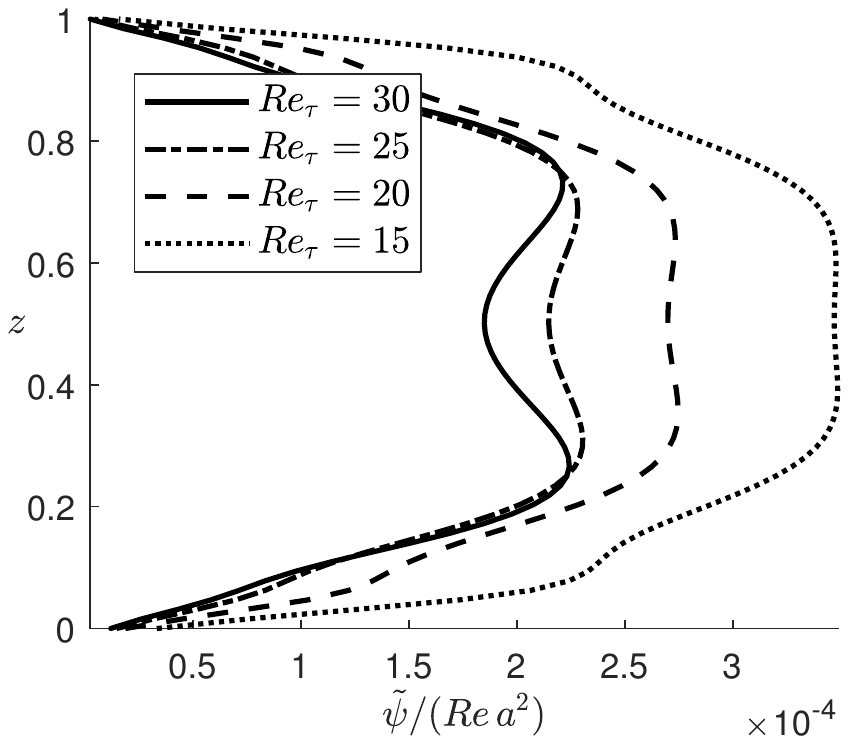}%
}\subfigure[$\thy=\pi/2$, $\etanil= 0.0938$]{
\includegraphics[width=.33\columnwidth]{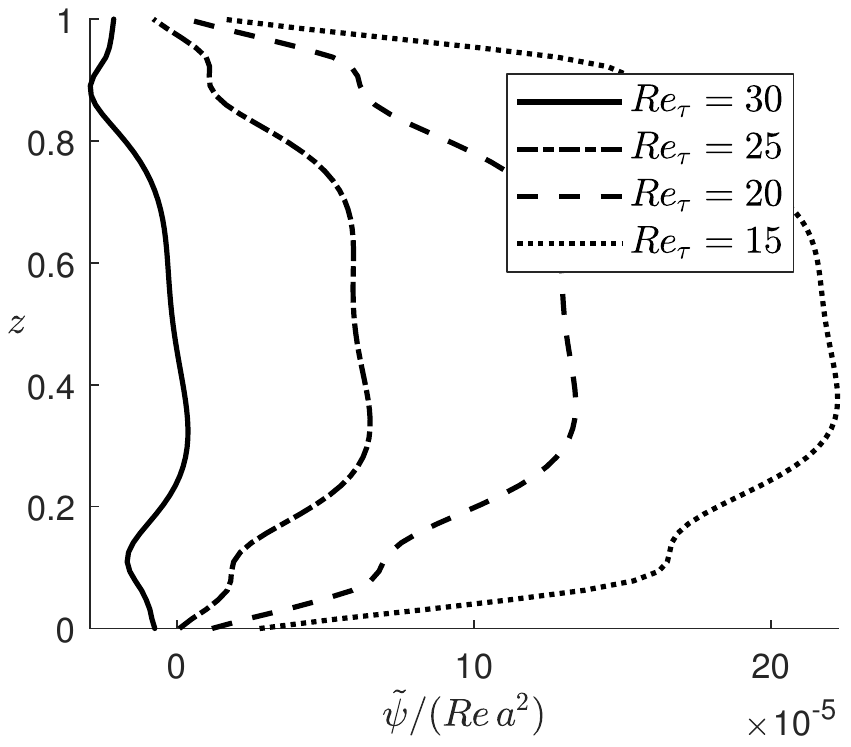}%
}
\caption{
\arev{Double wall-bounded flow.}
Normalised stream functions $\tilde \psi/(\Rey\,\etanil^2)$ estimated from the steady states of lattice--Boltzmann simulations using \eqref{eq:tilde_psi}.
$\knil \approx 2\pi$, $\theta \approx \pi/8$.
}%
\label{fig:LBM_psi}%
\end{figure}

\begin{figure}%
\centering
\subfigure[$\thy=0$, $\etanil = 0.0391$]{
\includegraphics[width=.33\columnwidth]{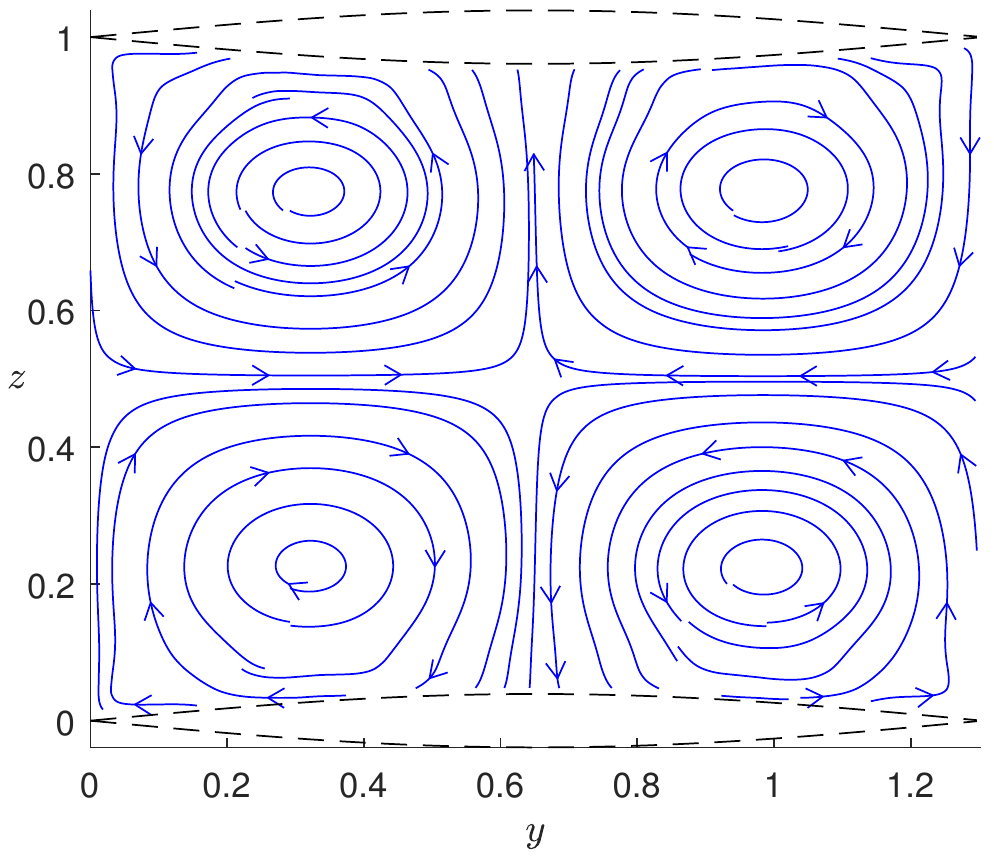}%
}%
\subfigure[$\thy=0$, $\etanil = 0.0625$]{
\includegraphics[width=.33\columnwidth]{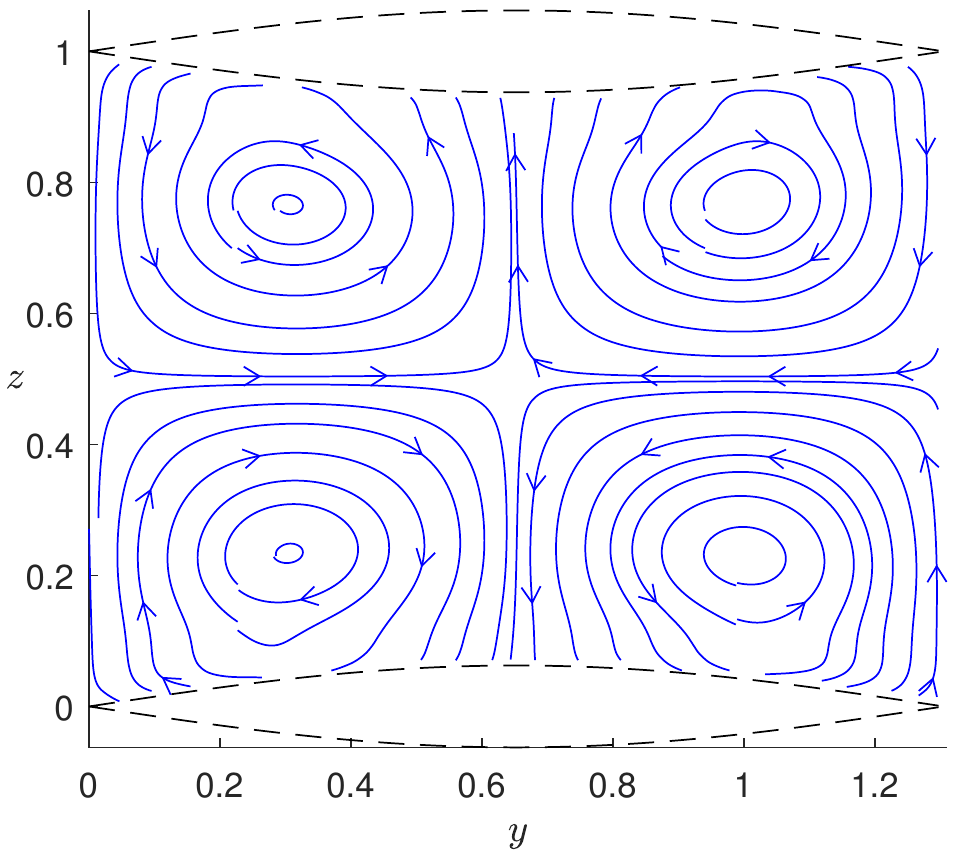}%
}%
\subfigure[$\thy=0$, $\etanil = 0.0938$]{
\includegraphics[width=.33\columnwidth]{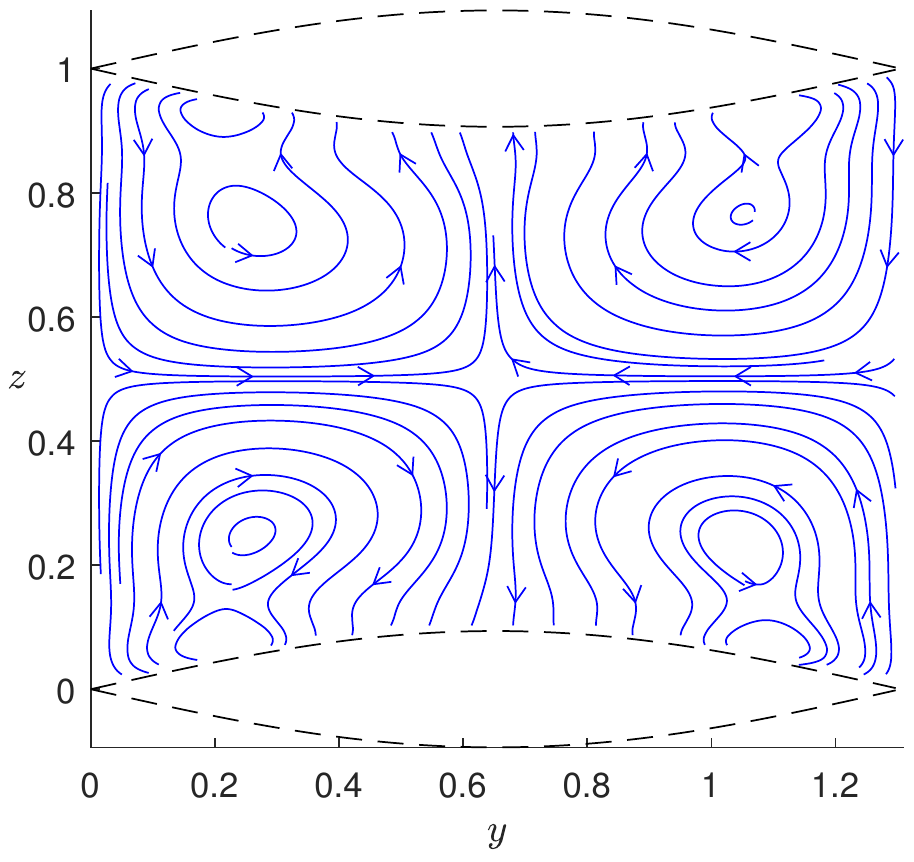}%
}
\\
\subfigure[$\thy=\pi/2$, $\etanil = 0.0391$ ]{
\includegraphics[width=.33\columnwidth]{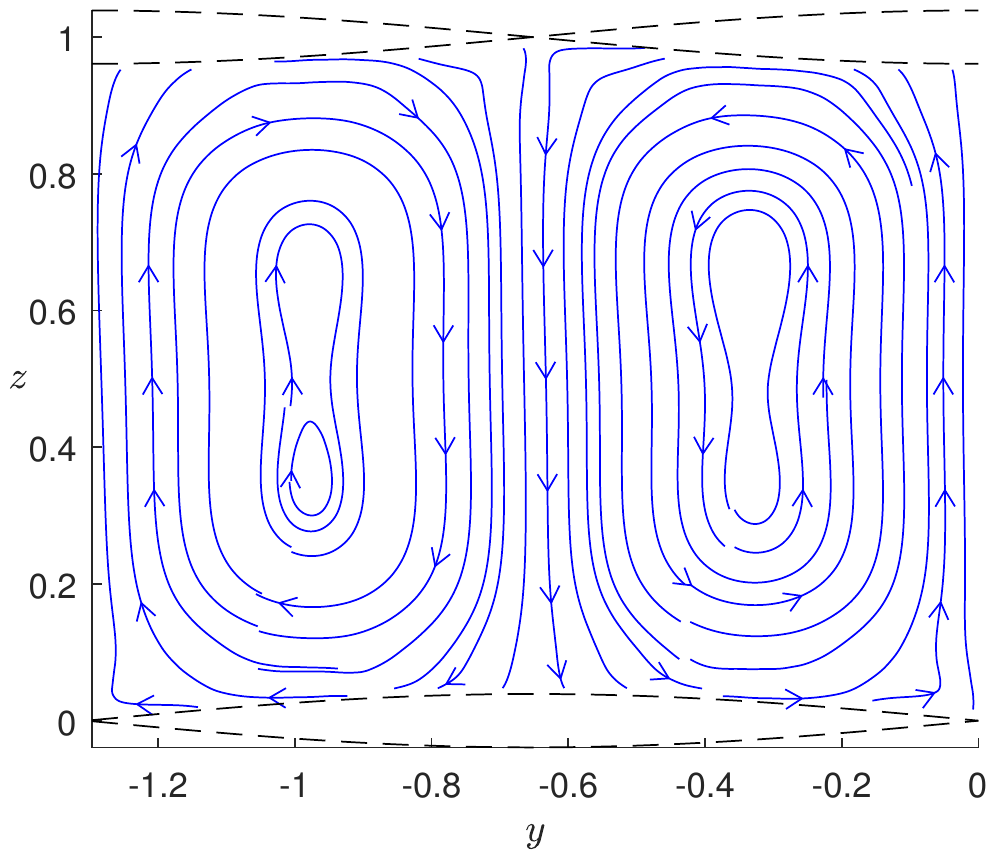}%
}%
\subfigure[$\thy=\pi/2$, $\etanil = 0.0625$ ]{
\includegraphics[width=.33\columnwidth]{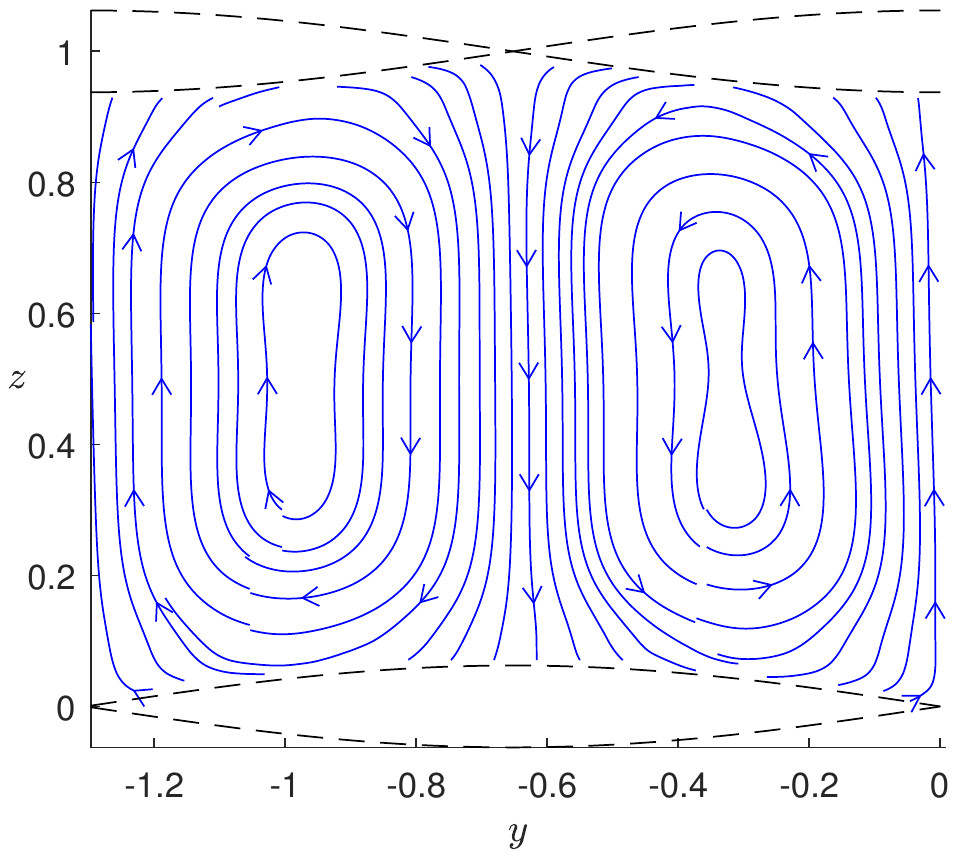}%
}%
\subfigure[$\thy=\pi/2$, $\etanil = 0.0938$]{
\includegraphics[width=.33\columnwidth]{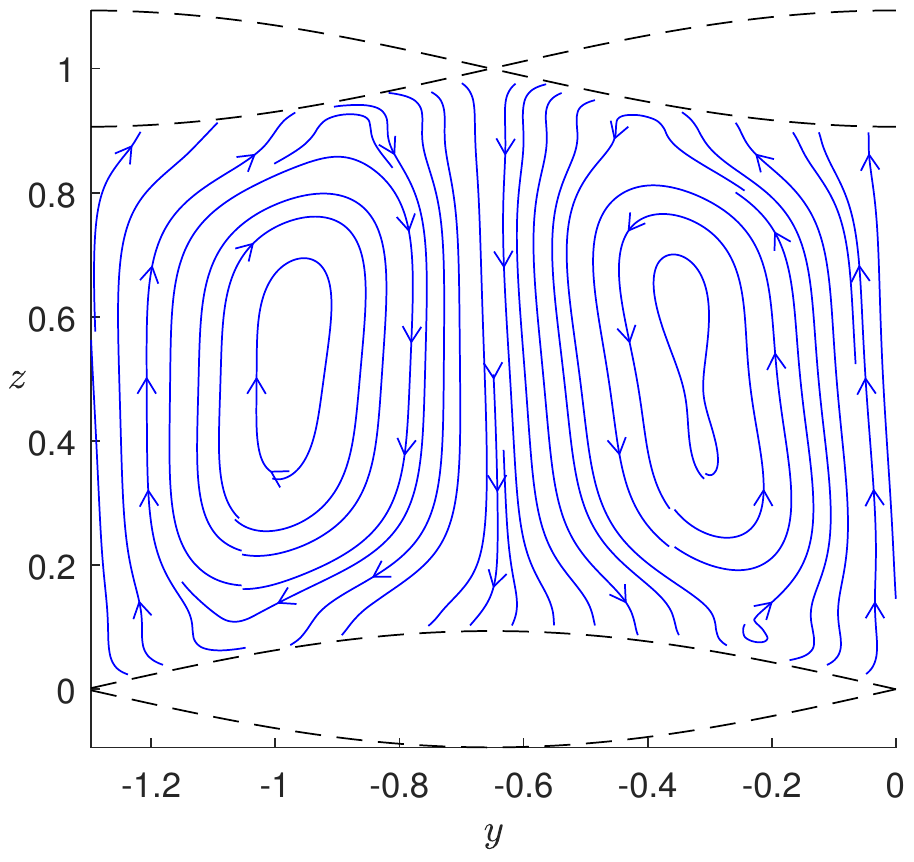}%
}
\caption{
Numerically computed streamlines of the two-dimensional velocity field obtained when averaging along the streamwise dimension.
$Re_\tau = 20$, $\knil \approx 2\pi$, $\theta \approx \pi/8$.
}%
\label{fig:LBM_streamline}%
\end{figure}

\arev{In order to compare the simulated flow field to our theoretical results, we compute an approximation of $\psi$ from the streamwise averaged velocity field based on the principle of volume flux.
Provided the vortices are not skewed ($\thy$ is a right or straight angle),
\begin{equation}
\tilde\psi(z) = \int_{0}^z \!\dd \xi\, \ol{\p v}(y_0,\xi),
\label{eq:tilde_psi}
\end{equation}
approximates $\psi$,
$\ol{\p v}$ being the streamwise mean of the spanwise velocity component.
Spanwise position $y_0$ is the spanwise coordinate of the vortex centre.
We here assume  $y_0=\pi/4\kynil$.
}
Simulation results for a range of wall Reynolds numbers, $\Rey_\tau = (h/2)u_* /\nu$ where $u_*=\sqrt{\tau\_w/\rho}$, are shown in figure~\ref{fig:LBM_psi} and \ref{fig:LBM_streamline}.
The wall Reynolds number is here the appropriately fixed parameter (a measure of the flow forcing) and
relates to $\Rey=h U_0/\nu$ as $\Rey=\Rey_\tau^2$ in the case of  flat-walled, laminar Poiseuille flow.
The figures show, when compared to the viscous, stationary results of figure~\ref{fig:parabolic_delta:theta0}--%
\ref{fig:parabolic_streamlines}, qualitative agreement between simulation and theory in terms of vortex orientation and rotational direction. 
Normalised steady-state quantities $\p\psi/(\Rey\, \etanil^2)$ and $\p u/(\Rey\, \etanil)^2$ are in the theory invariant to the Reynolds number, 
although Reynolds number dependency is evident in the simulation results, 
notably in the $\thy=\pi/2$ case where the value of the Reynolds number affects the degree to which vortices are merged to cover the cross-sections or exist as separate, co-rotating vortex pairs.
\arev{
It should also be noted that, contrary to the  $\thy=0$ topography, the  $\thy=\pi/2$ topography is not antisymmetric about the horizontal plain $z=1/2$.
This generates some minor vertical asymmetry in each vortex due to dynamic effects.
The asymmetry differs between the four vortices in each spanwise period;
we here display the vortex centred at $y=\pi/4\kynil$.
}

\begin{figure}%
\centering
\subfigure[$\Rey_\tau=15$, $\thy=0$]{
\includegraphics[width=.49\columnwidth]{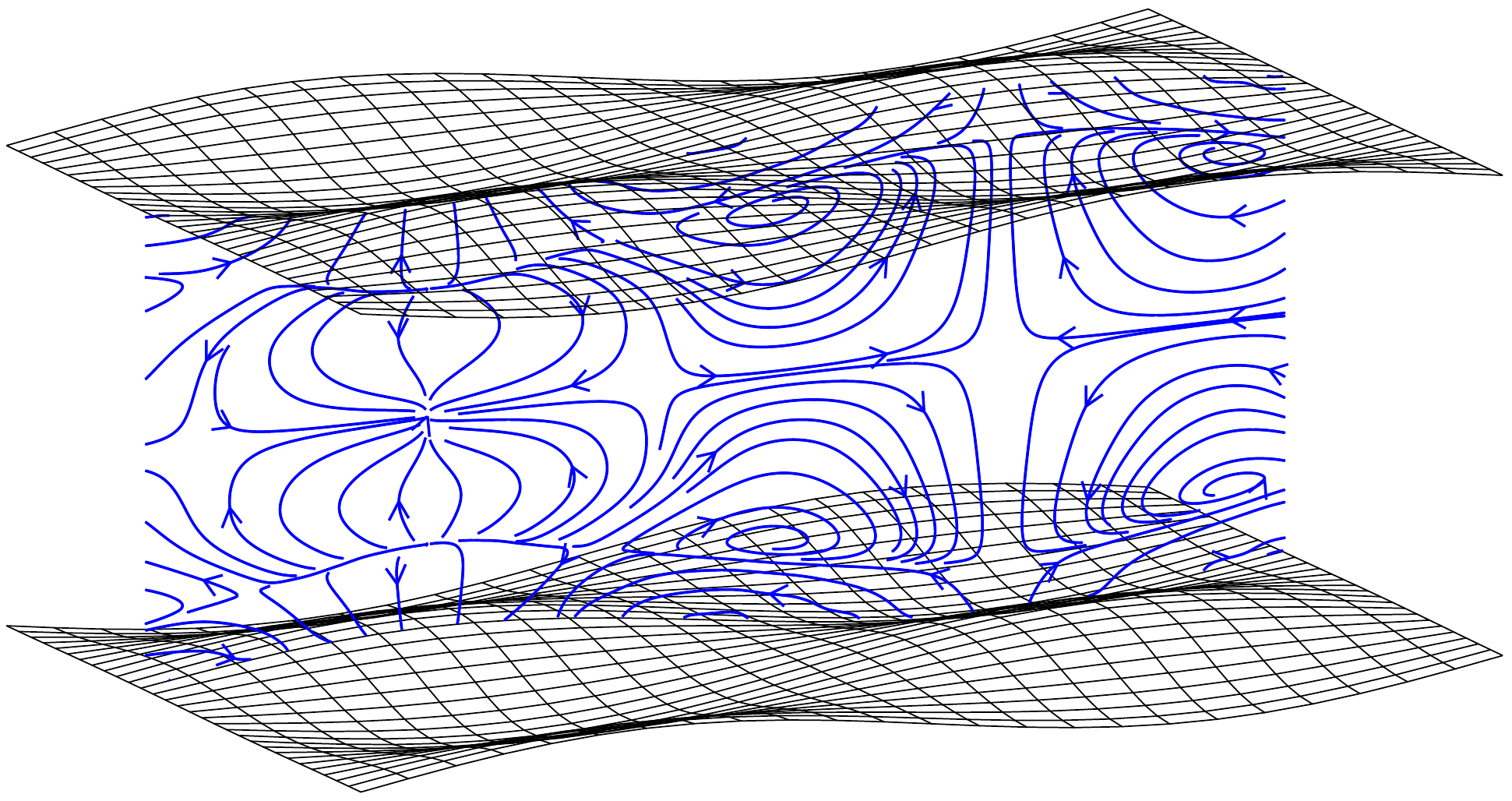}%
}%
\subfigure[$\Rey_\tau=30$, $\thy=0$]{
\includegraphics[width=.49\columnwidth]{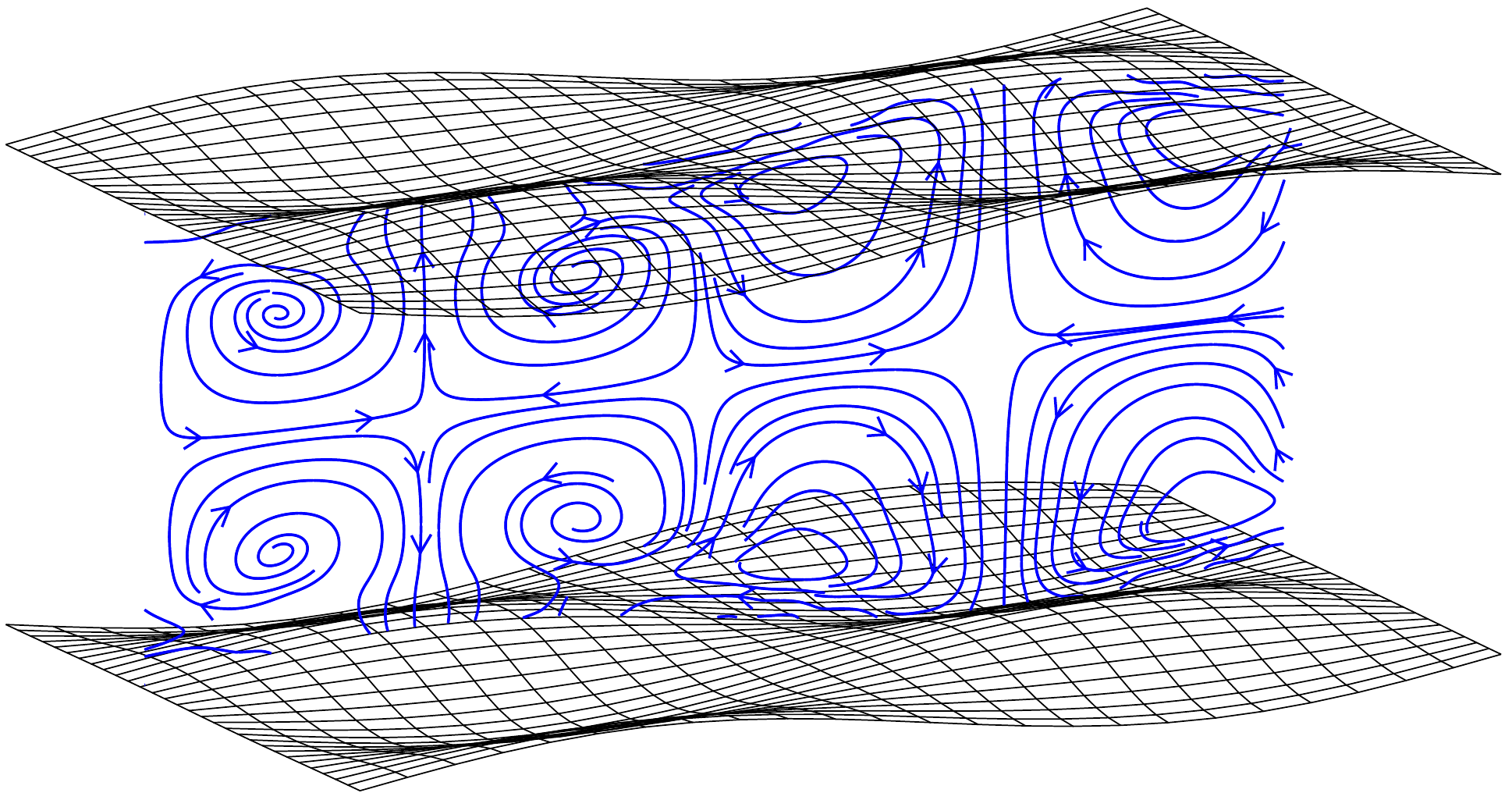}%
}%
\\
\subfigure[$\Rey_\tau=15$, $\thy=\pi/2$]{
\includegraphics[width=.49\columnwidth]{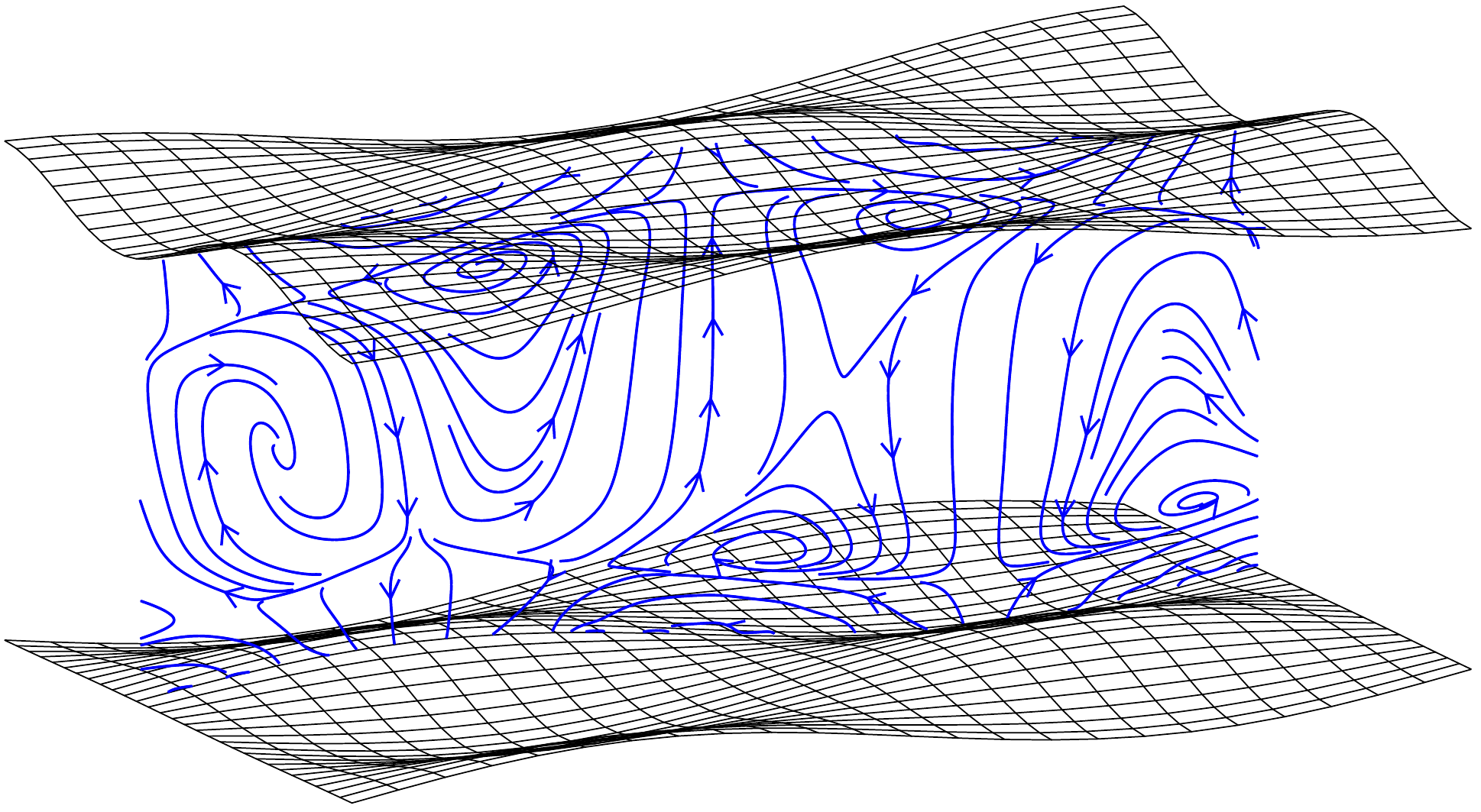}%
}%
\subfigure[$\Rey_\tau=30$, $\thy=\pi/2$]{
\includegraphics[width=.49\columnwidth]{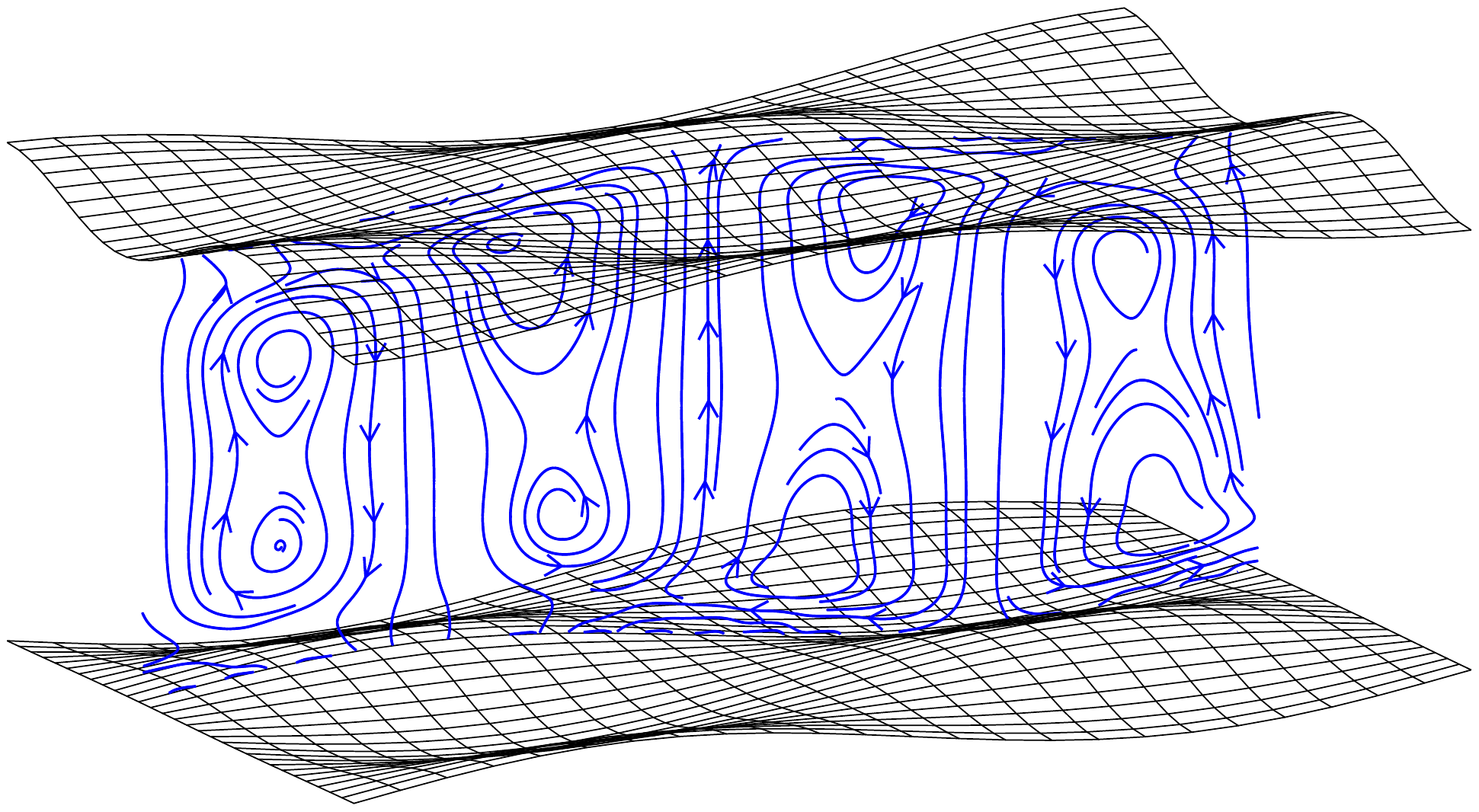}%
}%
\caption{
Streamlines slices of full three-dimensional flow field.
$\etanil\approx 0.0625$, $\knil \approx 2\pi$, $\theta \approx \pi/8$.
Slice at $x=\pi/6\kxnil$.
}%
\label{fig:full_flow_field}%
\end{figure}

\arev{Figure~\ref{fig:full_flow_field} shows slices of the velocity field in the $yz$-plane at a prescribed $x$-location. 
These are not averaged in the streamwise direction and so give an impression of the circulation strength relative to the first-order undulating motion. 
The two types of motion are in the cases studied found to be of the same order of magnitude, with first-order motion dominating near peaks and troughs, and circulation dominating in the intervening regions.
The slices shown in the plots  of figure~\ref{fig:full_flow_field} are at the somewhat arbitrary location $x=\pi/6\kxnil$.  
Circulation intensity relative to the undulating motion increases with increasing Reynolds numbers.
}

Quantitatively, the simulations most closely resemble the no-slip boundary condition cases where the value for $\delta$ is small.
This dependency escalates with increased wall undulation amplitude.
For the sake of briefness and clarity
we have not attempted to adjust the simulation parameters to better accommodate our theoretical construct. 
Indeed, the undulation amplitudes used in the normalisation are the actual amplitudes and not an estimate of the streamsurface amplitudes seen by the flow above some displacement thickness.
(Recall that the circulation is theoretically proportional to these amplitudes squared.)
Likewise, the duct height $h$ is taken as the average distance between the actual walls.
\\

\begin{figure}%
\centering
\subfigure[Theoretical, viscous, no-slip]{
\includegraphics[width=.4\columnwidth]{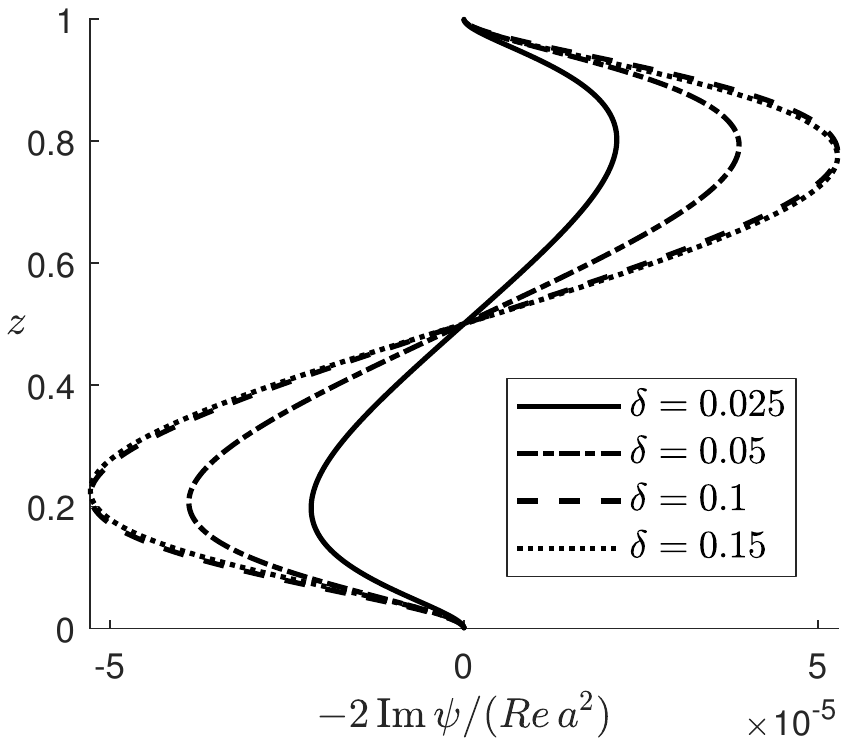}%
}
\subfigure[Simulated, $\etanil = 0.0625$]{
\includegraphics[width=.4\columnwidth]{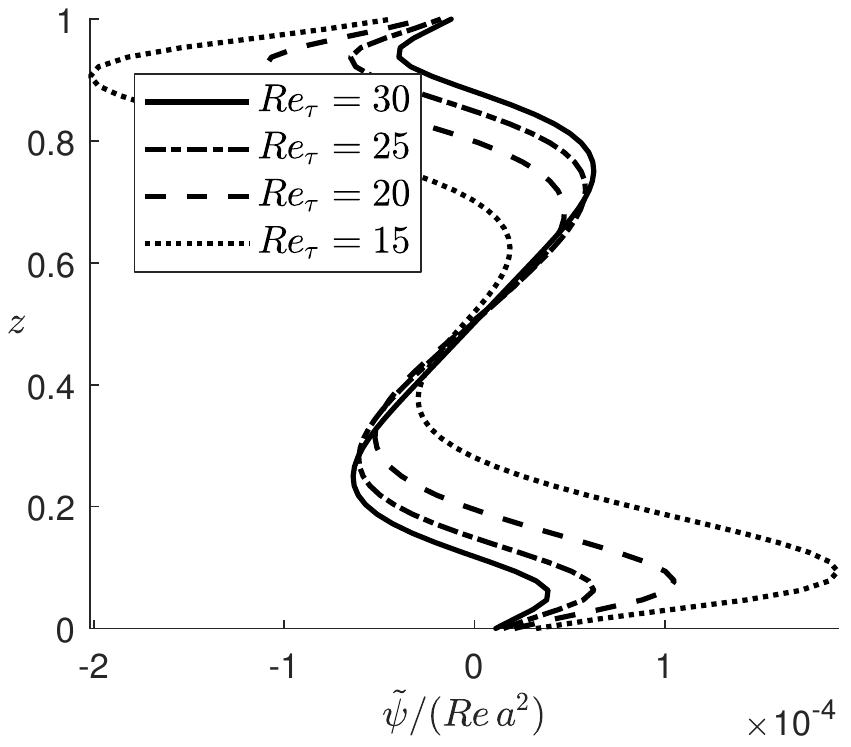}%
}
\caption{
`Rotating the boundary topography'; $\theta = 3\pi/8$.
This will, according to theory, cause rotation in the opposite direction relative to the boundary undulations.
$\knil=\pi$, $\thy = 0$.}%
\label{fig:theta3pi8}%
\end{figure}

Earlier, in section~\ref{sec:results:free_surface}, we made the observation that large values of $\theta$ (i.e., boundary undulations that are stretched out in the streamwise direction rather than the spanwise) can generate vortices which rotate in the opposite direction relative to the boundary topography (pushing fluid away from peaks and troughs as opposed to in towards them).
Simulation results for such a case, where $\theta = 3\pi/8$ and $\knil=\pi$, are presented in figure~\ref{fig:theta3pi8} along with some corresponding theoretical stream functions.
We observe that the rotation is indeed reversed in the bulk interior of the cross-section (compared with e.g. figure~\ref{fig:LBM_psi}), although thin additional   vortices appear close to the walls which still rotate in an unreversed direction. 
This may be an artefact of the dynamic stresses caused by the undulating boundaries which has not been incorporated into our kinematic theory. 
(The only means with which other harmonics are filtered out 
field is through the averaging in the streamwise direction.)
\arev{
Viscosity serves to modify the first-order wave field
and give a spanwise perturbation to the primary flow.
Dynamic effects may generate circulation with its own rotational preference, 
as observed in turbulent flows by \citet{%
anderson_2015_secondary_flow_due_to_roughness_patches%
,chan_2018_pipe_egg_carton_roughness_secondary_flow%
,yang_2018_topography_drive_secondary_flow%
,vanderwel_2015_spacing_of_lego_strips_and_secondary_flwo%
,kevin_2017_secondary_flow_herringbone_riblets%
}.
Presumably, a dynamic rotational effect is always present, either assisting or impeding the Langmuir circulation.
}
\\

Secondary vortex motion, aligned with the flow, can be generated via spanwise intermittent roughness patches \citep{anderson_2015_secondary_flow_due_to_roughness_patches,willingham_2015_secondary_flow_due_to_roughness_patches}
or streamwise-aligned obstacles \citep{yang_2018_topography_drive_secondary_flow,vanderwel_2015_spacing_of_lego_strips_and_secondary_flwo,kevin_2017_secondary_flow_herringbone_riblets,sirovich_1997_drag_reduction_passive_mech_Nature}.
\citet{anderson_2015_secondary_flow_due_to_roughness_patches} demonstrated that these structures are related to Prandtl's secondary flow of the second kind, driven and sustained by spatial gradients in the Reynolds-stress components.

\begin{figure}%
\centering
\includegraphics[width=.75\columnwidth]{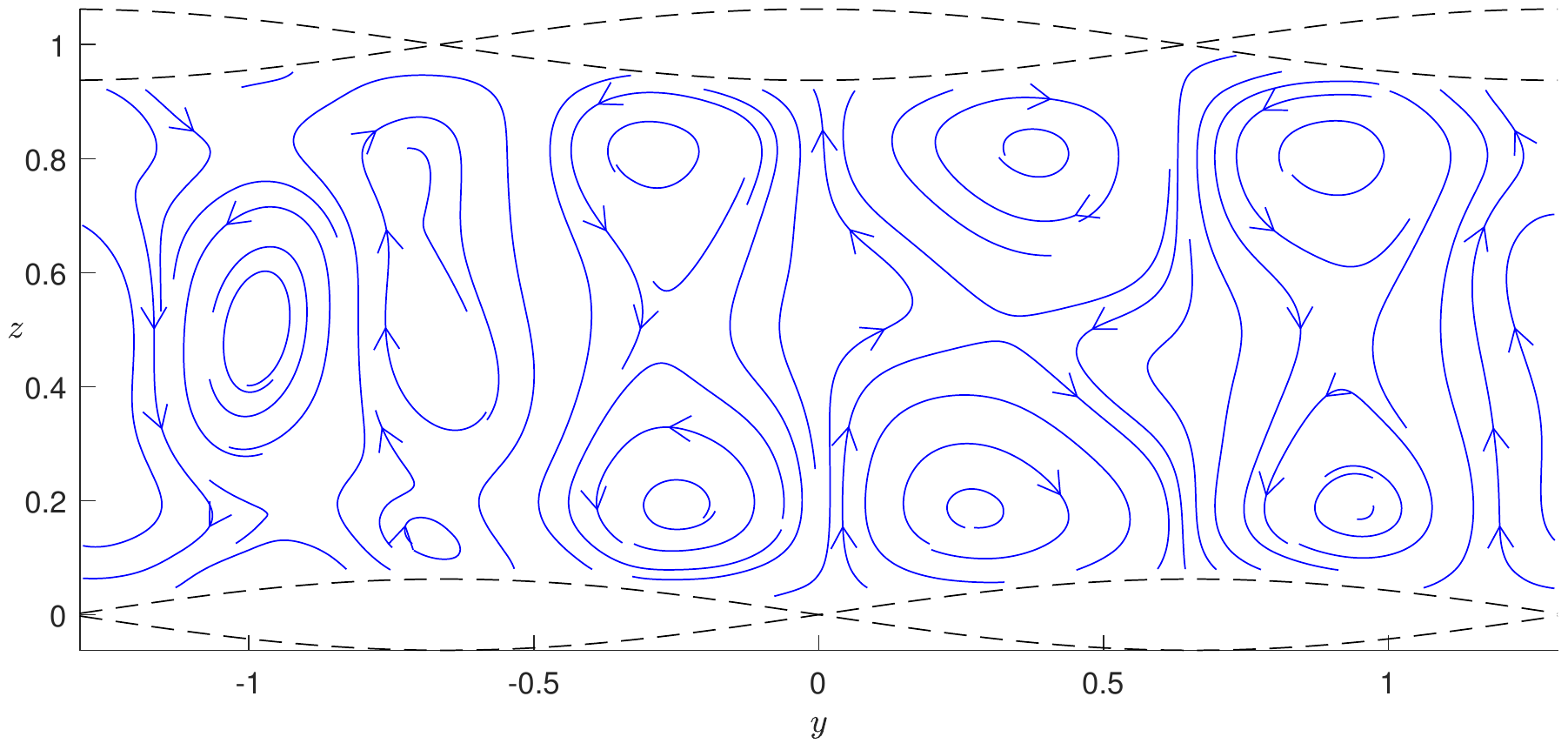}%
\caption{
Unstable vortex field. 
$\Rey_\tau = 40$, $\thy=\pi/2$, $\theta = \pi/8$, $\knil = 2\pi$.
}%
\label{fig:unstable_vortex}%
\end{figure}

Our simulations start showing signs of vortex instability
at wall Reynolds numbers above 30.
An example is shown in figure~\ref{fig:unstable_vortex}.
The degree of instability also increases with increased boundary undulation amplitude and seems more unstable for $\thy=\pi/2$ than $\thy=0$ as the former generate vortices which are more stretched out over the cross-section. 
Vortex instability manifests as the spawning of higher-wavenumber vortices.
These are naturally less steady as the Reynolds number increases.
Longitudinal vortex instability is a well-studied phenomenon since their discovery by \citep{craik_1977_CL2_stability},
and their effect on strongly sheared flows such as those considered here thoroughly investigated by Phillips and co-workers
\citep{phillips_1994_Langmuir_stability,phillips_1996_family_of_zq_CL2_instability,phillips_1996_Langmuir_qoverz_profile,phillips_2014_Langmuir_shallow_water}.
Vortex instability is likely to play a decisive role in the evolution of flows over undulating topographies as we move towards the turbulent regime.

\section{Summary \arev{and closing remarks}}
\label{sec:summary} 
\arev{
We have studied the creation of streamwise vortices in the boundary layer of a shear flow over a wall with a particular topography. 
The roll-like flow structures are closely analogous to Langmuir circulation created through a 
strong shear `CL1'-type mechanism for
the algebraic instability of longitudinal vortices.
Instead of a shear flow beneath the wavy free surface normally associated with Langmuir circulation, the same mechanism can be 
imposed on a flow by 
furnishing walls with a criss-crossing wavy patten.
Under sufficient influence of viscosity,
the flow structures  due to the CL1-instability
}
will stabilise and form coherent longitudinal vortices whose transverse size is matches the wavelength of the chosen wall topography. 
This mode of circulation requires no external forcing other than maintaining the ambient shear flow, but functions by redirecting the vorticity inherently present in the current itself.

Assuming a shear profile of arbitrary form $U(z)$ when unperturbed, 
a resonant wave--current interaction is computed in two limiting cases: transient inviscid circulation and steady-state viscous circulation. 
The vortical motion in the cross-flow plane grows proportional to $a^2 t$ in the early, essentially inviscid stage ($a$: wave-pattern amplitude, $t$: time), and is proportional to $Re \, a^2$ in the latter, viscous steady state. 
The instability is algebraic in nature, growing linearly with 
time in the former case and to the Reynolds number in the latter.  
The streamwise velocity perturbations are respectively proportional to these quantities squared.

After the onset of the instability, at first purely kinematically driven and essentially inviscid, viscosity 
serves to push the vortex centres out away form the undulating boundaries.
Vortex intensity is observed to be greatest 
in the wavelength range where vortex height and width are comparable.
Circulation is also typically strongest when the half-angle $\theta$ between the crossing plane waves making up the sinusoidal wall corrugations
lies in the range $10^\circ<\theta<25^\circ$.
A reversal of the direction of circulation 
is possible within the range  $\theta>45^\circ$, albeit with weaker rotation strength.
Several weak vortices can coexist vertically in the region where this switch of rotational direction takes place.
For a channel flow between two similarly corrugated walls, the
circulation generated near each wall can merge into single vortices spanning the cross-section, provided the two boundaries are suitably shifted relative to each other.
Such a merger is more pronounced the closer the boundaries are situated relative to vortex wavelength.

The investigation has been supplemented with laminar flow simulations using a lattice--Boltzmann method. 
These simulations show that the forced circulation phenomenon does indeed take place also in laminar flows over no-slip boundaries, and that the resulting circulation is qualitatively similar to the results of the perturbation theory 
even though the latter conceptually assumes a current velocity at the walls.
The laminar simulations furthermore show that circulation is quickly established
relative to the time it takes for the primary flow to stabilise.
The circulation intensity is comparable to the intensity of 
the first-order motion.
Dynamic effects, excluded from the theoretical model, were here also observed in near-wall regions. 
Quantitatively, a degree of uncertainty is related to the closure assumption of a displacement height $\delta$ adopted in the theory.

\arev{
The practical potential of the mechanism here studied is as yet unexplored and subject to conjecture. 
Literature, as presented in section~\ref{sec:introduction}, suggests that longitudinal vortices can be beneficial in terms of turbulent drag reduction, 
although the 
divers mechanisms hitherto employed to create such 
vortices differ fundamentally from the Langmuir mechanism suggested herein. 
Therefore, studies of the continued development of CL1-type Langmuir circulation in the turbulent regime are desirable, particularly regarding the stability of the generated structures for increasing Reynolds numbers. 
The optimal scale of descried Langmuir vortices is an open question as vortices could be made to exist 
either as a tiny structure
deep within the boundary layer, or 
as large structures reaching into the bulk of the flow,
depending on the scale of the wall undulations.
As indicated by the literature,  
such vortices could conceivably act to 
stabilise the transition to turbulence or to reduce drag in fully developed turbulence, respectively. 
}

\section*{Declaration of Interests} The authors report no conflict of interest.

\section*{Acknowledgements}
This work was funded by the Research Council of Norway (programme FRINATEK), grant number 249740.

\appendix
\section{Integration constants for the viscous steady state problem}
\label{sec:AB}
Determining the integration coefficients in \eqref{eq:wO2_sol_viscous} subject to the full-slip/full-slip  boundary conditions, $w(0)=w''(0)=w(1)=w''(1)=0$,
yields
\begin{subequations}
\begin{equation}
w(z) = - \frac{2\kappa w\cross(1)- \nabla^2 w\cross(1)\coth\kappa}{2\kappa\sinh\kappa} \sinh \kappa z
-\frac{\nabla^2 w\cross(1)}{2\kappa\sinh\kappa}  z \cosh \kappa z
+  w\cross(z),
\label{eq:coeff_wO2:full_slip}
\end{equation}
where $\nabla^2 w\cross = w\cross''-\kappa^2w\cross$.
We have combined the $\exp(\pm \kappa z)$ terms in 
\eqref{eq:wO2_sol_viscous}
into hyperbolic functions.
Subjected to the no-slip/no-slip conditions, $w(0)=w'(0)=w(1)=w'(1)=0$,
\eqref{eq:wO2_sol_viscous} reads
\begin{align}
w(z) &=\frac{(\kappa \cosh\kappa + \sinh\kappa) \sinh\kappa z
- \kappa z [\kappa \cosh \kappa(1-z) + \sinh \kappa \cosh \kappa z]
}{\kappa^2-\sinh^2 \kappa} w\cross(1)
\nonumber\\&
+\frac{\kappa z \sinh \kappa(1-z) - (1-z)\sinh\kappa\sinh\kappa z
}{\kappa^2-\sinh^2 \kappa}w\cross'(1)
+ w\cross(z).
\label{eq:coeff_wO2:no_slip}
\end{align}
Mixed conditions, with no-slip at the lower boundary and full-slip at the upper, $w(0)=w'(0)=w(1)=w''(1)=0$, gives
\begin{align}
w(z) &=
\frac{
\cosh\kappa\sinh\kappa z - \kappa z \cosh \kappa(1-z)
}{\kappa-\cosh \kappa \sinh\kappa}w\cross(1)
\nonumber\\&
-\frac{\kappa(1-z)\sinh\kappa\sinh\kappa z-\kappa^2z\sinh\kappa(1-z)}
{\kappa-\cosh \kappa \sinh\kappa}\frac{\nabla^2 w\cross(1)}{2\kappa^2}
+ w\cross(z).
\label{eq:coeff_wO2:mixed}
\end{align}%
\label{eq:coeff_wO2}%
\end{subequations}%
Derivatives of $w\cross\of z$ are, from \eqref{eq:wcO2_sol_viscous} and Leibnitz's rule,
\begin{equation}
w\cross^{(n)}\of z = \frac{\Rey}{2\kappa^3 }
\int_0^z \!\dd \zz\, \mc R\of {\zz} \kappa^n G^{(n)}[ \kappa (z-\zz)]
; \qquad[n=0,1,2,3]
\label{eq:dwcross}
\end{equation}
since $G^{(n)}(0) = 0$ for $n\leq3$. Parenthesised superscript is here the derivative order.

Finally, horizontal velocity components \eqref{eq:u_O2} subjected to the full-slip condition $u'(0)=u'(1)=0$, no-slip condition $u(0)=u(1)=0$ or mixed condition $u(0)=u'(1)=0$ get the homogeneous part
\begin{equation}
\sum_\pm d_u^\pm  \rme^{\pm\kappa z} =
-\frac{\Rey}{\kappa}\int_0^1 \!\dd \xi\, U'(\xi) w(\xi)
\times\begin{cases}
\frac{\cosh\kappa(1-\xi)}{\sinh\kappa} \cosh\kappa z;  &[\text{full-slip}],
\\[1.5ex]
\frac{ \sinh\kappa(1-\xi)}{\sinh\kappa} \sinh\kappa z; &[\text{no-slip}],
\\[1.5ex]
\frac{\cosh\kappa(1-\xi)}{\cosh\kappa} \sinh\kappa z; &[\text{mixed}].
\end{cases}
\label{eq:coeff_uO2}
\end{equation}

The spanwise velocity component $v$ is the $z$-derivative of the stream function (in either real or wave space) 
while the pressure can most easily be retrieved from the spanwise momentum, yielding
\begin{equation}
p = -\frac{1}{\kappa^2}
(\pp_t-\Rey^{-1}\nabla^2)w'
- \frac{(\kxnil^2-\kynil^2)\pnil^2+(\pnil')^2}{\kxnil^2U^2}.
\label{eq:p_O2}
\end{equation}
This is an explicit expression with necessary $w$-derivatives given by
\eqref{eq:dwcross}.

\bibliographystyle{jfm}
\bibliography{refs_wave}

\end{document}

%% file: figures/drawing_doublewall.tex
\begin{tikzpicture}[scale=2.25,font=\normalsize] 

\def\q{(1/4)}; 

\def\d{.2} 
\def\yb{.1} 
\def\ys{.13} 
\def\x{.4}

\begin{scope}
\clip (0,-\d) rectangle (2.5,1+\d);

\draw[thick,name path=ground,fill=lightgray] 
(0,-\d)-- (0,0) sin +(\x,\yb) cos +(\x,-\yb) sin +(\x,-\yb) cos +(\x,\yb) 
						sin +(\x,\yb) cos +(\x,-\yb) sin +(\x,-\yb) cos +(\x,\yb)
						-- (2.5,-\d);

\draw[thick,name path=ceil,fill=lightgray]
(-.1,1+\d)--
 (-.1,1-\ys)  cos +(\x,\ys)
sin +(\x,\ys) cos +(\x,-\ys) sin +(\x,-\ys) cos +(\x,\ys) 
						sin +(\x,\ys) cos +(\x,-\ys) sin +(\x,-\ys) cos +(\x,\ys)
						sin +(\x,\ys) cos +(\x,-\ys) sin +(\x,-\ys) cos +(\x,\ys)
						--(3,1+\d);
						
\draw[dashed] (0,0)-- (3,0);		
\draw[dashed] (0,1)-- (3,1);			
						
\end{scope}


\draw[name path global=U,-,domain=0:1,samples=100,mylabel=at 0.5  right with {$~U(z)$}] plot  ({4*(1-\x)*\x},{(1+2*\d)*\x-\d});
\foreach \z in {0.2,0.3,...,.9}{
\draw[red,->] (0,{(1+2*\d)*\z-\d})--({4*(1-\z)*\z},{(1+2*\d)*\z-\d});
};

\draw[->] (0,0)-- (1.0,0) node[above]  {$x$};
\draw[->] (0,0)-- (0,1.25) node[left]  {$z$};

\draw[-,font=\small] (0,0)-- (-.05,0) node[left]  {$0$};
\draw[-,font=\small] (0,1)-- (-.05,1) node[left]  {$1$};

\draw[<->] (1.25,-\d) -- (1.25,0) node[right, midway] {$\delta$};
\draw[<->] (1.25,1) -- (1.25,1+\d) node[right, midway] {$\delta$};
\draw[<->] (1.9,0) -- (1.9,\yb) node[right, midway] {$\petab$};
\draw[<->] (2.2,1) -- (2.2,1+\ys) node[right, midway] {$\petas$};


\end{tikzpicture}


%% file: figures/drawing.tex
\begin{tikzpicture}[scale=2.25,font=\normalsize] 

\def\q{(1/4)}; 

\def\d{.2} 
\def\yb{.1} 
\def\ys{.13} 
\def\x{.4}

\begin{scope}
\clip (0,-\d) rectangle (2.5,1+\d);

\draw[thick,name path=ground,fill=lightgray] 
(0,-\d)-- (0,0) sin +(\x,\yb) cos +(\x,-\yb) sin +(\x,-\yb) cos +(\x,\yb) 
						sin +(\x,\yb) cos +(\x,-\yb) sin +(\x,-\yb) cos +(\x,\yb)
						-- (2.5,-\d);

\draw[thick]
(0,1)
sin +(\x,\ys) cos +(\x,-\ys) sin +(\x,-\ys) cos +(\x,\ys) 
						sin +(\x,\ys) cos +(\x,-\ys) sin +(\x,-\ys) cos +(\x,\ys);

\draw[dashed] (0,0)-- (3,0);		
\draw[dashed] (0,1)-- (3,1);						
				
\end{scope}

\draw[name path global=U,-,domain=0:1,samples=100,mylabel=at 0.6 above right with {$~U(z)$}] plot  ({(\x)^\q},{-\d+(1+\d)*\x});
\foreach \z in {0.2,0.3,...,1.1}{
\draw[red,->] (0,{-\d+(1+\d)*\z})--({\z^\q},{-\d+(1+\d)*\z});};

\draw[<->] (1.25,-\d) -- (1.25,0) node[right, midway] {$\delta$};
\draw[<->] (1.9,0) -- (1.9,\yb) node[right, midway] {$\petab$};
\draw[<->] (1.9,1) -- (1.9,1+\ys) node[right, midway] {$\petas$};

\draw[->] (0,0)-- (1.0,0) node[above]  {$x$};
\draw[->] (0,-\d)-- (0,1+.25) node[left]  {$z$};

\draw[-,font=\small] (0,0)-- (-.05,0) node[left]  {$0$};
\draw[-,font=\small] (0,1)-- (-.05,1) node[left]  {$1$};


\begin{scope}[shift={({4*\x},1)}]
\begin{scope}[scale=.075,rotate=+25]
\draw[-] (0,0)--({1/sqrt(2)},1)--({-1/sqrt(2)},1)--(0,0);
\draw[-] (-1,0)--(1,0);
\draw[-] (-.75,-.2)--(.75,-.2);
\draw[-] (-.5,-.4)--(.5,-.4);
\draw[-] (-.25,-.6)--(.25,-.6);
\end{scope}
\end{scope}

\end{tikzpicture}

%% file: figures/drawing_wallshift.tex
\begin{tikzpicture}[scale=1.75,font=\normalsize] 
\def\y{.1} 
\def\x{.4}
\def\d{.1}
\def\H{.75}
\def\yz{.4}

\draw[<->]  (0-2.5,\yz-.2) node[above]{$z$}--(0-2.5,0-.2)--  (\yz-2.5,0-.2)node[right]{$y$};

\begin{scope}[shift={(-2.5,0)}]
\clip (.1,-\d) rectangle (2.4,\H+\d);

\draw[thick,fill=lightgray] 
(-0,-\d)-- (0,0) sin +(\x,\y) cos +(\x,-\y) sin +(\x,-\y) cos +(\x,\y) 
						sin +(\x,\y) cos +(\x,-\y) sin +(\x,-\y) cos +(\x,\y)
						-- (3.5,-\d);
\draw[thick,dashed] 
(-0,-\d)-- (0,0) sin +(\x,-\y) cos +(\x,+\y) sin +(\x,+\y) cos +(\x,-\y) 
						sin +(\x,-\y) cos +(\x,+\y) sin +(\x,+\y) cos +(\x,-\y)
						-- (3.5,-\d);

\draw[thick,fill=lightgray] 
(0,\H+\d)-- (0,\H) sin +(\x,-\y) cos +(\x,+\y) sin +(\x,+\y) cos +(\x,-\y) 
						sin +(\x,-\y) cos +(\x,+\y) sin +(\x,+\y) cos +(\x,-\y)
						-- (3.5,\H+\d);
						
\draw[thick,dashed] 
(0,\H+\d)-- (0,\H) sin +(\x,+\y) cos +(\x,-\y) sin +(\x,-\y) cos +(\x,+\y) 
						sin +(\x,+\y) cos +(\x,-\y) sin +(\x,-\y) cos +(\x,+\y)
						-- (3.5,\H+\d);

\node[right] at (.1,\H/2) {$\vartheta=0$};								
\end{scope}

\begin{scope}[]
\clip (.1,-\d) rectangle (2.4,\H+\d);

\draw[thick,fill=lightgray] 
(-0,-\d)-- (0,0) sin +(\x,\y) cos +(\x,-\y) sin +(\x,-\y) cos +(\x,\y) 
						sin +(\x,\y) cos +(\x,-\y) sin +(\x,-\y) cos +(\x,\y)
						-- (3.5,-\d);
\draw[thick,dashed] 
(-0,-\d)-- (0,0) sin +(\x,-\y) cos +(\x,+\y) sin +(\x,+\y) cos +(\x,-\y) 
						sin +(\x,-\y) cos +(\x,+\y) sin +(\x,+\y) cos +(\x,-\y)
						-- (3.5,-\d);

\draw[thick,fill=lightgray] 
(0,\H+\d)-- (0,\H-\d)  cos +(\x,+\y) sin +(\x,+\y) cos +(\x,-\y) 
						sin +(\x,-\y) cos +(\x,+\y) sin +(\x,+\y) cos +(\x,-\y)
						-- (3.5,\H+\d);
						
\draw[thick,dashed] 
(0,\H+\d)-- (0,\H+\d) cos +(\x,-\y) sin +(\x,-\y) cos +(\x,+\y) 
						sin +(\x,+\y) cos +(\x,-\y) sin +(\x,-\y) cos +(\x,+\y)
						-- (3.5,\H+\d);

\node[right] at (.1,\H/2) {$\vartheta=\pi/2$};								
\end{scope}

\begin{scope}[shift={(2.5,0)}]
\clip (.1,-\d) rectangle (2.4,\H+\d);

\draw[thick,fill=lightgray] 
(-0,-\d)-- (0,0) sin +(\x,\y) cos +(\x,-\y) sin +(\x,-\y) cos +(\x,\y) 
						sin +(\x,\y) cos +(\x,-\y) sin +(\x,-\y) cos +(\x,\y)
						-- (3.5,-\d);
\draw[thick,dashed] 
(-0,-\d)-- (0,0) sin +(\x,-\y) cos +(\x,+\y) sin +(\x,+\y) cos +(\x,-\y) 
						sin +(\x,-\y) cos +(\x,+\y) sin +(\x,+\y) cos +(\x,-\y)
						-- (3.5,-\d);

\draw[thick,fill=lightgray] 
(0,\H+\d)-- (0,\H) sin +(\x,\y) cos +(\x,-\y) sin +(\x,-\y) cos +(\x,\y) 
						sin +(\x,\y) cos +(\x,-\y) sin +(\x,-\y) cos +(\x,\y)
						-- (3.5,\H+\d);
						
\draw[thick,dashed] 
(0,\H+\d)-- (0,\H) sin +(\x,-\y) cos +(\x,+\y) sin +(\x,+\y) cos +(\x,-\y) 
						sin +(\x,-\y) cos +(\x,+\y) sin +(\x,+\y) cos +(\x,-\y)
						-- (3.5,\H+\d);

\node[right] at (.1,\H/2) {$\vartheta=\pi$};								
\end{scope}

\end{tikzpicture}

%% file: figures/drawing_real_life.tex
\begin{tikzpicture}[scale=3,font=\normalsize] 

\def\x{.4}
\def\rot{atan(\y/\x)}

\begin{scope}[]
\clip (1,-\y-.025) rectangle (2.5,.5);

\draw[thick,fill=lightgray] 
(-0,-2*\y)-- (0,0) sin +(\x,\y) cos +(\x,-\y) sin +(\x,-\y) cos +(\x,\y) 
						sin +(\x,\y) cos +(\x,-\y) sin +(\x,-\y) cos +(\x,\y)
						-- (3.5,-2*\y);
\draw[thick,dashed] 
(0,\d) sin +(\x,\y) cos +(\x,-\y) sin +(\x,-\y) cos +(\x,\y) 
						sin +(\x,\y) cos +(\x,-\y) sin +(\x,-\y) cos +(\x,\y)
						-- (3.5,-\d);

\begin{scope}[shift={({4*\x},0)}]
\begin{scope}[scale=1,rotate=\rot] 

\draw[<->] (0,0) -- (0,\d) node[rotate=\rot,midway,left]{$\delta$};

\def\stretchx{.75}
\def\stretchy{1.5}
\draw[name path global=U,-,domain=0:.4,samples=100] plot  ({\stretchx*4*(1-\x)*\x},{\stretchy*(1+2*\d)*\x});
\foreach \z in {0.05,0.1,...,.4}{
\draw[red,->] (0,{\stretchy*(1+2*\d)*\z})--({\stretchx*4*(1-\z)*\z},{\stretchy*(1+2*\d)*\z});
};

\draw[dotted](0,0)--(0,1); 

\end{scope}
\end{scope}

\end{scope}

\end{tikzpicture}

%% file: figures/drawing_transformed_flat.tex
\begin{tikzpicture}[scale=3,font=\normalsize] 

\def\x{.4}

\begin{scope}[]
\clip (1,-\y-.025) rectangle (2.5,.5);

\draw[thick,dashed,fill=lightgray] 
(-0,-2*\d)-- (0,0) -- (3.5,0)-- (3.5,-2*\d);
\draw[solid] 
(0,\d) -- (3.5,\d);

\draw[thick,dotted] 
(0,\d) sin +(\x,\y) cos +(\x,-\y) sin +(\x,-\y) cos +(\x,\y) 
						sin +(\x,\y) cos +(\x,-\y) sin +(\x,-\y) cos +(\x,\y);

\begin{scope}[shift={({4*\x},0)}]
\draw[<->] (0,0) -- (0,\d) node[midway,left]{$\delta$};

\def\stretchx{.75}
\def\stretchy{1.5}
\draw[name path global=U,-,domain=0:.4,samples=100,mylabel=at 0.5  right with {$~U(z)$}] plot  ({\stretchx*4*(1-\x)*\x},{\stretchy*(1+2*\d)*\x});
\foreach \z in {0.05,0.1,...,.4}{
\draw[red,->] (0,{\stretchy*(1+2*\d)*\z})--({\stretchx*4*(1-\z)*\z},{\stretchy*(1+2*\d)*\z});
};

\draw[<->,thick] (0,\d+.3) node[left]{$z$} -- (0,\d) -- (.3,\d) node[below]{$x$};

\end{scope}

\end{scope}

\end{tikzpicture}